\documentclass[12pt]{article}
\input epsf
\newcommand{\Poin}{Poincar{\'e}}
\def\Im{\mathop{\rm Im}\nolimits}
\def\Re{\mathop{\rm Re}\nolimits}

\def\Tr{{\rm Tr}\hskip 1pt}
 \newcommand{\ft}[2]{{\textstyle\frac{#1}{#2}}}
 \newcommand{\Ka}{K{\"a}hler}
\newsavebox{\zzzbar}
\sbox{\zzzbar}
  {\setlength{\unitlength}{0.9em}
  \begin{picture}(0.6,0.7)
  \thinlines
  \put(0,0){\line(1,0){0.6}}
  \put(0,0.75){\line(1,0){0.575}}
  \multiput(0,0)(0.0125,0.025){30}{\rule{0.3pt}{0.3pt}}
  \multiput(0.2,0)(0.0125,0.025){30}{\rule{0.3pt}{0.3pt}}
  \put(0,0.75){\line(0,-1){0.15}}
  \put(0.015,0.75){\line(0,-1){0.1}}
  \put(0.03,0.75){\line(0,-1){0.075}}
  \put(0.045,0.75){\line(0,-1){0.05}}
  \put(0.05,0.75){\line(0,-1){0.025}}
  \put(0.6,0){\line(0,1){0.15}}
  \put(0.585,0){\line(0,1){0.1}}
  \put(0.57,0){\line(0,1){0.075}}
  \put(0.555,0){\line(0,1){0.05}}
  \put(0.55,0){\line(0,1){0.025}}
  \end{picture}}
\newcommand{\Zbar}{\mathord{\!{\usebox{\zzzbar}}}}
\def\unity{{\mathchoice {\mathrm{1\mskip-4mu l}} {\mathrm{ 1\mskip-4mu l}}
{\mathrm{ 1\mskip-4.5mu l}} {\mathrm{ 1\mskip-5mu l}}}}
\newcommand{\rmi}{{\rm i}}
\newcommand{\rmd}{{\rm d}}
\newcommand{\rme}{{\rm e}}
 \newcommand{\Nphi}{{\cal N} }
 \newcommand{\Omchi}{\Omega }
\newcommand{\mxi}{m}
\newcommand{\nxi}{n}
\newcommand{\igk}{\rmi\vec\gamma \cdot \vec k}
\newcommand{\gR}[2]{P_Rg^{-1}{} _#1^#2}
\newcommand{\gL}[2]{P_Lg^{-1}{} _#1^#2}
\newcommand{\gI}[2]{g^{-1}{} _#1^#2}
\newcommand{\Yrho}{Y}

\csname @addtoreset\endcsname{equation}{section}
\begin{document}
\begin{titlepage}
\begin{flushright}
CERN-TH/2000-118 \\ KUL-TF-2000/15\\ CITA-2000-12\\ hep-th/0006179
\end{flushright}
\vspace{.5cm}
\begin{center}
\baselineskip=16pt {\LARGE    Superconformal Symmetry, Supergravity and Cosmology}\\
\vfill

\

{\large Renata Kallosh,$^{1,\dagger}$ Lev Kofman,$^2$\\[3mm] Andrei
Linde,$^{1,\dagger}$
and Antoine Van Proeyen $^{3,\ddag}$ } \\
\vfill

\

{\small $^1$ Theory Division, CERN, CH-1211 Gen{\`e}ve 23, Switzerland
\\ \vspace{6pt}
$^2$ CITA, University of Toronto, 60 St George Str, Toronto, ON M5S
3H8, Canada
\\ \vspace{6pt}
${}^3$ Instituut voor Theoretische Fysica, Katholieke
 Universiteit Leuven,\\
Celestijnenlaan 200D B-3001 Leuven, Belgium }
\end{center}
\vfill
\begin{center}
{\bf Abstract}
\end{center}
{\small We introduce the general  $N=1$ gauge theory superconformally
coupled to supergravity. The theory has  local $SU(2,2|1)$ symmetry and
no dimensional parameters. The superconformal origin of the
Fayet--Iliopoulos (FI) terms is clarified. The phase of this theory with
spontaneously broken conformal symmetry gives various formulations of
$N=1$ supergravity interacting with matter, depending on the choice of
the $R$-symmetry fixing.

 We have found that the locally superconformal theory is useful
for describing the physics of the early universe with a conformally flat
FRW metric. Few applications of superconformal theory to cosmology
include the study of \textit{i)}  particle production after inflation,
particularly the nonconformal helicity-$\ft12$ states of
gravitino,
\textit{ii)} the super-Higgs effect in cosmology and  the derivation of
the equations for the gravitino interacting with any number of chiral and vector multiplets
in the gravitational background with varying scalar fields, \textit{ iii)}
the weak-coupling limit of supergravity $M_P\rightarrow \infty$ and
gravitino--goldstino equivalence. This explains why gravitino production
in the early universe is not suppressed in the limit of weak
gravitational coupling.

We discuss the possible existence of an unbroken phase of the
superconformal theories, interpreted as a strong-coupling limit of
supergravity $M_P\rightarrow 0$.

 } \vspace{2mm} \vfill \hrule width 3.cm
{\footnotesize
\noindent $^\dagger$ On leave of
absence from Stanford University until 1 September 2000\\
 \noindent $^\ddag$ Onderzoeksdirecteur, FWO, Belgium
}
\end{titlepage}
\tableofcontents{}

\parskip 5pt
\section{Introduction}

Over the last few years  M-theory and string theory have focused mainly
on the superconformal theories and adS/CFT (anti-de~Sitter/conformal
field theory) correspondence \cite{malda}. In particular, IIB string
theory on $adS_5\times S^5$ is related to $SU(2,2|4)$ superconformal
symmetry. The relation between various anti-de~Sitter compactifications of
 M-theory and string theory and the relevant superconformal symmetries were
 described in detail in \cite{CKV}. In particular, one finds the $SU(2,2|1)$
superconformal algebra from the anti-de~Sitter compactification of the
string theory with ${1\over 4}$ of the unbroken supersymmetry. These
recent developments in M-theory and non-perturbative string theory
suggest that we should take a \emph{fresh look at the superconformal
formulation underlying the supergravity}. One of the basic features of
superconformal theory is that generically it treats the gravitational
coupling $M_P^2= \kappa^{-2}$ as some function of a set of scalars
$\kappa^{-2} \Rightarrow-\ft13 {\cal N}(X, \bar X)$, just like in string
theory the string coupling is given by a dilaton field, $g_s^{-2}=
\rme^{\phi}$. In the superconformal action the term $\ft16 {\cal N}(X,
\bar X)  R$ replaces the standard Einstein term $- \ft12 M_P^2 R$ of
supergravity.

The function of scalars, ${\cal N}(X, \bar X)$, codifies the K{\"a}hler potential.
 When the theory is in a Higgs
phase, $-\ft13 {\cal N}(X, \bar X) = M_P^2 + f(X, \bar X)$  and the
function $f(X, \bar X)$ can be gauged away using local conformal symmetry:
 the superconformal symmetry is
broken spontaneously and  supergravity with the dimensional parameter
$M_P$ is recovered.

In addition to the state with the spontaneously broken superconformal
symmetry, one  may   also speculate about the {\it unbroken phase of the
superconformal theory} where  the function ${\cal N}(X, \bar X)$ has a
vanishing vacuum expectation value and there are no dimensional parameters
in the theory. This phase of the superconformal theory can be considered
as the strong-coupling limit of supergravity $M_P^2 \to 0$. In such a
limit the theory may be completely different from classical supergravity,
which represents the weak-coupling limit of the Higgs phase of the
superconformal theory. Even though we do not have a clear constructive
approach to this phase of the theory at the moment, we do have  some
distinct   examples of the configurations, analogous to cosmic strings,
where at the core of the string ${\cal N}(X, \bar X)=0$.  We consider
this as an indication that the superconformal theory, in addition to the
Higgs phase where it is equivalent to supergravity, may have an unbroken
phase related to the strong-coupling limit of supergravity.

At present the low-energy phenomenology is described by $N=1$, $d=4$
supergravity \cite{supergravity,N1YMmsg,PvNPR,general}. Some preferable
choices of the \Ka\  potentials, superpotentials and Yang--Mills
couplings hopefully will be selected at the level of the  fundamental
theory. Until the fundamental theory of all interactions is well
understood, one may try to address the issues of particle physics and
cosmology in the context of the most general phenomenological $N=1$
supergravity--Yang--Mills--matter theory \cite{general}. This, in fact,
has been the case over the last almost 20 years. One can find the
Lagrangian describing this theory in many textbooks and review papers
\cite{Gates:1983nr,Nilles,BW,Binetruy,BL}.

The phenomenological $N=1 $ supergravity--Yang--Mills--matter theory
 was  derived in \cite{general,KugoUe} using local $SU(2,2|1)$ superconformal symmetry
only as a tool, within the framework of
\cite{SU221,superconformN1,conformforPoin}. The transition from the
superconformal supergravity to the usual Poincar{\'e} supergravity occurs
after the conformal compensator field (which we will call the conformon)
becomes fixed. Until very recently, the opinion was that the only role of
superconformal theory is to provide the tensor calculus to derive
supergravity, and that this theory has no interesting physics in it.
Therefore, the textbook description of  the phenomenological $N=1$
supergravity--Yang--Mills--matter theory \cite{Nilles,BW,Binetruy,BL},
with the notable exception of \cite{Gates:1983nr}, practically skips the
superconformal formulation of this theory.

  $N=2$  supergravity--Yang--Mills--matter theory with special
geometry \cite{dWVPspec,special,whatspk} was constructed later than the
$N=1$ theory. This theory has been used extensively during recent years,
in particular with application to BPS black holes \cite{FKS}. It is
associated with low-energy string theory compactified on Calabi--Yau
manifolds, and instead of functions one has to define sections of
appropriate line bundles over the  \Ka\  manifold. This construction has
various advantages, in particular it treats all scalars in the
superconformal version of the theory on an equal footing. This means that
there is no distinction between the physical scalars and the conformal
compensator scalar before gauge-fixing of conformal symmetry is
performed. All available formulations of $N=1$ theory make a particular
choice for the compensator even before conformal symmetry is gauge-fixed
\cite{general,KugoUe,Rel}. We will find that some features of $N=2$
theory may be implemented into $N=1$ theory, which may eventually lead to
a better understanding of this theory. In fact, only very recently the
power of superconformal pre-Poincar{\'e} $N=2$ supergravity was
demonstrated in calculations of quantum corrections to the supersymmetric
black hole entropy \cite{MCW}.

{\it The first purpose of this paper is to present a detailed derivation of
 $SU(2,2|1)$-symmetric
superconformal theory}. In previous derivations \cite{general,KugoUe},
where it was only a tool, some part of the  $SU(2,2|1)$ symmetry was
broken at an early stage. We will first present the action with full
superconformal symmetry before explicitly indicating how we gauge-fix the
dilations, $R$-symmetry and $S$-supersymmetry, leading to Poincar{\'e}
supergravity.

 This formulation of the theory has several advantages. For example,
it simultaneously incorporates two different formulations of
phenomenological supergravity depending on the gauge-fixing of the
$R$-symmetry. The first  formulation, which is more standard, corresponds
to \cite{general}, where the Lagrangian depends not on two functions, the
\Ka\ potential ${\cal K}(z, z^*)$ and the superpotential $W(z)$,
 but only on one combination ${\cal
G}(z, z^*) = -{\cal K} (z, z^*) - \ln |M_P^{-3}W|^2$ and has  real fermion
mass terms. The other one, closer to \cite{BW,Binetruy}, has a
non-singular dependence on the superpotential $W$, but complex fermion
masses in general. It is important to have both versions of the theory
under control, especially in situations where the superpotential $W(z)$
may vanish, which often happens in cosmological applications.

The new formulation will allow us to give a detailed explanation of the
superconformal origin of Fayet--Iliopoulos (FI) terms by including gauge
transformations of the conformon field as
  first suggested in \cite{Rel}.

In addition to providing a new perspective on the old problems of
supergravity, the superconformal formulation appears to be most suitable
one for investigation of cosmology. Indeed, the
Friedmann--Robertson--Walker (FRW) universe is conformally flat. This
means, in particular, that the metric of an FRW universe can be
transformed into the form $\rm{d}s^2 = a^2(\eta) (\rm{d}\eta^2 -
\rm{d}x^2)$, where $\eta$ is conformal time, $a(\eta)$ is the scale
factor of the universe. If the theory is conformally invariant, then
the   scale factor $a(\eta)$ can be absorbed into a redefinition of the
metric and fields. These redefined fields do not depend on $a(\eta)$, so
the theory of particles in an expanding curved universe reduces to a much
simpler theory in a fixed flat Minkowski space. This considerably
simplifies the investigation of the behaviour of particles and fields
during the expansion of the universe. Ideally, one may perform the
investigation of all processes in the early universe within the framework
of superconformal theory, where expansion of the universe does not show
up, and then one may switch to the standard formulation with the fixed
Planck mass at the very end of the calculations. This is a new and very
exciting possibility; it would be hard to anticipate this possibility by
looking at the lengthy and extremely complicated Lagrangian of
phenomenological supergravity which appears after the breaking of the
superconformal invariance of the original theory.

In the beginning of our investigation we did not fully recognize this
possibility. We were working within the traditional framework of
phenomenological supergravity, making only occasional use of the
underlying superconformal invariance. Therefore, we were quite surprised
when we realized that in order to obtain some of our results it was
necessary to make certain field redefinitions in phenomenological
supergravity, which eventually brought us back to the original
superconformal formulation.

One of the problems that we were trying to address, was the issue of
conformal invariance of the gravitino and the possibility of non-thermal
gravitino production  in the early universe.

Many observable properties of the universe are to a large extent
determined by the underlying  conformal properties  of the fields. One
may consider inflaton scalar field(s) $\phi$ which drive inflation,
inflaton fluctuations which generate cosmological metric fluctuations,
gravitational waves generated during inflation,   photons in the cosmic
microwave background (CMB) radiation which propagate (almost) freely from
the last scattering surface, etc. If the conformal properties of any of
these fields were different, the universe would look quite different too.
For example, the theory of the usual massless electromagnetic field is
conformally invariant. This implies, in particular, that the strength of
the magnetic field in the universe decreases as $a^{-2}(\eta)$. As a
result, all vector fields become exponentially small after inflation.
Meanwhile the theory of the inflaton  field(s)  should not be conformally
invariant, because otherwise these fields would rapidly disappear and
inflation would never happen.

Superconformal supergravity is particularly suitable for studying the
conformal properties of various fields, because within this framework all
fields initially are conformally covariant; this invariance becomes
spontaneously broken only when one uses the gauge   $-\ft13 {\cal N}(X,
\bar X) = M_P^2$.

The issue of conformal invariance of the gravitino remained rather
obscure for a long time. One could argue that a massless gravitino should
be conformally invariant. Once we introduce the scalar field driving
inflation, the gravitino acquires a mass $m_{3/2} = \rme ^{{\cal
K}/2}|W|/M_P^2$. Thus, one could expect that conformal invariance of
gravitino equations should be broken only by the small gravitino mass
$m_{3/2}$, which is suppressed by the small gravitational coupling
constant $M_P^{-2}$. This is indeed the case for the gravitino component
with helicity $\pm \ft32$. However, breaking of conformal invariance for
the gravitino component with helicity $\pm \ft12$, which appears due to
the super-Higgs effect, is much stronger. In the first approximation in
the weak gravitational coupling, it is related to the chiral fermion mass
scale \cite{GravProd}.

The difference between the two gravitino components becomes especially
important when one studies gravitino production after inflation. It is
usually assumed that gravitinos have mass $m_{3/2} \sim 10^2$--$10^3$~GeV.
Such particles  decay very late, which leads to disasterous cosmological
consequences unless the ratio of their number density $n_{3/2}$ to the
entropy density $s$ is extremely small. In particular, the ratio
$n_{3/2}/s$  should be smaller than ${\rm O}(10^{-15})$ for gravitinos
with mass ${\rm O}(100)$~GeV \cite{gravitinos,Moroi}.

The standard thermal mechanism of gravitino production involves scattering
of particles at high temperature in the early universe.  To avoid
excessive production of gravitinos one must assume that the reheating
temperature of the universe after inflation was smaller than
$10^8$--$10^9$~GeV \cite{gravitinos,Moroi}.

However, gravitinos can also be produced during  the oscillations of the
inflaton field at the end of inflation. The theory of the production of
the gravitino with helicity $\pm \ft32$ is relatively straightforward,
and the effect typically is not very large, because it appears mainly due
to the non-adiabatic change of the small gravitino mass during the scalar
field oscillations and the expansion of the universe
\cite{Maroto:2000ch,Lemoine:1999sc}.  This effect disappears in the limit
of the small gravitational coupling, $M_P \to \infty$.

One could expect that the same should happen for the gravitinos with
helicity $\pm \ft12$. However, we have found, in models with one chiral
multiplet, that  the gravitinos with helicity $\pm \ft12$ can be produced
as abundantly as normal matter particles not belonging to the
gravitational multiplet, i.e.\ the rate of their production does not
vanish in the limit $M_P \to \infty$ \cite{GravProd}.

More exactly, we have found that if one considers the underlying globally
supersymmetric theory with one chiral multiplet, then the chiral fermion
$\chi$ has  mass $W_{,\phi\phi} = {\partial^2 W\over
\partial \phi^2}$, where $\phi = z M_P$. This mass oscillates during
the oscillations of the scalar field $\phi$, which would lead to
production of fermions $\chi$ in the globally supersymmetric theory. In
supergravity, these fermions enter the definition of goldstinos, which
are eaten by gravitinos and give rise to the gravitino component with
helicity $\pm \ft12$. This is the deep reason why, as we have found, the
probability of production of the helicity-$\ft12$ gravitinos does not
vanish in the limit $M_P \to \infty$ and coincides with the probability
of the production of the chiral fermions $\chi$ (and, correspondingly, of
the  goldstino production) in the underlying globally supersymmetric
theory  \cite{GravProd}. This surprising result was confirmed in the
paper by Giudice,  Tkachev and  Riotto \cite{Giudice:1999yt}, and  in
subsequent papers by several other authors
\cite{Lyth:1999ph}--\cite{Bastero-Gil:2000je}.

However, until now a complete supergravity treatment of the gravitino
production was achieved only for models with one chiral multiplet. Even in
this case the theory of gravitino production was very complicated, and it
was not quite clear how one could study the realistic models with many
chiral and vector multiplets.

 One of the ideas was to use the gravitino--goldstino
correspondence in the hope that the leading effect can be found by the
relatively simple investigation of production of goldstinos in a globally
supersymmetric theory, instead of a direct investigation of the gravitino
production. Indeed, this idea worked well for the case of one multiplet
\cite{GravProd,Giudice:1999yt,Maroto:1999vd},  but there was no proof that
the same method will apply in the general case as well.

Equations for the goldstino in the theory without vector multiplets but
with any number of chiral multiplets were derived in \cite{GTR} within
the framework of globally supersymmetric theories, neglecting expansion of
the universe. The hope was expressed that the gravitino--goldstino
equivalence theorem \cite{fayet,casalbuoni} will justify such derivation
as representing the equation for the helicity $\pm \ft12$ of the
gravitino in the limit of $M_P\rightarrow \infty$ in supergravity.

However, the relevance of the equivalence theorem for scattering
amplitudes in the high-energy limit \cite{fayet,casalbuoni} to the
problem of gravitino production by the oscillating scalar field in a
cosmological background in the theory with many chiral and vector
multiplets was not quite obvious. In fact, one did not even have a clear
picture of the super-Higgs effect in cosmology, which was essential for
the understanding of the goldstino--gravitino correspondence, and it was
rather non-trivial to give a precise definition of what we mean by the
gravitino interactions in the limit $M_P\rightarrow \infty$.

As we will see, the concept of the goldstino in the background with
time-dependent metric and scalars, and the proper definition of the
$M_P\rightarrow \infty$ limit of supergravity is greatly simplified in
the context of the superconformal theory. It turned out that the {\it
original fields of the superconformal theory} rather than the fields used
in phenomenological supergravity are those which should be held fixed in
this limit.

Therefore, after deriving two versions of standard supergravity, before
starting with gravitino equations,  we will perform specific
modifications of the theory so that the basic fields will again be those
of superconformal theory. This includes, in particular, a number of
rescalings. All of them have one simple purpose: to use as basic
variables the fields of the underlying superconformal theory. This form
of supergravity is suitable for considering the limit in which gravity
decouples and  a globally supersymmetric theory appears.

These modifications of the usual supergravity will allow us to achieve
 {\it the second purpose of this paper:  to
generalize the super-Higgs effect for cosmology, and to derive the gravitino
field equations in the theory with any number of chiral and vector  multiplets}.

In Minkowski space, the supersymmetry breaking and super-Higgs effect
occur only if  the gravitino has non-vanishing mass $m_{3/2} = \rme
^{{\cal K}/2}|W|/M_P^2$, because the generator of the supersymmetry
transformations is proportional to $m_{3/2}^2$. This parameter depends on
$\phi$, so it changes during the oscillations of the scalar field $\phi$,
and it may vanish at some stages of the evolution of the universe.
However, we will show that the criterion for supersymmetry breaking and
the existence of the super-Higgs effect in the post-inflationary universe
with the energy density provided by the scalar field is not $m_{3/2}^2
\not = 0$, but rather $H^2+m_{3/2}^2 \not = 0$. \footnote{It is tempting
to say that in the cosmological background, the gravitino may eat the
goldstino even if it is massless, $m_{3/2}=0$. However, as we will see,
the equation of motion for the gravitino with helicity $\pm \ft12$ in the
cosmological background is more complicated than the standard equation
for a particle with mass $m_{3/2}$.} This criterion is {\it always}
satisfied, so that the super-Higgs effect will always take place in a
universe with the energy density dominated by the scalar fields. This is
an important conclusion that will allow us to use the unitary gauge, in
which the goldstino is eaten by the gravitino throughout our calculations.

The system of equations for the gravitino in the theory with many chiral
and vector multiplets is rather complicated. However, they will be given
in a form in which the limit $M_P\rightarrow \infty$ can be taken
straightforwardly. This will clarify the status of the
gravitino--goldstino equivalence theorem in the limit $M_P\rightarrow
\infty$ and the relation of supergravity to the globally supersymmetric
theories. The difference with \cite{fayet,casalbuoni} is that we will not
be working with the $S$-matrix elements in the flat background, but with
classical equations for bosons and fermions in the cosmological setting.
We will therefore not compare the matrix elements, but the form of the
highly nonlinear field equations, relevant for the production of
particles in the early universe. This equivalence theorem will explain
why the effect of non-thermal production of gravitinos in the early
universe, in general, is not suppressed by the inverse powers of $M_P$.

This theorem should be used with care since its results are easy to
misinterpret. It is very useful when the process of particle production
occurs so fast that one can neglect expansion of the universe. However,
in the theories with many chiral and vector multiplets, the definition of
the goldstino gradually changes during the expansion of the universe. In
the beginning, the goldstino is associated with a certain combination of
chiral fermions, the superpartners of the scalar fields driving inflation
(inflatino). In the end of the process, the goldstino may be associated
with a completely different combination of chiral fermions. Therefore, one
could argue that gravitino (inflatino) production in the very early
universe may be irrelevant for the calculation of the final number of
gravitinos produced, because all chiral fermions produced at the
beginning of this process may not give any contribution to the goldstino
eaten by the gravitino at the end of the process \cite{Nilles99}.

This issue turns out to be rather complicated. The goldstino--gravitino
correspondence is a useful tool in regimes when the expansion of the
universe can be neglected. However, in order to describe the change
between different definitions of the goldstino due to the expansion of the
universe one should go beyond the limit $M_P \to \infty$, which may
invalidate the argument given above. In this case instead of using the
equivalence theorem one should study gravitino equations in the unitary
gauge. As we will see, in the epoch where the definition of goldstino
changes because of the expansion of the universe, the number of gravitinos
may also change because they can mix with other fermions. A detailed
investigation is necessary in order to find out whether this leads to a
depletion of the gravitinos produced at the first stages of the process,
or, vice~versa, to their additional production due to the non-adiabaticity
related to the change of the nature of the goldstino.

 In our paper we will develop the framework which can be used in order
to address this question, as well as other questions related to the
gravitino production. We will derive equations for the gravitino in
theories with an arbitrary number of chiral and vector multiplets, and
analyse some limiting cases where their solutions can be obtained.

The paper is organized as follows. In section~\ref{ss:examplesCon} we
give an example of the local conformally symmetric theory and introduce
the concept of the conformon field. We use this example, in particular, to
explain the conformal properties of the gravitino. We  discuss the
cosmic-string-type configuration of superconformal theory. There is an
infinite gravitational coupling at the core of the string.

 Section~\ref{ss:superconfAction} presents the $SU(2,2|1)$  symmetric
action of the Yang--Mills and chiral multiplets superconformally coupled
to supergravity. In  section~\ref{ss:presAction}, the conformal and
chiral weights of all fields are organized in a table. The generic
cosmic-string-type configurations with vanishing ${\cal N}$ at the core
are discussed. The superconformal origin of the \Ka\ potential and
superpotential is explained and the $S$- and $Q$-supersymmetry
transformations of all fields are given. The auxiliary fields are
eliminated on shell. We comment  on the quantization in the case of
topologically non-trivial \Ka\ manifolds. Section~\ref{ss:gfKa} explains
the gauge-fixing of superfluous symmetries which leads to supergravity.
In particular, the conformon field is gauge-fixed to give $M_P$. Here, as
well as in appendix~\ref{app:elimA}, the origin of the \Ka\ geometry is
clarified.

Section~\ref{ss:gaugeIsom} is about the Yang--Mills part of the theory,
showing that any gauged isometry can be obtained in the superconformal
formulation by including transformations of the conformon multiplet. In
section~\ref{ss:Killing}, Killing isometries, which act on all scalars of
the theory, including the conformon field, are given. The Killing
potentials, which encode the  isometries, may include some constant parts
related to the gauge symmetries of the conformon. This provides the
superconformal origin of the Fayet--Iliopoulos terms ($D$-terms). The new
mechanism of the generation of the FI terms is explained in
section~\ref{ss:mechD}, and some examples are given.

Section~\ref{ss:derPhenL} shows the derivation of the phenomenological
Lagrangian. First, in section~\ref{ss:Kastructure}, the \Ka\ structure
and the superpotential are derived from the superconformal structures.
Section~\ref{ss:Goldstino} gives a first discussion of the goldstino,
which is still natural to do in the superconformal context.
Section~\ref{ss:LagU1} presents the full phenomenological Lagrangian,
where the $R$-symmetry of the local $SU(2,2|1)$ is still not gauge-fixed.
The two possibilities of the gauge-fixing of this remaining $U(1)$
symmetry are explained in this section: either the \Ka\ symmetric gauge,
or that which is non-singular when the superpotential is vanishing. Thus
it is easy to obtain both theories
 from one action in
(\ref{phenomL}). Finally, in section~\ref{ss:rigidlimit}, we perform
various rescalings of the fields of the theory in view of a rigid limit,
which bring us back to
the original fields from which the superconformal theory was build and
which will be used in the rest of the paper.

Section~\ref{ss:eqsGravitino} is dedicated to the gravitino. In
section~\ref{ss:simplAction}, the relevant part of the action is given
which shows clearly the mixing between the gravitino, chiral fermions and
gaugino. Field equations without specifying a gauge-fixing for local
supersymmetry are derived in section~\ref{ss:fieldEqsGr}.
Section~\ref{ss:MasterGrav} offers a master gravitino field equation and
the constraints. We discuss the choice of the unitary gauge where the
goldstino vanishes in the case of the non-constant gravitational and
scalar backgrounds.

In section~\ref{ss:confInspBG} we develop our formalism in application to
cosmology. We first specify the assumptions related to cosmology in
section~\ref{ss:Assumptions}. Section~\ref{ss:superHiggsCosm} describes
the super-Higgs effect in cosmology with  time-dependent scalars and a
conformally flat time-dependent metric. In section~\ref{ss:constraintsGr}
 the important
constraints for the gravitino are given in the unitary gauge, where the
goldstino vanishes. The system of equations for the gravitino  interacting
with other fermionic fields of the theory, chiral fermions as well as
gauginos, is derived in section~\ref{ss:dynEqGr}.  In
section~\ref{ss:longGr} we give a useful second-order form of the
equation for the helicity-$\pm \ft12$ gravitino    interacting with other
fermions.

Section~\ref{ss:equivTh} presents the goldstino--gravitino equivalence
theorem explaining why the effect of the helicity-$\pm \ft12$ gravitino is
not suppressed in the limit of large $M_P$.

In section~\ref{ss:2mult} we consider the cases with one or two
multiplets, and make a number of simplifications. Some parts of this
section consider only real  scalar fields. In other parts the limit of
these equations at large $M_P$ is taken.

Section~\ref{ss:creation} is devoted to gravitino production after
inflation. We give an overview of the previous investigation  of the
theories with one chiral multiplet, and then discuss various issues
related to gravitino production in realistic models with several
multiplets.

In the discussion section we give a short overview of the results.

Appendix~\ref{app:notations} presents the notation, including a
clarification of the rules of complex conjugation. The consequences of a
conformal metric are reviewed in appendix~\ref{app:conf_metric}.
Appendix~\ref{app:elimA} gives the steps needed to go from a conformal
form of the action to the K{\"a}hler geometry. It is formulated such that it
applies more generally than for $N=1$ theories. It also discusses the
relation with Sasakian manifolds, and at the end gives the connection
between the conformal chiral-covariant derivatives and K{\"a}hler-covariant
derivatives. Appendix~\ref{app:calcmu} gives a detailed calculation of
the quantity $\mu $, which appears in the solution of the gravitino
equations, in the case of one chiral multiplet.

A short account of some of our results has been given in \cite{GravProd}.

\section{Simple examples of local conformal symmetries supported by a conformon
 field} \label{ss:examplesCon}

The action for general Yang--Mills--matter--supergravity theories with
$N=1$ local supersymmetry was derived in \cite{general} starting from the
superconformal symmetry. The superconformal symmetry was used mainly  as
a technical tool for the derivation of the Poincar{\'e} supergravity with
smaller symmetry \cite{conformforPoin}. It appears now that the
superconformal form of the action (before the gauge-fixing of extra
symmetries is performed) provides a natural framework within which to
address the issue of the conformal properties of the gravitino. For our
present purpose it is important to look at the gauge-fixing of the local
dilatational symmetry. The mechanism can be explained using a simple
example: an arbitrary gauge theory with Yang--Mills fields $W_\mu $
 coupled to fermions $\lambda $ and
gravity:
\begin{equation}
S^\mathit{conf}= \int \rmd ^4 x\, \sqrt {g }\left(\ft12 (\partial_\mu
\phi)\, (\partial_\nu \phi) \,g^{\mu\nu} - \ft1{12} \phi^2 \, R
 -\ft{1}{4}\Tr F_{\mu\nu} g^{\mu \rho }g^{\nu \sigma }F_{\rho \sigma }
-\ft{1}{2} \bar \lambda \not\!\! {\cal D} \lambda\right)\,.
\label{conf}\end{equation} The field $\phi$ is referred to as a conformal
compensator.
 The
last two terms in the action represent super-Yang--Mills theory coupled to
gravity. The action is conformal invariant under the following local
transformations:
\begin{equation}
 g_{\mu\nu}'= \rme ^{-2\sigma(x) } g_{\mu\nu}\,, \qquad
\phi'=\rme ^{\sigma (x)}\phi \,,   \qquad  W'_\mu=W_\mu \,,\qquad \lambda
' =\rme ^{\frac{3}{2}\sigma (x)}\lambda
 \,.\label{scaletransf}
\end{equation}

The gauge symmetry (\ref{scaletransf}) with one local gauge parameter can
be gauge-fixed. If we choose the $\phi=\sqrt{6}M_P$ gauge\footnote{Note
that one has to take a scalar field with ghost-like sign for the kinetic
term to obtain the correct kinetic term for the graviton.  This does not
lead to any problems since this field disappears after the gauge-fixing
$\phi=\sqrt{6}M_P$.}, the $\phi $-terms in (\ref{conf}) reduce to the
Einstein action, which is no longer conformally invariant:
\begin{equation}
S^\mathit{conf}_{\mbox{\scriptsize{\it gauge-fixed}}}\sim \int \rmd ^4 x
\, \sqrt {g } \left( -\ft{1}{2}M_P^2 R
 -\ft{1}{4} F_{\mu\nu} g^{\mu \rho }g^{\nu \sigma }F_{\rho \sigma }
+\ft{1}{2} \bar \lambda \not\!\! D \lambda\right)\,.
\end{equation}
Here $M_P \equiv M_\mathit{Planck}/\sqrt{8\pi} \sim 2 \times
10^{18}$~GeV. In this action, the transformation (\ref{scaletransf}) no
longer leaves the Einstein action invariant. The $R$-term transforms with
derivatives of $\sigma(x)$, which in the action (\ref{conf}) were
compensated by the kinetic term of the compensator field. However, the
actions of the Yang--Mills sector of the theory, i.e.\ spin-$\ft12$ and
spin-1 fields interacting with gravity, remain conformal invariant. Only
the conformal properties of the gravitons are affected by the removal of
the compensator field.

Now consider a supersymmetric version of this mechanism.

We will give the action with local superconformal symmetry associated
with the gauging of the $SU(2,2|1)$ algebra \cite{SU221}. These include:
\textit{i)} general coordinate symmetry, local Lorentz symmetry, local
$Q$-supersymmetry; and \textit{ii)} chiral $U(1)$ symmetry, local
dilatations, special conformal symmetry and  $S$-supersymmetry. The
second group are local symmetries of the superconformal group which are
not part of the super-Poincar{\'e} algebra, and are gauge-fixed. This leads
to the final form of the action of Poincar{\'e} supergravity in
equations~(4.16)--(4.20) of \cite{general}, which has only symmetries of
the first group: i) general coordinate symmetry, local Lorentz symmetry
and local $Q$-supersymmetry.

Consider first the example of one chiral multiplet conformally coupled to
supergravity. This multiplet will play the role of a conformal
compensator. The superconformal multiplet has gauge fields corresponding
to each symmetry in $SU(2,2|1)$. However, some of these gauge fields are
dependent, in the same sense that the spin connection, the gauge field of
Lorentz rotations, is dependent on the metric. Moreover, we can
immediately gauge-fix special conformal transformations by eliminating
the gauge field of dilatations,  and $S$-supersymmetry by removing the
spinor field of the chiral multiplet. Therefore, the remaining action
still has super-Poincar{\'e}, as well as local dilatations and chiral $U(1)$
symmetry. The action invariant under these symmetries is (omitting terms
quartic in fermion fields)
\begin{equation}
S^\mathit{grav}= \int \rmd ^4 x\left\{  \sqrt {g }({\cal D}_\mu  \phi)\,
({\cal D}_\nu \phi^*) \,g^{\mu\nu}  - \ft{1}{6}|\phi|^2\left[ \sqrt {g
}R+ \sqrt {g }\bar \psi_\mu R^\mu+
\partial _\mu \left( \sqrt {g }\bar \psi \cdot \gamma \psi ^\mu
\right)\right] \right\}\!,
 \label{grav}
\end{equation}
where $D_\mu \phi =\partial _\mu \phi +\frac{\rmi}{3} A_\mu \phi $,
and $A_\mu $ is the gauge field for the chiral $U(1)$ symmetry.
Furthermore,
\begin{equation}
  R^\mu=e^{-1}\varepsilon^{\mu\nu\rho\sigma}\gamma_5\gamma
_\nu{\cal D} _\rho \psi _\sigma  = \gamma ^{\mu \rho \sigma }{\cal D}
_\rho \psi _\sigma\,,
  \label{defR}
\end{equation}
where the gravitational covariant derivative includes the spin
connection, chiral $U(1)$ field and Christoffels\footnote{With this
definition one has ${\cal D}_\rho e_\mu^a = 0$.}
\begin{equation}
{\cal D}_\mu\psi_\nu = \left((\partial_\mu + \ft14 \omega_\mu^{mn} \gamma
_{mn}+ \ft12\rmi\gamma_5 A_\mu )\delta_\nu^\lambda -
\Gamma_{\mu\nu}^\lambda \right)\psi_\lambda\,. \label{PvNRRel2}
\end{equation}
Note that in the curl $ {\cal D}_{[\mu}  \psi_{\nu]}$ the term with
Christoffels drops. Considering just the local dilatations,  the
first two terms in (\ref{grav}) combine to an invariant, as well as
the last two terms.

This local dilatation with parameter $\sigma(x)$ and chiral symmetry with
parameter $\Lambda(x)$ can be represented by transformations
\begin{eqnarray}
 g_{\mu\nu}'= \rme ^{-2\sigma(x) } g_{\mu\nu}\,, \qquad
\phi'=\rme ^{\sigma (x)-\ft13\rmi\Lambda (x)}\phi \,,\nonumber\\
  \psi'_\mu = \rme ^{-\frac{1}{2}\left[\sigma (x)+\rmi\gamma_5 \Lambda(x)\right]}\psi
  _\mu\,,
  \qquad A_\mu' =A_\mu + \partial_\mu \Lambda(x)\,.\label{conftrex2}
\end{eqnarray}
One can gauge-fix the local dilatation and chiral $U(1)$ symmetry by
choosing
\begin{equation}
 \phi= \phi^* = \sqrt{3}M_P \,.\label{gsdilphi}
\end{equation}

As soon as the compensator field $\phi (x)$ is fixed to give us the
normal gravity and supergravity theory, the conformal transformations of
the gravitino, as well as of the metric, no longer form a symmetry. Under
conformal rescaling of the gravitino, the action will have non-invariant
terms. The overall scaling does not match, and there are terms depending
on derivatives of the scaling function due to the absence of the last
term of (\ref{grav}), which is now a total derivative. Changing the
conformal weight of the gravitino, these variations with $\partial \sigma
$ still remain.

One could try to maintain the conformal invariance by absorbing $\phi $
in $\psi _\mu $ as $\tilde \psi _\mu =\phi \psi _\mu $, which gives
$\tilde \psi $ conformal weight $+\ft12$, rather than $-\ft12$ in
(\ref{conftrex2}). The action, which is still conformal invariant, is (we
do break the chiral invariance by taking $\phi $ as real)
\begin{equation}
 - \ft{1}{6} \int \rmd ^4 x\, \left[ \sqrt {g }  \bar
{\tilde \psi}_\mu \tilde R^\mu-2 \left( \partial _\mu \ln \phi
\right) \left( \sqrt {g }\bar{\tilde  \psi} \cdot \gamma \tilde \psi
^\mu \right)\right]\,. \label{gravitinoconf}
\end{equation}
At first sight, one may think that the kinetic term of the gravitino is
conformal invariant by giving it conformal weight $\ft12$, as is the case
for $\tilde \psi $. Indeed, the first term does not depend on $\phi $, and
thus it is not affected by the breaking of the conformal invariance by
the gauge-fixing (\ref{gsdilphi}). However, only the sum of the two terms
in (\ref{gravitinoconf}) is a conformal invariant, and this formula thus
indicates where the conformal invariance is broken. Clearly, there will
be the usual source of the deviation from conformal symmetry via mass
terms, which will appear through spontaneous breaking of supersymmetry.
In this respect, the gravitino is not different from any other field.
However, here we see that even in the absence of mass terms there are new
features.

The gravitino field equation that follows from the superconformal
action is
\begin{equation}
\tilde R^\mu -  \gamma^\mu \tilde \psi ^\nu \partial _\nu \ln \phi
+\gamma  \cdot \tilde  \psi \partial _\mu \ln \phi=0\,.
\end{equation}
In the FRW cosmological problems only time derivatives of the
 scalar fields are
 important, therefore in $\tilde \psi ^\nu\partial _\nu \ln \phi  $
  only the term $\tilde \psi ^0\partial _0 \ln \phi $ is relevant.
After gauge-fixing, the conformal symmetry will be broken for
configurations for which either
\begin{equation}
 \gamma  \cdot  \psi \neq 0 \qquad \rm or  \qquad  \psi_0\neq 0\,.
\end{equation}
Only such terms will be sensitive to the absence of the terms $\partial
_0 \ln \phi$ due to gauge-fixing when $\phi \phi ^*=3 M_P^2$. The
gravitino in the general theory with spontaneously broken supersymmetry
will be massive. The states of a free massive spin-$\ft32$ particle were
studied by Auvil and Brehm in \cite{AB} (see also \cite{Moroi} for a nice
review). A free massive gravitino has $\gamma  \cdot  \psi =0$.
Helicity-$\pm 3/2$ states are  given by transverse space components of the
gravitino,  $\psi_i^T$. Helicity-$\pm \ft12$ states are given by the time
component of the gravitino field, $\psi_0$. In the cases when the
gravitino interacts with gravity and other fields, we will find that
$\gamma  \cdot  \psi \neq 0$ and will be related to $\psi_0$. Thus the
consideration of superconformal symmetry leads us to a conclusion that
helicity-$\pm \ft12$ states of the gravitino are not conformally coupled
to the metric. When these states are absent, the $\pm \ft32$ helicity
states are conformally coupled (up to the mass terms, as usual). Thus the
conformal properties of the gravitino are simple,  as is  known for
scalars:
 if the action has an additional term
${1\over 12} \phi^2 R$, the massless scalars are conformal. If this term
is absent, the scalars are not conformal. Note that both of these
statements are derivable from the superconformal action (\ref{grav}). We
will see the confirmation of this prediction in the solutions of the
gravitino equations in section~\ref{ss:eqsGravitino}.

Thus we explained here a simple reason why the gravitino is not
conformal. From this consideration one concludes:  the linearized
equation of motion for the gravitino interacting with all other fields of
supergravity in an FRW conformally flat metric will not decouple from the
scale factor of the metric, as happens for the graviton and contrary to
the case of the massless Dirac and Yang--Mills (YM) fields.

Before going further and considering the total supergravity Lagrangian
containing other matter fields, we would like to make a comment. In this
section we {\it assumed} that the classical background field  $\phi(x)$
does not vanish. Only in this case can one gauge-fix the local dilatation
and chiral $U(1)$ symmetry in such a way as to make this field constant,
$\phi= \phi^* = \sqrt{3}M_P$, and recover the standard Poincar{\'e}
supergravity.

Whereas the assumption  $\phi(x) \not = 0$ is quite legitimate, and we
are going to use it throughout the paper, one may also contemplate the
existence of another phase of the original superconformal theory, where
$\phi(x) = 0$, which corresponds to the strong-coupling limit $\kappa^2 =
M_P^{-2} \to \infty$. In such a phase, the superconformal symmetry will be
unbroken, unlike in the `Higgs phase'  $\phi= \phi^* = \sqrt{3}M_P$.

The existence of such a phase may not be very unnatural. Indeed, let us
write the field $\phi$ as ${1\over \sqrt 2}(\phi_1 + \rmi\phi_2)$. There
is no obvious reason to assume that the complex vector $(\phi_1,\phi_2)$
must have the same direction everywhere in the whole universe prior to
the gauge-fixing, which aligns the field $\phi$ and makes it real, $\phi=
\phi^* = \sqrt{3}M_P$.

 Consider, for example, a cosmic-string-type configuration
$\phi \sim |\phi (x)| \rme ^{\rmi n \theta}$, where $\phi(x) \not = 0$ at
a large distance from the $z$-axis. Then for topological reasons the field
$\phi$ must vanish on some string(s) stretched along the $z$-axis. The
vector field is equal to $A_\mu= {3\rmi\over 2 |\phi|^2}
({\phi^*\partial_\mu \phi - \phi\,
\partial_\mu \phi^* })$. Therefore,  the asymptotic value of
the vector field  for the string configuration is given by $A_\mu = 3 \rmi
\partial_\mu \ln {\phi\over |\phi|}$. It looks like a `pure gauge', but
it cannot be gauged away by a regular gauge transformation for $n\neq 0$.
The choice of the unitary gauge $\phi= \phi^* = \sqrt{3}M_P$ is possible
everywhere except on the  string(s) containing quantized magnetic flux,
just like the usual cosmic strings described in \cite{vilenkinshellard}.
Indeed, if we integrate around the closed path encircling the string
configuration, we find that it contains quantized magnetic flux $[F]
=\oint {\bf A} \cdot {\bf dl}= {6\pi n}$.

In this paper we will not give  a detailed discussion of the possible
existence of cosmic strings associated with supergravity  (see, however,
the next section). Even if such strings are present in the early universe,
the distance between them should become exponentially large because of
inflation. Thus    one may assume that   one has $\phi \not = 0$ in the
observable part of the universe, which allows us to use the gauge $\phi=
\phi^* = \sqrt{3}M_P$ and recover usual supergravity from the
superconformal one. Still the possibility of having two different phases
of superconformal supergravity and the issue of topological defects in
this theory deserves separate investigation.

\section{Local superconformal action of $N=1$ gauge theories}
\label{ss:superconfAction}

We will now present the full action for $n+1$ chiral multiplets and some
number of Yang--Mills vector multiplets superconformally coupled to
supergravity\footnote{It was shown in \cite{Rel} that other matter
multiplet representations or other sets of auxiliary fields do not lead
to different theories from those obtained with chiral multiplets and
corresponding auxiliary fields.}.
 This action with super-\Poin\ symmetry was derived first
in \cite{general}. Conformal methods were helpful for its construction,
but the conformal symmetry was broken rather soon; it was only used as a
tool, as an intermediate step in formulating $N=1$ supergravity. Soon
thereafter, it was shown in \cite{KugoUe} that several steps of the
construction could be simplified by taking cleverer gauge choices.
However, also in that work the conformal invariance was broken at an early
stage, which we want to avoid. Several improvements followed, especially
concerning the structure of the \Ka\ manifold, i.e.\ its global structure
and isometries \cite{BagWitN1,Bagger}. A singular limit, independence of
auxiliary fields and Fayet--Iliopoulos terms were treated in \cite{Rel}.
Although all of these developments were based on a superconformal
approach, the conformal action has not yet been written, which is the
first purpose of this section. Another basic difference with the work
\cite{N1YMmsg,general,KugoUe} is that we treat all $n+1$ chiral
multiplets (a conformon and the physical chiral multiplets) on an equal
footing. In this way, our description of the structure of the $N=1$
theory will be close to that used for $N=2$ supergravity where one has
special geometry \cite{dWVPspec,special,whatspk}.

\subsection{Presentation of the superconformal action}
\label{ss:presAction}
The fields in the theory are\footnote{For a review on the methods
used here, see \cite{Karpacz}.}
\begin{equation}
  \begin{array}{|c|ccc|ccc|ccc|}
\hline
 &   e_\mu ^a &\psi _{\mu L}&A_\mu  & X_I &\Omchi _I& h_I & W_\mu ^\alpha&\lambda _L^\alpha   & D^\alpha   \\
 \hline
 w  & -1 &-\ft12&0&1 &\ft32 & 2 & 0 &\ft32 & 2  \\
 c & 0 &-\ft12&0& -\ft13&\ft16 & \ft23  & 0 & -\ft12&0  \\
\hline
\end{array}
 \label{tablefieldswc}
\end{equation}
The first part contains the independent fields of the Weyl multiplet,
vielbeins $e_\mu^a$, gravitino $\psi_\mu$ and $U(1)$ gauge field $A_\mu
$, which gauge $SU(2,2|1)$, together with the gauge field of dilatation
which is absent in the action due to the special conformal
symmetry\footnote{All fields in (\ref{tablefieldswc}) are invariant under
the special conformal symmetry, once one considers general coordinate
transformations as an independent symmetry. The fact that it acts on the
gauge field of dilatations $b_\mu $ as $\delta b_\mu =e_\mu^a \Lambda
_{Ka}$ then implies the absence of $b_\mu $ from an invariant
action.\label{fn:bmu}  }, and composite fields. All fields in the table
transform under  a local dilatation symmetry with a dilatation weight
$w$, and local chiral symmetry with  a chiral weight $c$:
\begin{equation}
\Phi' = \rme ^{w \sigma (x)+\rmi c\Lambda (x)} \Phi\,.\label{defwcweights}
\end{equation}
An exception to this rule is the gauge field of chiral symmetry,
$A_\mu $,
 which thus transforms
as $\delta A_\mu =\partial _\mu \Lambda $. The diagonal transformation
rule (\ref{defwcweights}) is obtained on the fermion fields after
defining their left and right components $\lambda _L$ and $\lambda _R$
(see appendix~\ref{app:notations}). These chiral fermions have opposite
chiral weights.

For the fermion fields of the chiral multiplets, we can indicate the
chirality by the position of the index $I$. The components $\Omchi _I$
are the left-handed ones, while $\Omchi^I$ are right-handed. Similarly
for the bosons in these multiplets we use the notation $X^I \equiv(
X_{I})^*$ and $h^I \equiv( h_{I})^*$. The latter are the last components
of these multiplets, which will play the role of auxiliary fields, and
whose real and imaginary parts are often denoted as $F$ and $G$. The
index $I$ takes values $I=0,1, \ldots , n$. One of the chiral multiplets,
the conformon multiplet, will be used for the gauge-fixing of local
dilatation, chiral symmetry and special supersymmetry as in the cases we
studied above.

The vectors $W_\mu^\alpha$ are gauge fields for an arbitrary gauge
group, with gauge group index $\alpha$. The gaugino
$\lambda^\alpha$ and  auxiliary scalars $D^\alpha$ are in the adjoint
of this group. The chiral multiplets may transform under this gauge
group as will be shown below.

Observe, that in this conformal set-up we have chosen to take for all
physical scalars the Weyl weight $w=1$. Originally \cite{general,KugoUe},
the scalars were taken to be of Weyl weight $w=0$, except for the
conformon scalar, which was taken to be $w=1$. This is a matter of
choice, and one can go from one formulation to the other by field
redefinitions. However, to be as close to a conformal invariant action as
possible, the Weyl weight $w=1$ (and correspondingly $w=\ft{3}{2}$ for
the spin-$\ft{1}{2}$ fields) is the most natural choice.

The $SU(2,2|1)$-invariant Lagrangian of $N=1$ supergravity coupled to
$n+1$ chiral multiplets and Yang--Mills vector multiplets
superconformally has no dimensional parameters. It consists of  3 parts,
each of which is conformally invariant separately:
\begin{equation}
{\cal L} =  [\Nphi (X,X^*)]_D + [\mathcal{W}(X)]_F + \left[ f_{\alpha
\beta } (X) \bar \lambda_L ^\alpha \lambda_L ^\beta\right] _F
\label{symbL}
\end{equation}
It thus depends on 3 functions which should transform in a
homogeneous way under dilatations and chiral transformations:
\begin{equation}
  \begin{array}{|c|ccc|}
\hline
 &   \Nphi (X, X^*) & {\cal  W}(X) & f_{\alpha \beta } \\
 \hline
 w  & 2 & 3 & 0 \\
 c & 0 &-1  & 0  \\
\hline
\end{array}
 \label{tablewcFunct}
\end{equation}
 The function $\Nphi (X, X^*)$ will be related to
 the \Ka\ potential. The holomorphic function ${\cal W}(X)$
encodes the superpotential. The holomorphic function $f_{\alpha
\beta}(X)$ encodes the kinetic terms for the vector multiplet fields.
Derivatives of these functions will be indicated by, for example, $
\Nphi_I$ for a derivative with respect to $X^I$, or $f_{\alpha \beta }^I$
for a derivative with respect to $X_I$. The homogeneity which follows from
(\ref{tablewcFunct})
 implies relations
\begin{eqnarray}
&&\Nphi =X^I \Nphi _I = X_I \Nphi ^I = X_I \Nphi ^I{}_J X^J \,,\qquad
\Nphi _I=  X_J \Nphi ^J{}_I \,, \nonumber\\
 &&   X_J \Nphi ^{JI}=0\,, \qquad
 X_I \Nphi ^I{}_{JK} = \Nphi _{JK}\,,\qquad  X_K\Nphi ^{IK}_J=0\,.
 \label{N}
\end{eqnarray}
The latter relation will be especially important, see
appendix~\ref{app:elimA}. This way of working was developed first for
$N=2$ supergravity \cite{dWVPspec}. In $N=2$ the function $\Nphi $ is,
furthermore, restricted to the real part of a holomorphic function.

The full action is
\begin{eqnarray}
\lefteqn{  [\Nphi ]_De^{-1}=\ft16\Nphi (X, X^*)\left[ R + \bar \psi_\mu
R^\mu+ e^{-1}
\partial _\mu (e \bar \psi \cdot \gamma \psi ^\mu)+{\cal L}_{SG,\mathit{torsion}}
\right]}\nonumber\\ &-&\Nphi _I{}^J(X, X^*)\left[({\cal D}_\mu X^I)
  ({\cal D}^\mu X_J) + \bar \Omchi_{J}
 \not\!\! {\cal D} \Omchi^I+ \bar \Omchi^I
 \not\!\! {\cal D} \Omchi_J -h_J h^I\right]\nonumber\\
 &+& \left\{-\Nphi _J{}^{IK} \bar\Omchi_{I} \Omchi_{K} h^{J}+\Nphi _K{}^{IJ}
 \bar\Omchi_{I}(\not\!\! {\cal D} X_J) \Omchi^K \right.\nonumber\\
 &+&\Nphi _J{}^I \bar \psi_{\mu L}(\not\!\! {\cal D} X^J)
 \gamma^\mu \Omchi_{I}-\ft23\Nphi ^I\bar \Omchi_{I}
 \gamma ^{\mu\nu}\hat {\cal D}_\mu \psi_{\nu L}
 -\ft{1}{8}e^{-1}\varepsilon ^{\mu \nu \rho \sigma }
 \bar \psi _\mu \gamma _\nu \psi _\rho \Nphi^I{\cal D}_\sigma X_I
 \nonumber\\
 &+& \left. \ft12\Nphi ^I k_{\alpha I}(-\rmi D^\alpha +  \bar
 \psi_L
 \cdot \gamma \lambda_R^\alpha) - 2 \Nphi _I{}^{J} k_{\alpha J}
 \bar \lambda _R^\alpha \Omchi^I  + {\rm h.c.}  \right\}\nonumber\\
 &+&\Nphi _J{}^I\left( \ft{1}{8}e^{-1}\varepsilon ^{\mu\nu\rho\sigma}\bar \psi _\mu \gamma
 _\nu \psi _\rho \bar \Omchi ^J\gamma _\sigma \Omchi _I
 -\bar \psi _\mu \Omchi ^J\,\bar \psi _\mu \Omchi _I\right)
  +\Nphi^{IJ}_{KL}\bar \Omchi _I \Omchi _J\,\bar \Omchi ^K \Omchi ^L
\nonumber\\ \nonumber\\ {} \lefteqn{[\mathcal{W}]_Fe^{-1}= {\cal
W}^I h_I -{\cal  W}^{IJ}\bar \Omchi _{I}\Omchi _{J}
 +{\cal  W}^I\bar \psi_R\cdot
\gamma  \Omchi _{I}+\ft12 {\cal  W}\bar \psi _{\mu R} \gamma ^{\mu \nu
}\psi _{\nu R}+ {\rm h.c.} } \nonumber\\ \nonumber\\ \lefteqn{\left[
f_{\alpha \beta }\bar \lambda_L ^\alpha \lambda_L ^\beta\right]
_Fe^{-1}=}\nonumber\\ &&\Re f_{\alpha \beta}(X)\left[ -\ft14 F_{\mu \nu
}^\alpha F^{\mu \nu \,\beta } -\ft12 \bar \lambda ^\alpha
\not\!\!\hat{\cal D}\lambda ^\beta +\ft12 D^\alpha D^\beta +\ft18 \bar
\psi _\mu \gamma ^{\nu \rho }\left( F_{\nu \rho }^\alpha+ \hat F_{\nu
\rho }^\alpha \right) \gamma ^\mu \lambda ^\beta \right] \nonumber\\
&+&\rmi \ft 14\Im f_{\alpha \beta}
 F_{\mu \nu }^\alpha \tilde F^{\mu \nu \,\beta }+\rmi \ft 14 \left( {\cal D}_\mu \Im f_{\alpha \beta}\right)
\bar \lambda ^\alpha \gamma _5\gamma ^\mu \lambda ^\beta \nonumber\\
&+&\left\{ \ft12f^I_{\alpha \beta}(X)\left[\bar \Omchi _{I}\left(
-\ft12\gamma ^{\mu \nu } \hat F_{\mu \nu }^{-\alpha }+\rmi D^\alpha
\right)\lambda _L^\beta  -\ft12\left(  h_I+\bar \psi _R\cdot\gamma \Omchi
_I\right) \bar \lambda _L^\alpha\lambda _L^\beta\right]\right.
\nonumber\\ &&\left. +\ft14 f^{IJ}_{\alpha \beta }\bar \Omchi _{I}\Omchi
_J \bar \lambda _L^\alpha
 \lambda _L^\beta + {\rm h.c.} \right\}
\label{fullaction}
\end{eqnarray}
where
\begin{eqnarray}
   {\cal D}_\mu &=&\partial _\mu -\rmi cA_\mu -W_\mu ^\alpha \delta _\alpha
   +\ft14 \omega _\mu {}^{ab}(e)\gamma _{ab}\,,\nonumber\\
\hat{{\cal D}}_\mu &=&\partial _\mu -\rmi cA_\mu -W_\mu ^\alpha
\delta _\alpha
   +\ft14 \hat{\omega} _\mu {}^{ab}(e,\psi )\gamma _{ab}\,,\nonumber\\
\hat{\omega} _\mu {}^{ab}(e,\psi )&=&\omega _\mu
{}^{ab}(e)+\ft14\left(2 \bar \psi _\mu \gamma ^{[a}\psi ^{b]}+\bar
\psi ^a\gamma _\mu \psi ^b\right)\,, \nonumber\\ \hat{F}^\alpha_{\mu
\nu }&=&F^\alpha_{\mu \nu }+\bar \psi _{[\mu }\gamma _{\nu ]}\lambda
^\alpha\,,\qquad F^\alpha_{\mu \nu }=2\partial _{[\mu }W^\alpha _{\nu
]}+W_\mu ^\beta W_\nu ^\gamma f_{\beta \gamma }^\alpha\,, \nonumber\\
\tilde F^{\mu \nu\,\alpha  }&=& \ft12 e^{-1}\varepsilon ^{\mu \nu
\rho \sigma } F^\alpha_{\rho \sigma }\,,\nonumber\\ \hat{F}^{-\alpha
}_{\mu \nu }&=&\ft12\left(\hat{F}^\alpha_{\mu \nu }
-\tilde{\hat{F}}{}^\alpha_{\mu \nu }\right) = F^{-\alpha }_{\mu \nu
}-\ft14 \bar \psi _{\rho L}\gamma _{\mu \nu }\gamma ^\rho \lambda
^\alpha _R +\ft14 \bar \psi_R\cdot \gamma\gamma _{\mu \nu }\lambda
^\alpha _R   \,.
   \label{covcalD}
\end{eqnarray}
Here $\delta _\alpha $ are the Yang--Mills transformations, see below,
and $A_\mu$, as in the example of the previous section, is the gauge
field of the $U(1)$ R-symmetry of the superconformal algebra. The terms in
the covariant derivative with the spin connection $\omega $ appear only
for fermions. The gauge field for the local special conformal symmetry
$b_\mu$ can be removed from the covariant derivative as explained in
footnote~\ref{fn:bmu}. The covariant derivative ${\cal D}_\mu \lambda
^\alpha ={\cal D}_\mu\lambda ^\alpha_L+ {\cal D}_\mu\lambda ^\alpha_R$
contains $\ft\rmi 2 A_\mu (P_L-P_R)\lambda ^\alpha= \rmi \ft 1 2 A_\mu
\gamma _5\lambda ^\alpha $.

The torsion terms in the first line of (\ref{fullaction}) are
\cite{supergravity}
\begin{equation}
  {\cal L}_{SG,\mathit{torsion}}=\ft1{16}\left[ 2(\bar \psi _\mu \gamma
  _\nu\psi_\rho)(\bar \psi^\nu \gamma^\mu\psi^\rho)+(\bar \psi _\mu \gamma
  _\nu\psi_\rho)^2-4(\bar \psi \cdot \gamma \psi _\mu )^2\right]\,.
\label{LSGtorsion}
\end{equation}

It remains to explain the notation related to the Yang--Mills symmetry.
In \cite{general} the conformon, called a compensator multiplet at that
time, was used as a separate multiplet, which did not participate in the
Yang--Mills transformations. In \cite{Bagger} a more general gauge group
was considered, also obtaining the Fayet--Iliopoulos terms,
 using the Killing vectors as the main ingredient.
Consequently, in \cite{Rel} it was shown that the Fayet--Iliopoulos terms
are obtained if the compensator multiplet transforms under the gauge
group. This was inspired by similar results in $N=2$ supergravity. We
will show in section~\ref{ss:gaugeIsom} that unifying the conformon with
all the other chiral multiplets, one can obtain the gauging of all the
Killing symmetries as in \cite{Bagger}. Therefore, the gauge
transformations are given here in terms of the conformal multiplets:
\begin{eqnarray}
\delta _\alpha  X_I=k_{\alpha I}(X)&\,, \qquad&
  \delta _\alpha  X^I=k_\alpha{}^I(X^*)\,,\nonumber\\
   \delta _\alpha \Omchi _I= k_{\alpha I}{}^J\Omchi _J&\,,\qquad&
   \delta _\alpha \Omchi ^I=k_\alpha {}^I{}_J\Omchi ^J\,.
\label{Killingv}
\end{eqnarray}
The vectors $k_{\alpha I}(X)$ are holomorphic and should be of
dilatational weight~1, i.e.\ using our notation that adding an index
indicates a derivative,
\begin{equation}
  k_{\alpha IJ}=k_\alpha {}^{IJ}=0\,,\qquad
   k_{\alpha I}{}^JX _J=k_{\alpha I}\,,\qquad
    k_\alpha{}^I{}_JX ^J=k_\alpha {}^I   \,.
\label{holKX}
\end{equation}
The commutators define the structure constants\footnote{$f$ is used
for structure constants and for the functions in (\ref{symbL}), but
the difference should be clear from the indices.}:
\begin{equation}
  k_{\beta I}{}^Jk_{\alpha J}-k_{\alpha  I}{}^Jk_{\beta  J}=f_{\alpha \beta
  }^\gamma k_{\gamma I}\,,
\label{structurec}
\end{equation}
which determine the transformations of the fields in the vector
multiplets, e.g.
\begin{equation}
\delta _\gamma  \lambda ^\alpha =\lambda ^\beta f_{\beta \gamma
}^\alpha \,. \label{deltalambda}
\end{equation}
The functions $\Nphi$, ${\cal W}$ and $f_{\alpha \beta }$ should be
invariant or covariant, i.e.\ \begin{equation}
  \Nphi^Ik_{\alpha I}+\Nphi_Ik_\alpha {}^I={\cal W}^Ik_{\alpha I}=0\,,\qquad
  f_{\alpha \beta }^I k_{\gamma I}+2f_{\delta (\alpha }f_{\beta )\gamma }^\delta
  =\rmi c_{\alpha \beta }\,,
\label{invfunct}
\end{equation}
where $c_{\alpha \beta }$ are real constants.

These Yang--Mills symmetries commute with the $SU(2,2|1)$ superconformal
symmetries. The bosonic part of the superconformal symmetry group
consists of the general coordinate transformations and Lorentz rotations,
the Weyl and chiral transformations, given through the weights in
table~\ref{tablefieldswc}, and the special conformal transformations
that have been used to eliminate $b_\mu $ as only transforming field.
The fermionic part consists of $Q$- and $S$-supersymmetries. Omitting
the fields that will be auxiliary, the transformations of the
independent fields are
\begin{eqnarray}
\delta e_\mu ^a&=&\ft12\bar \epsilon\gamma ^a\psi _\mu\,, \nonumber\\
\delta \psi _\mu  & = & \left( \partial _\mu  +\ft14 \omega _\mu
{}^{ab}(e,\psi )\gamma _{ab} +\ft 12\rmi A_\mu \gamma _5\right)\epsilon
-\gamma _\mu \eta\,, \nonumber\\
\delta X_I&=&\bar \epsilon _L\Omega _I\,,\nonumber\\
 \delta \Omega _I & = & \ft12 \gamma ^\mu
\left( {\cal D}_\mu X_I-\bar \psi _\mu \Omega _I\right)  \epsilon _R
+ \ft12 h_I\epsilon _L+X_I\eta _L\,, \nonumber\\
\delta W_\mu ^\alpha &=&-\ft12\bar \epsilon \gamma _\mu \lambda ^\alpha\,, \nonumber\\
\delta \lambda^\alpha  &=&\ft14\gamma ^{\mu \nu }\hat{F}_{\mu \nu
}^\alpha\epsilon +\ft12\rmi\gamma _5\epsilon  D^\alpha\,,
\label{confsusytransf}
\end{eqnarray}
where $\epsilon $ and $\eta $ are the parameters of $Q$- and
$S$-supersymmetry, respectively.

One may eliminate auxiliary fields, still maintaining the superconformal
invariance. The auxiliary fields are  $h_I$, $D^\alpha$ and $A_\mu $. The
values of these auxiliary fields are
\begin{eqnarray}
-h^J\Nphi_J{}^I&=&{\cal W}^I -\Nphi ^I{}_{JK} \bar\Omchi^{J} \Omchi^{K}
-\ft14f^I_{\alpha \beta }\bar \lambda_L^\alpha \lambda_L^\beta
\nonumber\\ (\Re f_{\alpha \beta })D^\beta  &=& {\cal P}_\alpha+  {\cal
P}^F_\alpha \nonumber\\ {\cal P}_\alpha&=& \rmi \ft 12\left[ \Nphi^I
k_{\alpha I}- \Nphi_I k_\alpha{}^I\right] =\rmi\Nphi^I k_{\alpha I}=-\rmi
\Nphi_I k_\alpha{}^I \nonumber\\ {\cal P}^F_\alpha&=&-\rmi\ft
12f^I_{\alpha \beta }\bar \Omchi _I\lambda ^\beta
+\rmi \ft 12f_{\alpha \beta I}\bar \Omchi ^I\lambda ^\beta    \nonumber\\
A_\mu &=&A_\mu ^B+A_\mu ^F\label{fullAmu}\\ A_\mu
^B&=&\frac{3\rmi}{2\Nphi}\left[ \Nphi^I\,\hat{\partial} _\mu X_I-
\Nphi_I\,\hat{\partial} _\mu X^I\right]= \frac{3\rmi}{2\Nphi}\left[
\Nphi^I\,{\partial} _\mu X_I- \Nphi_I\,{\partial} _\mu
X^I\right]-\frac{3}{\Nphi}W_\mu ^\alpha {\cal P}_\alpha \nonumber\\
A_\mu ^F&=&\frac{3\rmi}{2\Nphi}\left[\Nphi_I\bar \psi_{\mu R}\Omchi ^I
-\Nphi^I\bar \psi_{\mu L}\Omchi_I-\Nphi_I{}^J\bar \Omchi _J\gamma _\mu
\Omchi ^I -\ft34(\Re f_{\alpha \beta }) \bar \lambda ^\alpha
  \gamma _\mu \gamma _5\lambda ^\beta \right]\,,\nonumber
\end{eqnarray}
where $ \hat{\partial }_\mu $ denotes a derivative with only a YM
connection. We split the values of $D^\alpha $ and $A_\mu $ into a boson
and a fermion part. For the former, we introduce the notation ${\cal
P}_\alpha $ by default for the bosonic part, which will play an important
role below. Re-inserting the values of the auxiliary fields in the
action, some simplifications occur, such that the full action can be
written as\footnote{Note that after the elimination of the auxiliary
fields, the action is still fully superconformal invariant. The only
change is that now the commutator of two supersymmetries closes in new
transformations of the superconformal algebra only modulo field
equations.}
\begin{eqnarray}
e^{-1}{\cal L}&=&\ft16\Nphi\left[ R + \bar \psi_\mu R^\mu+ e^{-1}
\partial _\mu (e \bar \psi \cdot \gamma \psi ^\mu)+{\cal L}_{SG,\mathit{torsion}}
\right] \nonumber\\ &-&\Nphi _I{}^J\left[({\cal D}_\mu X^I)
  ({\cal D}^\mu X_J) + \bar \Omchi_{J}
 \not\!\! {\cal D} \Omchi^I+ \bar \Omchi^I
 \not\!\! {\cal D} \Omchi_J \right]\nonumber\\
&+&(\Re f_{\alpha \beta})\left[ -\ft14 F_{\mu \nu }^\alpha F^{\mu \nu
\,\beta } -\ft12 \bar \lambda ^\alpha \not\!\!\hat{\cal D}\lambda ^\beta
\right] +\rmi \ft 14(\Im f_{\alpha \beta})\left[
 F_{\mu \nu }^\alpha \tilde F^{\mu \nu \,\beta }- \hat{\partial }_\mu\left(
\bar \lambda ^\alpha \gamma _5\gamma ^\mu \lambda
^\beta\right)\right] \nonumber\\
 &-&\Nphi _I{}^Jh_J h^I-\ft12(\Re f_{\alpha \beta}) D^\alpha D^\beta  \nonumber\\
&+&\ft18(\Re f_{\alpha \beta})\bar \psi _\mu \gamma ^{\nu \rho
}\left( F_{\nu \rho }^\alpha+ \hat F_{\nu \rho }^\alpha \right)
\gamma ^\mu \lambda ^\beta  \nonumber\\ &+& \left\{\Nphi _K{}^{IJ}
\bar\Omchi_{I}(\hat{\not\! \partial}  X_J) \Omchi ^K +\Nphi _J{}^I
\bar \psi_{\mu L}(\not\!\! {\cal D} X^J)
 \gamma^\mu \Omchi_{I}
- \ft14f^I_{\alpha \beta} \bar \Omchi _{I}\gamma ^{\mu \nu } \hat
F_{\mu \nu }^{-\alpha } \lambda _L^\beta\right.\nonumber\\
 &&\ -\ft23\Nphi ^I\bar \Omchi_{I}  \gamma ^{\mu\nu}\hat {\cal D}_\mu \psi_{\nu L}
+\ft12 \bar  \psi_R  \cdot \gamma\left( \Nphi _I k_\alpha{}^I
\lambda_L^\alpha +2{\cal  W}^I  \Omchi _{I}\right) \nonumber\\
 && \ + \ft12 {\cal  W}\bar \psi _{\mu R} \gamma ^{\mu \nu }\psi _{\nu R}
 -{\cal  W}^{IJ}\bar \Omchi _{I}\Omchi _{J}
  - 2 \Nphi _I{}^{J} k_{\alpha J}  \bar \lambda _R^\alpha \Omchi^I
 \nonumber\\
&&\left.\ - \ft14f^I_{\alpha \beta}\bar \psi _R\cdot\gamma  \Omchi _I
\bar \lambda _L^\alpha \lambda _L^\beta
 +\ft14 f^{IJ}_{\alpha \beta }\bar \Omchi _{I}\Omchi _J \bar \lambda _L^\alpha
 \lambda _L^\beta + {\rm h.c.} \right\}\label{confact2}\\
 &+&\Nphi _J{}^I\left( \ft{1}{8}e^{-1}\varepsilon ^{\mu\nu\rho\sigma} \bar
\psi _\mu \gamma_\nu \psi _\rho \bar \Omchi ^J\gamma _\sigma \Omchi _I
 -\bar \psi _\mu \Omchi ^J\,\bar \psi _\mu \Omchi _I\right)
  +\Nphi^{IJ}_{KL}\bar \Omchi _I \Omchi _J\,\bar \Omchi ^K \Omchi ^L
 +\ft 19\Nphi (A_\mu ^F)^2  \,.\nonumber
\end{eqnarray}
In this action the auxiliary fields have the values from (\ref{fullAmu}).
The covariant derivatives have $U(1)$ connection with the bosonic part
$A_\mu ^B$ only. Thus, explicitly,
\begin{eqnarray}
{\cal D}_\mu X_I & = & \hat{\partial }_\mu X_I +\ft 13\rmi A_\mu ^B
X_I \,,\qquad \hat{\partial }_\mu X_I  =  \partial _\mu X_I-W_\mu
^\alpha k_{\alpha I}\,, \nonumber\\ {\cal D}_\mu \Omchi _I &=&\left(
\partial _\mu +\ft14 \omega _\mu {}^{ab}(e)\gamma _{ab}
 -\ft 16\rmi A^B_\mu\right)\Omchi _I-W_\mu ^\alpha k _{\alpha I}{}^J\Omchi
 _J\,,
   \nonumber\\
{\cal D}_\mu \lambda ^\alpha &=&\left( \partial _\mu +\ft14 \omega
_\mu {}^{ab}(e)\gamma _{ab} +\ft 12\rmi A^B_\mu \gamma _5\right)
\lambda ^\alpha  -W_\mu ^\gamma \lambda ^\beta f_{\beta \gamma
}^\alpha \,,    \nonumber\\ R^\mu &=&\gamma ^{\mu \rho \sigma }{\cal
D}_\rho \psi _\sigma \,,\qquad {\cal D}_{[\mu }\psi _{\nu ]}=\left(
\partial _{[\mu} +\ft14 \omega _{[\mu} {}^{ab}(e)\gamma _{ab} +\ft
12\rmi A^B_{[\mu} \gamma _5\right)\psi _{\nu ]}\,. \label{covderAB}
\end{eqnarray}
The $\hat{{\cal D}}_\mu $ differs as before from ${\cal D}_\mu $ in
that its spin connection contains $\psi $-torsion, see
(\ref{covcalD}). To arrive at some of the simplifications, observe
that
\begin{equation}
   \Nphi^I{\cal D}_\mu X_I=\Nphi_I{\cal D}_\mu X^I=
   \ft12\left( \Nphi^I\partial _\mu X_I+\Nphi_I\partial _\mu
   X^I\right)\,.
\label{NIDXI}
\end{equation}
The expression (\ref{confact2}) has the following structure. In the first
line it is $\Nphi(X,X^*)$ times the pure supergravity action. The second
line contains the kinetic terms of the chiral multiplets, which need the
first line in order to be conformally invariant, similar to what we saw
in the examples in section~\ref{ss:examplesCon}. The third line contains
the kinetic terms of the vector multiplets. The potential is contained in
the fourth line, after inserting the bosonic part of the solution of the
field equations for the auxiliary fields. Lines 5 and 6 contain
derivative interactions between the fields. The first term in the seventh
line will later disappear by gauge choices of $S$. The eighth line has
mass terms for the fermions, first of all for the gravitino. For the
other fields, the mass matrix has also contributions from the auxiliary
fields in line~4. The last two lines have only 4-fermion interactions.

Finally, we draw the readers attention to a \emph{global issue}, similar
to the discussion
 of the cosmic string at the end of section~\ref{ss:examplesCon}. The
 vanishing of the conformon field $|\phi|$ in that example is now
 generalized to the vanishing of the function ${\cal N}(X, \bar X)$.
 We
have a compact $U(1)$ group, and therefore the cohomology class of
 the 2-form gauge field strength is
quantized: if fields transform as $\psi\rightarrow U\psi$, then the gauge
field $A_\mu$, normalized so that it transforms as $\partial_\mu + \rmi
A_\mu \rightarrow U^{-1} (\partial_\mu + \rmi A_\mu) U$, has a field
strength $F_{\mu\nu}=\partial_\mu A_\nu -\partial_\nu A_\mu$, of which
the integral over an arbitrary 2-cycle is
\begin{equation}
[F]\equiv \int F_{\mu\nu}\,\rmd x^\mu\,\rmd x^\nu = 2\pi n\,,\qquad n\in
\Zbar\,. \label{Fquant}
\end{equation}
This is analogous to the quantization of the magnetic charge; a similar
effect for a topologically non-trivial  K{\"a}hler manifold is
 nicely explained in \cite{BagWitN1,BagBonn}
(see the next section). In more mathematical language this means that the
transformation functions should define a complex $U(1)$ line bundle,
whose first Chern class should be of integer cohomology \cite{Chern}. If
there are several fields with different $U(1)$ charges, they each define a
bundle, and should be multiples of a basic unit. The most stringent
condition is given by the bundle having this unit charge.

We now apply this consideration to the $U(1)$ transformations. If there
is a string with vanishing ${\cal N}$, then there are non-trivial loops
around this string and we have to care about the quantization of the
$U(1)$ curvature. This generalized cosmic string suggests an example when
the gravitational coupling constant $\kappa ^2=-3/{\cal N}$ goes to
infinity, and the Planck mass goes to zero.
\subsection{Gauge-fixing and \Ka\ geometry}\label{ss:gfKa}

The action (\ref{confact2}) has full superconformal symmetry. We will now
break the symmetries which are not present in the super-Poincar{\'e} group.
The gauge-fixing of local dilatations, will replace the modulus of the
conformon scalar, the non-physical scalar, which is part of $X_I$, with a
dimensionful constant, the Planck mass.

In this process, the relation will be established between the
superconformal action (\ref{confact2}) and the action describing a \Ka\
manifold. Furthermore, the $U(1)$ symmetry of the superconformal algebra
will be related to the \Ka\ symmetry.

The gauge choice of dilatations is chosen \cite{KugoUe} such that the
first term of (\ref{confact2}) becomes the standard Einstein term, as
in the examples in section~\ref{ss:examplesCon}:
\begin{equation}
D\mbox{-gauge:}\quad   \Nphi =-3M_P^2\,. \label{gfdil}
\end{equation}
This condition breaks the dilatations, and at the same time defines a
submanifold of the $n+1$ complex-dimensional space of the $X_I$.
Apart from the real condition (\ref{gfdil}), there is still the $U(1)$
symmetry such that the effective manifold is complex $n$ dimensional.

To clarify the structure one first performs a change of variables of
the $n+1$ variables $X_I$ to $\Yrho $, which will be the conformon
scalar, and $n$ physical scalars $z_i$, which are Hermitian coordinates
for parametrizing the \Ka\ manifold in the
 \Poin\ theory. One defines
\begin{equation}
  X_I=\Yrho\, x_I(z_i)\,,
\label{Xrhoxz}
\end{equation}
where $x_I(z_i)$ are arbitrary functions, as long as $\partial^i x_I$ is
a matrix of rank $n$. We use $\partial ^i$ for $\frac{\partial }{\partial
z_i}$ and $\partial_i$ for $\frac{\partial }{\partial z^i}$. The modulus
of $\Yrho $ is determined by the dilatational gauge condition, while its
modulus is the $U(1)$ gauge degree of freedom, which we can later choose
conveniently. The freedom of the arbitrary functions $x_I(z_i)$ is useful
to obtain a global atlas of charts to cover the \Ka\ manifold, whose
global structure is contained in (\ref{gfdil}).

The appearance of the \Ka\ structure of the manifold is identical to
the corresponding steps in $N=2$ special \Ka\ geometry. We review
this in a common language (and without breaking the $U(1)$
transformations) in appendix~\ref{app:elimA}. There, we show that the
kinetic terms of $X_I$ give rise to the \Ka ian sigma model with \Ka\
potential and metric
\begin{eqnarray}
  {\cal K}(z,z^*)&=&-3 \ln \left[-\ft13x^I(z^*) \Nphi_I{}^J(z,z^*)
  x_J(z)\right]\,,
  \nonumber\\
g^i{}_j\equiv  \partial ^i\partial _j {\cal K}&=& -3 (\partial
^iX_I)(\partial _jX^J)\partial
  ^I\partial _J\ln \Nphi\,.
\label{definK}
\end{eqnarray}
The $\partial ^i$ derivatives in the second expression can be taken
with constant $\Yrho$. The \Ka\ metric is thus the pullback of
$\partial _I\partial ^J\ln \Nphi$ to the surface (\ref{gfdil}).

In these variables the gauge-fixing of the dilatational invariance
(\ref{gfdil}) is given by
\begin{equation}
  \Yrho \Yrho ^*  \exp \left(- \ft{1}{3}{\cal K}\right) =M_P^2 = - \ft13{\cal N}
  \,.
\label{rho}
\end{equation}

The \Ka\ invariance has its origin in the non-uniqueness of the
splitting (\ref{Xrhoxz}). This creates an invariance under a
redefinition
\begin{equation}
  \Yrho'= \Yrho\, \rme ^{\ft13\Lambda _\Yrho (z)}\,,\qquad  x'_I=x_I\,
   \rme ^{-\ft13\Lambda _\Yrho   (z)}
\label{firstKatr}
\end{equation}
for an arbitrary holomorphic function $\Lambda _\Yrho (z)$. This
redefinition changes the \Ka\ potential to
\begin{equation}
{\cal K}'={\cal K}+\Lambda_\Yrho  (z)+\Lambda _\Yrho ^*(z^*)\,.
\label{KatrK}
\end{equation}
The properties of (\ref{tablewcFunct}) imply that ${\cal W}$ is of
the form (introducing the mass scale for later convenience)
\begin{equation}
   {\cal W}=\Yrho ^3 M_P^{-3}W(z)\,.
\label{superpotential}
\end{equation}
The \Ka\ transformations act therefore on the superpotential $W$ as
\begin{equation}
  W'=W\rme ^{-\Lambda _\Yrho (z)}\,,
\label{KatrW}
\end{equation}
leaving $ {\cal W}$ invariant.

The $U(1)$ invariance can be fixed by fixing the modulus of $\Yrho $. One
can again take a `clever choice'. For example, if the gravitino mass term
(the first term in the eighth line of (\ref{confact2})) is non-vanishing,
then one can choose a
\begin{equation}
 \mbox{\Ka\ symmetric } U(1)\mbox{-gauge:}\quad {\cal W}={\cal W}^*\,.
\label{U1gauge}
\end{equation}
This choice of the local $R$-symmetry  gauge-fixing leads to the action
of phenomenological $N=1$ supergravity as given in
\cite{general,Nilles,BL}. This gauge makes sense  only for ${\cal W}\neq
0$, as for ${\cal W}= 0$ the condition is empty.
If one is interested in theories where ${\cal W}=0$ in some
instances, one can use a
\begin{equation}
 \mbox{ non-singular at  ${\cal W}=0$ } U(1)\mbox{-gauge:}\quad \Yrho =\Yrho^*\,.
\label{U1gauge2}
\end{equation}
In this gauge  for $U(1)$, the theory
 is  non-invariant under the \Ka\ transformations
(\ref{firstKatr}). This implies that the remaining invariance is a
combination of chiral $U(1)$ and \Ka\ transformations. The action in this
form will be closer to the action of the phenomenological $N=1$
supergravity as given in \cite{BW,Binetruy}, where it was derived by the
superspace methods.

The nice feature of the $SU(2,2|1)$ superconformal action is that it
allows the derivation of these two forms of the action easily by
using these two gauges.

We still have to fix $S$-supersymmetry, to reduce the invariance group to
the super-Poincar{\'e} group. The dilatation gauge was chosen such that the
kinetic terms of the graviton do not mix with the scalars. A suitable
$S$-gauge avoids mixing of the kinetic terms of the gravitino with the
spin-$\ft12$ fields. That mixing occurs in the first term of the seventh
line of (\ref{confact2}). To eliminate them, we choose \cite{KugoUe}
\begin{equation}
  S\mbox{-gauge:}\quad \Nphi ^I \Omchi _{I}=0\,.
\label{Sgauge}
\end{equation}
As for the bosons, we now have to choose independent physical fermions.
We choose the fermions $\chi _i$, as they appear in the transformation
law of the $z_i$, such that
\begin{equation}
  \delta z_i=  M_P^{-1}\bar \epsilon \chi _i \,, \qquad
  \delta z^i= M_P^{-1}\bar\epsilon \chi^i \,.
\label{defdelzi}
\end{equation}
The relation is
\begin{equation}
  \Omchi _I= M_P^{-1}\Yrho\chi _i {\cal D}^ix_I= M_P^{-1}\chi _i\frac{1}{\Yrho ^*}
  \partial ^i\left( \Yrho \Yrho ^*x_i\right)\ \Rightarrow\
  \chi _i=M_P^{-1} g^{-1}{} _i^j\Yrho ^*({\cal D}_jx^J)
  \Nphi_J{}^I\Omchi_I\,,
\label{Omegatochi}
\end{equation}
where ${\cal D}^ix_I$ is also introduced in the appendix,
(\ref{DXI}). It satisfies
\begin{equation}
   {\cal D}^i x_I=\partial ^ix_I+\frac{1}{3}\left( \partial ^i{\cal K}\right)
   x_I=\frac{1}{\Yrho \Yrho ^*}\partial ^i\left( \Yrho \Yrho
   ^*x_I\right) \,,\qquad  \Nphi^I {\cal D}^ix_I=0\,.
\label{propDix}
\end{equation}
The latter equation implies that the definition (\ref{Omegatochi})
satisfies the $S$-gauge condition automatically. Note that this
definition, before the breaking of dilatational and $U(1)$
invariance, would imply that $\chi _i$ has (Weyl,chiral) weight
$(\ft12,\ft12)$ (while the scalars $z_i$ have $(0,0)$). This is how
the smallest charge $\ft16$ is avoided for the physical fields.

The quantization condition of the $U(1)$ curvature, can thus now be
formulated as a quantization due to \Ka\ transformations, as was
originally found in \cite{BagWitN1}. The smallest charge, determining the
most stringent quantization condition, seems according to
(\ref{tablewcFunct}) to be the chiral fermion $\Omchi _I$, which has
charge $\ft16$. However, we saw that some of these fields are still gauge
degrees of freedom. The remaining physical scalars $z$ have zero chiral
charge, while the spinors $\chi $ have charge $\ft12$, which is then the
remaining lowest one. Therefore, the condition~(\ref{Fquant}) should apply
to the integrals of
\begin{eqnarray}
F_{\mu \nu }^\mathit{quant}&=&\ft12(\partial _\mu A_\nu -\partial _\nu
A_\mu )= 3\rmi(\partial _{[\mu }
  X^I)( \partial _{\nu ]}X_J)\partial _I\partial ^J\ln \Nphi\nonumber\\
&=&3\rmi(\partial _{[\mu }z^i) (\partial _i X^I)( \partial _{\nu
]}z_j)(\partial ^jX_J)
\partial _I\partial ^J\ln \Nphi\nonumber\\
&=&-\rmi(\partial _{[\mu }z^i)( \partial _{\nu ]}z_j)\partial ^i\partial
_j {\cal K}\,.\label{Fmunuquant}
\end{eqnarray}
This curvature is the pullback of the \Ka\ 2-form
\begin{equation}
   \frac{\rmi}{2\pi }g_i^j\,\rmd z_j \wedge \rmd z^i\,.
\label{Ka2form}
\end{equation}
The bundle's first Chern class
\begin{equation}
c_1\in 2\Zbar\,.
\end{equation}
should thus be an (even) integer. In the mathematical literature
\cite{Chern,Wells}, K{\"a}hler manifolds of which the K{\"a}hler form is
of integer cohomology are called {\em K{\"a}hler manifolds of
restricted type} or {\em Hodge manifolds}. The action, as seen in
(\ref{KahlerL}), is proportional to the \Ka\ metric times $M_P^2$.
Constants in a non-trivial action should thus be $M_P^2$ times
integers.

\section{Gauge symmetries as isometries}\label{ss:gaugeIsom}
\subsection{Killing vectors and potentials}\label{ss:Killing}
We will now show that the Yang--Mills gauge transformations of the
scalars, which may also act on the conformon multiplet, are exactly
all the Killing isometries. Holomorphic Killing vectors $\xi_{\alpha
i}(z)$, with complex conjugate $\xi _{\alpha }{}^i(z^*)$, for the
\Ka\ metric $g_i^j$ satisfy the Killing equation
\begin{equation}
0=  g^j_k\partial _i\xi_\alpha{} ^k+g^k_i\partial^j\xi_{\alpha
k}+\left(
  \xi_{\alpha k}\partial ^k+\xi_\alpha{}^k \partial_k\right)g^j_i=
 \partial _i \partial^j\left( \xi_\alpha {}^k \partial _k {\cal K}
 +\xi_{\alpha k}\partial ^k{\cal K} \right) \,.
\label{Killingeq}
\end{equation}

In the conformal set-up, the transformations are defined from
transformations of the $X_I$, as in (\ref{Killingv}). The fact that the
functions $k_{\alpha I}(X)$ have (Weyl,chiral) weight $(1,-\ft13)$,
implies that these transformations can be expressed as
\begin{equation}
  \delta _\alpha \Yrho =\Yrho\, r_\alpha (z)\,,\qquad
  \delta _\alpha z_i= \xi_{\alpha i}(z)\,,
\label{delarhoz}
\end{equation}
where $r$ and $\xi $ are $n+1$ holomorphic functions for every
symmetry. They determine
\begin{equation}
  k_{\alpha I}=\Yrho \left[r_\alpha (z) x_I(z)+\xi_{\alpha i}(z)\partial
  ^ix_I(z)\right] \,.
\label{k_in_r_and_xi}
\end{equation}
The invariance of $\Nphi$ is (\ref{invfunct}), which reads
\begin{equation}
  0=\Nphi^Ik_{\alpha I}+\Nphi_Ik_\alpha{}^ I=\Nphi\left[ r_\alpha (z)
  +r^*_\alpha (z^*)-\ft{1}{3}\left(\xi _{\alpha i}
  \partial ^i {\cal K}(z,z^*) +\xi _\alpha {}^i\partial _i{\cal K}(z,z^*)\right)
  \right]\,,
\label{preKillingEqn}
\end{equation}
stating that the real part of $\xi _{\alpha i}(z)\partial ^i {\cal
K}(z,z^*) $ should be the real part of a holomorphic function, which is
then proportional to $r_\alpha (z)$. In other words, its derivative
$\partial _j\partial ^k$ should be zero. That leads precisely to the
Killing equation~(\ref{Killingeq}). This proves that the transformations
that are possible in the conformal framework are exactly the Killing
isometries!

Observe that $r_\alpha(z)$ describes the non-invariance of the \Ka\
potential:
\begin{equation}
  \delta _\alpha {\cal K}=\xi_{\alpha i}(z)\partial ^i{\cal K}(z,z^*)
  +\xi _\alpha {}^i\partial _i{\cal K}(z,z^*)=
  3(r_\alpha(z)+r_\alpha^*(z^*))\,.
\label{deltaalphaK}
\end{equation}
However, imaginary constants in $r_\alpha $ do not show up here. We also
find that the bosonic part of the value of the auxiliary field $D^\alpha $
is determined by
\begin{eqnarray}
 {\cal P}_\alpha(z,z^*)&=&\ft12\rmi M_P^2\left[\left(
 \xi_{\alpha i}(z)\partial ^i{\cal K}(z,z^*)-
\xi _\alpha {}^i\partial _i{\cal K}(z,z^*)\right)
-3 r_\alpha(z)+3 r_\alpha^*(z^*))\right]\label{bosD}\\ &=&
\rmi\,M_P^2\left(  \xi_{\alpha i}(z)\partial
^i{\cal K}(z,z^*)-3r_\alpha(z)\right) = \rmi\,M_P^2\left( -\xi _\alpha
{}^i\partial _i{\cal K}(z,z^*)+3r_\alpha^*(z^*)\right) \,. \nonumber
\end{eqnarray}
These real functions $ {\cal P}_\alpha(z,z^*)$,  called Killing
potentials, encode the transformations. Indeed, their derivatives
determine the Killing vectors:
\begin{equation}
\partial _i{\cal P}_\alpha(z,z^*)=\rmi\,M_P^2\,\xi _{\alpha j}g_i^j\,.
\label{Killvecfrpot}
\end{equation}

Invariance of other parts of the action demands
\begin{equation}
  \Yrho {\cal W}^I\,\xi _{\alpha i}\,\partial ^ix_I =-3r_\alpha {\cal
  W}\,,
\qquad \Yrho\,f_{\alpha \beta }^I \,\xi _{\gamma i}\,\partial ^ix_I
  +2f_{\delta (\alpha }f_{\beta )\gamma }^\delta =\rmi c_{\alpha
  \beta,\gamma   }\,,
\label{condinvWf}
\end{equation}
where $c_{\alpha   \beta,\gamma   }$ are real constants. Note that if
$r_\alpha \neq 0$, the transformation of the superpotential is
non-trivial. In terms of $W(z)$ it is
\begin{equation}
  \xi _{\alpha i}\partial ^iW=-3r_\alpha W\,.
\label{gtransfW}
\end{equation}
This statement is the expression that the superpotential should be
$R$-invariant in order to allow Fayet--Iliopoulos terms in supergravity
\cite{BarFerNanSte}. Indeed, we will now relate $r_\alpha \neq 0$ to FI
terms.

\subsection{New mechanism for $D$-terms}\label{ss:mechD}
In the usual context of K{\"a}hler geometry, Killing potentials are
determined up to constants (see, e.g., \cite{BW}, appendix~D), but here
these constants are determined by the functions $r_\alpha (z)$. This
important difference gives rise to a different way to understand FI terms.
In \cite{Bagger} the arbitrary constants are the cause of these FI terms,
while for us, as first recognized in \cite{Rel}, the gauge
transformations of the conformon multiplet, encoded in $r_\alpha (z)$,
are responsible for the FI terms.

First observe that $r_\alpha \neq 0$ signals the mixture of chiral
transformations and  gauge transformations with  index $\alpha $. Indeed,
after fixing the modulus of $\Yrho $ by some gauge choice, the remaining
invariance is the linear combination of gauge transformations that leaves
$\Yrho $ invariant, and this depends on $r_\alpha $. Another way to see
this is that the gravitino field couples to $A_\mu ^B$. This includes
higher-order couplings with other fields, and direct gauge couplings
between the vector $W_\mu^\alpha $ and the gravitino if the vacuum
expectation value of ${\cal P}_\alpha$ is non-zero. For unbroken gauge
symmetries, this happens when $\langle r_\alpha\rangle\neq  0 $.

We first give a simple example, with the trivial \Ka\ metric ${\cal
K}=z_iz^i$. The functions $x_I(z)$ are taken to be
\begin{equation}
  x_0=1\,,\qquad x_i=z_i\,.
\label{simpleparam}
\end{equation}
There is one gauge transformation, under which the scalars have a
charge $q_i$ in the sense that $\delta z_i= \rmi q_i z_i$. We
look for an imaginary constant in $r$, i.e.\ $r=\ft13\rmi\varpi M_P^{-2}$,
where the scaling with the Planck mass is done in view of the rigid limit
to be considered later. Then
the superpotential $ {\cal W}$ has to be $R$-invariant, and with
(\ref{superpotential}),
\begin{equation}
 q_i z_i \partial ^i W(z) =-M_P^{-2}\varpi W(z)\,.
\label{RinvariantW}
\end{equation}
If this is the case, the action is invariant and we find
\begin{equation}
   {\cal P}= -q_i M_P^2 z_iz^i +\varpi \,.
\label{Pexample1}
\end{equation}
For trivial kinetic terms for the vector ($f_{\alpha \beta }=1$), the
potential obtains a contribution $\ft12 {\cal P}^2$, which is the
Fayet--Iliopoulos cosmological constant $\ft12\varpi ^2$.

A second example with non-trivial \Ka\ potential is the one from
\cite{Bagger,BagBonn}. Here the \Ka\ potential is
\begin{equation}
  {\cal K}=-3\nu  \ln\left[ -\ft13(1+zz^*)^{-1/3}\right] \,,
\label{KaBagger}
\end{equation}
where $\nu $ is an arbitrary parameter.
The associated metric is
\begin{equation}
  g_z{}^z=\frac{\nu  }{(1+zz^*)^2}\,.
\label{metricBa2}
\end{equation}
The transformations are generated by the Killing potentials which
were given in \cite{Bagger,BagBonn} as
\begin{equation}
 {\cal P}_1=\frac\nu 2M_P^2\frac{z+z^*}
  {1+zz^*}\,,\qquad
  {\cal P}_2=-\frac{\rmi \nu }2 M_P^2 \frac{z-z^*}
  {1+zz^*}\,,\qquad
 {\cal P}_3=M_P^2 \frac{\nu zz^*} {1+zz^*}+\varpi\,,
\label{PDBag}
\end{equation}
where $\varpi $ is a parameter. One can then show that these generate
$SU(2)$ on $z$ for any value of $\varpi $, with transformation laws
\begin{equation}
  \delta _1z=\ft 12{\rmi}(z^2-1)\,,\qquad
  \delta _2z=\ft{1}{2 }(z^2+1)\,,\qquad
  \delta _3z =-\rmi z\,.
\label{deltazexB2}
\end{equation}
The same $SU(2)$ should be realized on $Y$, which fixes\footnote{We thank
Marco Zagermann for a remark on this issue.} $\varpi = - (1/2) \nu
M_P^2$, and the holomorphic functions $r_\alpha (z)$ are
\begin{equation}
  r_1=\ft 16\rmi\nu z\,,\qquad r_2=\ft16\nu z
  \,,\qquad r_3=\ft 13\rmi\varpi M_P^{-2}=-\ft 16\rmi\nu \,.
\label{rBag}
\end{equation}
These thus determine the transformations of the conformon field in this
example. The fact that $r_3$ and $ {\cal P}_3$ contain a constant is the
signature of the FI term for a a remaining $U(1)$ in the rigid
limit\footnote{ Note that if $z$ is of order $M_P^{-1}$, as we will
assume from section~\ref{ss:rigidlimit} onwards, the first two
transformations do not have a smooth limit $M_P\rightarrow \infty $, and
are thus not symmetries of the rigid limit. The third one is still
present in the rigid case, and $ {\cal P}_\alpha $ and $r_\alpha M_P^2$
have a finite limit, such that $\varpi $ gives the rigid FI term.}. The
quantization of (\ref{Fmunuquant}) implies here that $\nu $ should be an
integer.

We have thus shown that the superconformal tensor calculus, or
equivalently superspace methods, do allow the gauging and possible FI
terms for all Killing isometries, in contrast to the claim in a footnote
of \cite{Bagger}. For this result, we had to also include the conformon
multiplet in the gauge representation, as first remarked in \cite{Rel}.

\section{Derivation of phenomenological Lagrangians}
\label{ss:derPhenL}
\subsection{\Ka\ structure and potential}\label{ss:Kastructure}
The parametrizations (\ref{Xrhoxz}) and (\ref{Omegatochi}) are now used
to rewrite the Lagrangian in the form which shows its \Ka\ structure and
is closest to the form in \cite{general,BW,Binetruy}.
The formulae of appendix~\ref{app:elimA} 
are most useful for the translations. Further, note that from (\ref{rho})
we obtain
\begin{eqnarray}
g_i{}^j & = & \rme ^{{\cal K}/3}{\cal D}_ix^I\Nphi_I{}^J{\cal D}^jx_J
\nonumber\\ -3 & = & \rme ^{{\cal K}/3}\; x^I\; \Nphi_I{}^J\,x_J\,,
\label{forMatrixEqn}
\end{eqnarray}
and also using the last of equation~(\ref{propDix}), we arrive at an
$(n+1)\times (n+1)$ matrix equation
\begin{equation}
  \pmatrix{-3&0\cr 0&g_i{}^j}=\rme ^{{\cal K}/3}\pmatrix{x^I\cr {\cal
  D}_ix^I}\Nphi_I{}^J \pmatrix{x_J&{\cal D}^jx_J}\,.
\label{MatrixEqn}
\end{equation}
This equation shows that the $(n+1)\times (n+1)$ matrix
$\pmatrix{x_I&{\cal D}^ix_I}$ should be invertible. That is a requirement
on the choice of variables $z_i$. Moreover, as positive kinetic terms
imply that the matrix in the left-hand side should have the signature
$(-+\cdots +)$, the same should be true for $\Nphi_I{}^J$. The minus sign
finds it origin in the conformon scalar, absorbed in the vierbein.

The kinetic coupling of the vectors is now just a function of the
$z_i$, i.e.\ $f_{\alpha \beta }(z)$. The standard superpotential $W$
is defined via $ {\cal W}$ and $\Yrho$ in (\ref{superpotential}).

The potential consists of an $F$-term and a $D$-term:
\begin{eqnarray}
V&=&V_F+V_D\,,\nonumber\\
  V_F&=&\left. h^I\Nphi_I{}^Jh_J\right|_\mathit{bos}=
  {\cal W}^I\Nphi^{-1}{}_I{}^J{\cal W}_J\nonumber\\
&=&\rme ^{{\cal K}/3}\left[-\ft13{\cal W}^Ix_Ix^J {\cal W}_J+
 { \cal W}^I{\cal D}^ix_I  g^{-1}
 {}_i{}^j{\cal D}_jx^J{\cal W}_J\right]\nonumber\\
 &=&M_P^{-2}\rme ^{\cal K}\left[ -3WW^*+({\cal D}^iW)g^{-1}
 {}_i{}^j({\cal D}_jW^*)\right]\,,\nonumber\\
  V_D&=&\ft12\left.(\Re f_{\alpha \beta}) D^\alpha D^\beta\right|_\mathit{bos}=
 \ft{1}{2}(\Re f)^{-1\,\alpha \beta }  {\cal P}_\alpha {\cal P}_\beta\,.
\label{Vtotal}
\end{eqnarray}
To calculate the $F$-term of the potential, we went
from the coordinate basis $X_I$ to $(\Yrho ,z_i)$ and used
(\ref{MatrixEqn}) to find the inverse of $\Nphi_I{}^J$.
The covariant derivative on $W$ is
\begin{equation}
  {\cal D}^iW= \partial  ^i W+\left( \partial ^i {\cal K}\right) W\,.
\label{DiW}
\end{equation}
In general, the weight for the \Ka\ connection is defined in
appendix~\ref{app:elimA}. Note that in the last expression for $V_F$, the
first term is negative definite, while the second one is positive
definite. $V_D$ is positive definite,
\begin{equation}
  V_{F,+}=V+3{\cal W}{\cal W}^*=M_P^{-2}\rme ^{\cal K}({\cal D}^iW)g^{-1}
 {}_i{}^j({\cal D}_jW^*)\geq 0\,,\qquad V_D\geq 0\,,\qquad
  V_+=V_{F,+}+V_D\geq 0\,.
\label{V+}
\end{equation}

\subsection{Goldstino}\label{ss:Goldstino}
In applications which will follow, we will also fix the remaining \Poin\
supersymmetry. This will be done by setting a goldstino equal to zero. The
question is what is this goldstino? In the past, this has been looked at
for constant backgrounds \cite{general}, but in the cosmological
applications the scalar fields are time dependent in the background.
Therefore, we need a modification.

A first ingredient for this analysis is the part of the supersymmetry
transformation laws of the fermions involving scalars. In the conformal
theory, the fermionic transformations are given by (\ref{confsusytransf}).
The gauge-fixing of $S$-supersymmetry (\ref{Sgauge}) implies that the
parameter $\eta $ in these transformation laws is dependent on $\epsilon
$. We find for \Poin\ supersymmetry (only the part dependent on scalars)
\begin{equation}
 -2\Nphi\eta _L= 6M_P^2\eta _L=\Nphi^I \not\!\! {\cal D}X_I \epsilon _R
 +\Nphi^I h_I   \epsilon _L\,.
\label{etainPoin1}
\end{equation}
 Using (\ref{NIDXI}) with constant $\Nphi$ the first term
 vanishes, and for the second one we can use (\ref{fullAmu}) to
 derive
\begin{equation}
  \eta _L=-\ft12M_P^{-2}{\cal W}^*\epsilon _L\,.
\label{etainPoin2}
\end{equation}

When we consider the action (\ref{confact2}) in the $S$-gauge
(\ref{Sgauge}), there are (for vanishing vectors and up to quadratic
terms in fermions) two terms where gravitinos mix with the other
fermions, and these should give us a clue to which is the correct
goldstino. The terms are
\begin{equation}
  e^{-1}{\cal L}_\mathit{mix}=\Nphi _J{}^I \bar \psi_{\mu R}(\not\!\! {\cal D} X_I)
 \gamma^\mu \Omchi^J+\bar  \psi_R  \cdot \gamma\left(\ft12  \Nphi _I k_\alpha{}^I \lambda_L^\alpha
+{\cal  W}^I  \Omchi _{I}\right) +{\rm h.c.}  \label{Lmix}
\end{equation}
Let us therefore consider the following two spinors
\begin{equation}
 \upsilon^1_L=\ft12 \Nphi _I k_\alpha{}^I \lambda_L^\alpha
+{\cal  W}^I  \Omchi _{I}\,,\qquad \upsilon^2_L= \not\!\!{\cal D}X_I
\Nphi^I{}_J\Omega ^J\,. \label{Goldstino12}
\end{equation}
Under the superconformal transformation laws (\ref{confsusytransf}), they transform as
\begin{equation}
  \delta \upsilon^1_L=\ft12 \not\!\!{\cal D}{\cal W}\epsilon _R-\ft12
   V\epsilon _L +3{\cal W}\eta _L\,,\qquad
\delta \upsilon^2_L=-\ft12 \not\!\!{\cal D}{\cal W}\epsilon _R +
\ft12\not\!\!{\cal D}X_I \Nphi^I{}_J\not\!\!{\cal D}X^J \epsilon _L
+\not\!\!{\cal D}X_I\Nphi^I\eta _R   \,, \label{delvarpic}
\end{equation}
where we used the field equations for the auxiliary fields,
(\ref{fullAmu}), and $V$ is the potential (\ref{Vtotal}). We want to
obtain a combination whose variation is always non-zero for broken
supersymmetry. The first terms have an undetermined signature. Thus, even
before gauge-fixing any symmetry, it is clear that we have to consider as
the goldstino the field
\begin{equation}
  \upsilon= \upsilon^1+\upsilon^2\,,
\label{totvarpi}
\end{equation}
in order that the single derivative term cancels. This expression can
be written as
\begin{equation}
  \upsilon_L= \rmi\ft12\lambda_L^\alpha  {\cal P}_\alpha
+\chi _i M_P^{-4} \Yrho ^3  {\cal D}^iW +
M_P \not\!\hat{\partial }z_i \chi ^jg_j{}^i
 \,. \label{Goldstinof}
\end{equation}
After the $S$-gauge-fixing we can use (\ref{etainPoin2}) and
(\ref{NIDXI}) to obtain
\begin{equation}
    \delta \upsilon_L=\ft12  M_P^2 g^i{}_j \not\!\hat{\partial }z_i
    \not\!\hat{\partial }z^j \epsilon   _L-\ft12
    V_+ \epsilon   _L \,,
\label{delvarpi2}
\end{equation}
where $V_+$ is the positive-definite part of the potential given in
(\ref{V+}). When the scalars depend only on time,
 as we will assume in the cosmological models,
  the first term is (minus) the kinetic energy and has the same sign as the
  second term. Therefore, the variation is non-zero, and this is the
 goldstino.

The mixing terms in the action can then be rewritten as
\begin{eqnarray}
  e^{-1}{\cal L}_\mathit{mix}&=&2  \Nphi _J{}^I \bar \psi_{\mu R}\gamma ^{\nu \mu }
  \Omchi^J{\cal D}_\nu  X_I+\bar  \psi_R  \cdot \gamma\upsilon_L
+{\rm h.c.}  \nonumber\\
&=& 2M_Pg_j{}^i\bar \psi_{\mu R}\gamma ^{\nu \mu}\chi ^j\hat{\partial
}_\nu z_i+\bar  \psi_R  \cdot \gamma\upsilon_L+{\rm h.c.}
\label{Lmixf}
\end{eqnarray}

\subsection{Lagrangian and $U(1)$ gauge}\label{ss:LagU1}

Using the information on the \Ka\ structure, the action can be written as
\begin{eqnarray}
e^{-1}{\cal L}&=&-\ft12M_P^2\left[ R + \bar \psi_\mu R^\mu+{\cal
L}_{SG,\mathit{torsion}}\right]
-g_i{}^j\left[M_P^2(\hat{\partial }_\mu z^i)(\hat{\partial }^\mu
z_j)+
 \bar \chi _j  \not\!\! {\cal D} \chi^i+ \bar \chi^i
 \not\!\! {\cal D} \chi_j \right]\nonumber\\
&+&(\Re f_{\alpha \beta})\left[ -\ft14 F_{\mu \nu }^\alpha F^{\mu \nu
\,\beta } -\ft12 \bar \lambda ^\alpha \not\!\!\hat{\cal D}\lambda
^\beta \right] +\ft 14\rmi(\Im f_{\alpha \beta})\left[
 F_{\mu \nu }^\alpha \tilde F^{\mu \nu \,\beta }- \hat{\partial }_\mu\left(
\bar \lambda ^\alpha \gamma _5\gamma ^\mu \lambda
^\beta\right)\right] \nonumber\\
 &-&M_P^{-2} \rme ^{\cal K}\left[ -3 WW^*+({\cal D}^iW) g^{-1}
 {}_i{}^j({\cal D}_jW^*)\right] -\ft{1}{2}(\Re f)^{-1\,\alpha \beta }
 {\cal P}_\alpha {\cal P}_\beta
 \nonumber\\
&+&\ft18(\Re f_{\alpha \beta})\bar \psi _\mu \gamma ^{\nu \rho
}\left( F_{\nu \rho }^\alpha+ \hat F_{\nu \rho }^\alpha \right)
\gamma ^\mu \lambda ^\beta \nonumber\\
&+& \left\{M_Pg _j{}^i \bar
\psi_{\mu L}(\hat{\not\! \partial } z^j)
 \gamma^\mu \chi_i
 + \bar  \psi_R  \cdot \gamma\left[\ft12\rmi\lambda_L^\alpha
 {\cal P}_\alpha
+\chi _i \Yrho ^3M_P^{-4} {\cal D}^iW  \right]\right.
  \nonumber\\
 &&\ + \ft12 \Yrho ^3M_P^{-3}W\bar \psi _{\mu R} \gamma ^{\mu \nu }\psi _{\nu R}
- \ft14M_P^{-1}f^i_{\alpha \beta}\bar \chi _i\gamma ^{\mu \nu } \hat F_{\mu
\nu }^{-\alpha } \lambda _L^\beta \nonumber\\
 && \
 -\Yrho ^3M_P^{-5}({\cal D}^i{\cal D}^j  W)\bar \chi _i\chi _j
 +\ft 12\rmi(\Re f)^{-1\,\alpha \beta} {\cal P}_\alpha  M_P^{-1}f^i_{\beta \gamma }
\bar \chi _i\lambda ^\gamma
  - 2 M_P\xi _\alpha{}^i g_i{}^j  \bar \lambda ^\alpha \chi_j
 \nonumber\\
  &&\ + \ft14 M_P^{-5}\Yrho ^3({\cal D}^j W) g^{-1}{}_j{}^i
  f_{\alpha \beta i}\bar \lambda_R ^\alpha \lambda_R ^\beta
 \nonumber\\
&&\left.\ - \ft14 M_P^{-1}f^i_{\alpha \beta}\bar \psi _R\cdot\gamma  \chi _i
\bar \lambda _L^\alpha \lambda _L^\beta
 +\ft14  M_P^{-2}({\cal D}^i\partial ^jf_{\alpha \beta })\bar \chi _i\chi_j
  \bar \lambda _L^\alpha
 \lambda _L^\beta + {\rm h.c.} \right\}\label{phenomL}\\
 &+&g_j{}^i\left( \ft{1}{8}e^{-1}\varepsilon^{\mu\nu\rho\sigma}\bar
 \psi _\mu \gamma_\nu \psi _\rho \bar \chi ^j\gamma _\sigma \chi _i
 -\bar \psi _\mu \chi ^j\,\bar \psi^\mu \chi _i\right)
\nonumber\\ &+&M_P^{-2}\left( R_{ij}^{k\ell }-\ft12
g_i{}^kg_j{}^\ell\right) \bar \chi^i\chi ^j\bar \chi _k\chi
_\ell\nonumber\\ & +&\ft 3{64}M_P^{-2}\left( (\Re f_{\alpha \beta })
\bar \lambda ^\alpha
  \gamma _\mu \gamma _5\lambda ^\beta\right)^2
- \ft1{16}M_P^{-2}f^i_{\alpha \beta }\bar \lambda_L ^\alpha \lambda_L
^\beta g^{-1}{}_i{}^jf_{\gamma \delta j} \bar \lambda_R ^\gamma
\lambda_R ^\delta \nonumber\\ &+&\ft18(\Re f)^{-1\,\alpha \beta}
 M_P^{-2}\left( f^i_{\alpha \gamma  }\bar \chi _i\lambda ^\gamma- f_{\alpha
\gamma i }\bar \chi ^i\lambda ^\gamma \right) \left( f^j_{\beta
\delta}\bar \chi _j\lambda ^\delta -f_{ \beta \delta j}\bar \chi
^j\lambda ^\delta \right) \nonumber \,.
\end{eqnarray}
The fifth line is ${\cal L}_\mathit{mix}$ which could be written as in
(\ref{Lmixf}). We repeat where to find the definitions of the quantities
involved. $g_i{}^j$ is the \Ka\ metric, see (\ref{definK}). $F_{\mu \nu
}$ and $\hat{F}_{\mu \nu }$ are in (\ref{covcalD}). The covariant
derivative of $z$ is
\begin{equation}
  \hat{\partial }_\mu z_i= \partial _\mu z_i -W_\mu ^ \alpha (\delta
  _\alpha z_i)\,.
\label{hatdz}
\end{equation}
 $R^\mu$ and the covariant derivative of $\lambda ^\alpha $ are defined in
(\ref{covderAB}), where  $A^B_\mu $ can now be written as
\begin{equation}
  A_\mu ^B=\frac{1}{2}\rmi\left[ (\partial _i{\cal K})\partial _\mu z^i-(\partial
  ^i{\cal K})\partial _\mu z_i\right]+
  \frac{3}{2}\rmi\partial_\mu \ln\frac{\Yrho }{\Yrho ^*}
  +\frac{1}{M_P^2}W_\mu ^\alpha {\cal P}_\alpha \,.
\label{AmuBinz}
\end{equation}
$\hat{{\cal D}}_\mu $ differs as before from ${\cal D}_\mu $ in that its
spin connection contains $\psi $-torsion, see (\ref{covcalD}). The
covariant derivative on $\chi _i$ is
\begin{equation}
  {\cal D}_\mu \chi _i=\left( \partial _\mu +\ft14\omega _\mu
  {}^{ab}(e)\gamma _{ab}-\ft 12\rmi A_\mu ^B\right) \chi _i+\Gamma _i^{jk}\chi
  _j \hat{\partial }_\mu z_k  -W_\mu ^\alpha \left( \partial
  ^j\xi _{\alpha i}\right) \chi _j\,,
\label{Dchii}
\end{equation}
where the \Ka\ connection
\begin{equation}
  \Gamma_i^{jk} = g^{-1}{} _i^\ell \partial ^jg^k_\ell
\label{Kaconnection}
\end{equation}
has been used. Note that with our convention of raising and lowering
indices by complex conjugation, the non-vanishing connection coefficients
of the \Ka\ manifold are those with holomorphic indices $ \Gamma_i^{jk}$
and those with antiholomorphic indices: $\Gamma^i_{jk}$. That also
determines the \Ka\ curvature tensor
\begin{equation}
  R_{ij}^{k\ell }\equiv g_i^m\partial _j \Gamma _m^{k\ell}\,.
\label{Kacurv}
\end{equation}
To obtain the action in the form (\ref{phenomL}), we used the translation
formulae discussed at the end of appendix~\ref{app:elimA}.

The action (\ref{phenomL}) is invariant under the local \Poin\ group,
$Q$-supersymmetry and the gauge group with gauge fields $W_\mu ^\alpha $,
which are standard local symmetries of supergravity. In addition, we keep
here the local $R$-symmetry of the superconformal group, i.e.\ a local
$U(1)$ symmetry. Indeed, we did not choose a gauge for this $U(1)$, which
reflects itself in the presence of the phase of the complex field $\Yrho
$. Its modulus is fixed (see, e.g., (\ref{rho})), but its phase is left
arbitrary.

We already mentioned that when the gravitino mass  is non-vanishing ($
\Yrho ^3 W \neq 0$), it is convenient to take a real parameter in the
mass term. This is the choice (\ref{U1gauge}), which now reads
\begin{equation}
  \mbox{\Ka\ symmetric }U(1)\mbox{-gauge:}\quad
  \Yrho ^3 W=(\Yrho ^*)^3W^*\,.
\label{possU1gauge}
\end{equation}
With this choice, the contribution of the phase of $\Yrho $ in
(\ref{AmuBinz}) adds to the first term to combine in $\partial
_i{\cal G}=-\partial _i {\cal K}-\partial _i\ln W$, where ${\cal G}$ is the
invariant function
\begin{equation}
   \rme ^{-{\cal G}}=M_P^{-6} \rme ^{\cal K}|W|^2\,,
\label{invfcalG}
\end{equation}
which was introduced in \cite{N1YMmsg,general}. This combination of
the \Ka\ potential and superpotential occurs now because
(\ref{possU1gauge}) preserves the \Ka\ transformations of
(\ref{KatrK}) and (\ref{KatrW}), and (\ref{invfcalG}) is the
invariant combination. Also at other places this combination occurs,
e.g.
\begin{equation}
  \Yrho ^3W=M_P^6\rme ^{-{\cal G}/2}\,,\qquad
\Yrho ^3{\cal D}^iW=-M_P^6{\cal G}^i \rme ^{-{\cal G}/2}\,.
\label{combincalG}
\end{equation}
The resulting form of the action is that in \cite{general}. The advantage
of the general supergravity theory in this gauge is that the theory
depends only on one function ${\cal G}$. The disadvantage is that this
dependence includes $\ln W$ and therefore the action is singular at $W=0$.

However, our formulation allows us to easily choose any gauge for $U(1)$,
and so we can avoid the complications with the limit $W\rightarrow 0$, as
mentioned in the `Note added' to \cite{Rel} (and in \cite{BW} the action
has been given in this way).
 A simple choice  is to take
$\Yrho $ real, which, together with (\ref{rho}), means that
\begin{equation}
 \mbox{non-singular at }W=0\;  U(1)\mbox{-gauge:}\quad
 \Yrho^3 =  (\Yrho^* )^3=M_P^3 \rme ^{{\cal K}/2}\,.
\label{choise}
\end{equation}
The supergravity action in this gauge is that in (\ref{phenomL}), where
in lines six till eight in few places the values of $\Yrho^3$ as well as
$(\Yrho^* )^3$ have to be replaced by $M_P^3 \rme ^{{\cal K}/2}$. In the
rest of this paper we will use this as the supergravity Lagrangian. We
will explicitly present it (without 4-fermion terms)
 in section~\ref{ss:eqsGravitino}.

Also interesting is the form of $A_\mu ^B$ in the two gauge choices.
Using the K{\"a}hler-covariant gauge, one obtains
\begin{equation}
  A_\mu ^B=\ft12\rmi\left({\cal G}^i\hat{\partial }_\mu z_i-{\cal G}_i
  \hat{\partial }_\mu z^i\right) \,.
\label{ABKgauge}
\end{equation}
On the other hand, using the second gauge, one obtains
\begin{equation}
  A_\mu ^B=\ft12\rmi\left[(\partial_i{\cal K})  {\partial }_\mu z^i
  -(\partial ^i{\cal K}){\partial}_\mu z_i\right] +
  M_P^{-2}W_\mu ^\alpha {\cal P}_\alpha \,.
\label{AB2gauge}
\end{equation}
The last term is part of the covariant derivative in
(\ref{ABKgauge}). This expression makes clear that this term remains
in the limit $W\rightarrow 0$, as argued in \cite{Rel}.

\textit{From now on, we will adopt the gauge choice (\ref{choise}).} This
gauge condition breaks the $U(1)$ invariance, but it leaves invariant a
combination of this $U(1)$ with the \Ka\ transformations
(\ref{firstKatr}). We therefore find that the $U(1)$ transformations
contribute to the remaining \Ka\ transformations as
\begin{equation}
  \rmi\Lambda =\ft12 \left(\Lambda _\Yrho (z)-\Lambda^* _\Yrho
  (z^*)\right)\,.
\label{remainingU1}
\end{equation}
The remaining transformation of $\Yrho $ is then
\begin{equation}
  \Yrho '=\Yrho \exp\ft16\left(\Lambda _\Yrho +\Lambda^* _\Yrho\right)\,,
\label{rhoprime}
\end{equation}
consistent with (\ref{choise}) and the transformation of the \Ka\
potential (\ref{KatrK}). More details and
the \Ka\ covariant derivatives in this gauge are explained in
appendix~\ref{app:elimA}.
Most quantities transform under the resulting \Ka\ transformations as
\begin{equation}
  Q'=Q \rme ^{\rmi w_K \Lambda}  \,,
\label{Qwk}
\end{equation}
where $\Lambda $ is thus of the form (\ref{remainingU1}). In particular,
this is the case for the (complex) combination $m$ of $\rme ^{{\cal K}}$
and $W$,
\begin{equation}
  m  =\rme ^{{\cal K}/2}W\,,
\label{defm}
\end{equation}
which is related to the (real) gravitino mass,
\begin{equation}
  m_{3/2}  =|m|M_P^{-2},
\label{defmee}
\end{equation}
and for the fermions as we parametrize them now:
\renewcommand{\arraystretch}{1.5}
\begin{equation}
  \begin{array}{|c|cccccccc|}
\hline
 &m& m^*&  \psi _{\mu L}& \psi _{\mu R}& \chi _i & \chi ^i & \lambda _L^\alpha & \lambda _R^\alpha  \\  \hline
w_K  &-1 &1 &-\frac12&\frac12&\frac12& -\frac12& -\frac12 &\frac12 \\
\hline
\end{array}
 \label{tablefieldswk}
\end{equation}
\renewcommand{\arraystretch}{1}
The \Ka\ covariant derivatives of quantities transforming as $Q$ can
be written as
\begin{equation}
  {\cal D}_i Q=\partial _i Q +\ft12 w_K (\partial _i{\cal K}) Q \,,\qquad
{\cal D}^i Q=\partial^i Q -\ft12 w_K (\partial ^i{\cal K}) Q\,.
\label{Kacovder}
\end{equation}
For example, the covariant derivative on $W$, (\ref{DiW}), combines with
the derivative on the \Ka\ potential, and we have as in (\ref{Kacovder})
\begin{eqnarray}
&&m^i\equiv {\cal D}^i m\phantom{^*}  =   \rme ^{{\cal K}/2} {\cal D}^i
W\phantom{^*}=
 \partial ^i m\phantom{^*} +\ft12 (\partial ^i {\cal K}) m\phantom{^*}\,,\qquad
{\cal D}_i m\phantom{^*} \, =  \, \partial _i m\phantom{^*}
-\ft12 (\partial _i {\cal K}) m\phantom{^*} =0\,,\nonumber\\
&&m_i\equiv {\cal D}_i m^*  =   \rme ^{{\cal K}/2} {\cal D}_i W^*=
 \partial_i m^* +\ft12 (\partial _i {\cal K}) m^*\,,\qquad
{\cal D}^i m^* \, =  \, \partial ^i m^*
-\ft12 (\partial^i {\cal K}) m^* =0\,.\nonumber\\
&&\label{defmima}
\end{eqnarray}

\subsection{Rescalings for a rigid limit}\label{ss:rigidlimit}
The limit from a supergravity theory to a supersymmetry theory is not
always obvious. For $N=2$, a procedure has been investigated in
section~10.1 of \cite{Andrianopoli}, and this has been generalized in
section~2.3 of \cite{Billo}. It involves the expansion of the \Ka\
potential from a `classical point' in the space of the scalars.
In \cite{Billo} this was related to singular points of a Calabi--Yau
surface, where the latter degenerates such that the expansions around
these points give rise to rigid $N=2$ supersymmetric theories.  In
the general case only a subset of the scalars of the supergravity
theory appear in the rigid theory.

We will consider the simplest situation in $N=1$ supergravity, in which
all scalar fields appear in the rigid limit. Our aim is not to go to the
rigid limit, but to parametrize the scalar fields such that this limit
can be taken easily. We expand around the point $z=\stackrel{0}{z}$, with
$M_P^{-1}$ as the expansion parameter, and using $\phi _i$ for the fields
to be used below:
\begin{equation}
  z_i=\stackrel{0}{z}_i+M_P^{-1} \phi _i\,.
\label{zepsphi}
\end{equation}
The \Ka\ potential is expanded as
\begin{equation}
  {\cal K}= K_0 + M_P^{-1} \left(K_i\phi ^i+ K^i\phi _i\right) +
  M_P^{-2}  K(\phi ,\phi ^*,M_P^{-1} )\,,
\label{expK}
\end{equation}
where $K_0$ (real) and $K^i=(K_i)^*$ are constants and
$K(\phi ,\phi ^*,M_P^{-1})$ is regular at $M_P^{-1}=0 $.
The first terms are not physical, as they can be removed by \Ka\
transformations (\ref{KatrK}).
$K_i$ is proportional to
\begin{equation}
  K_i\propto \left. x_I\, \Nphi^I{}_J\,
 \partial_ix^J\right|_{z= \stackrel{0}{z}} \,.
\label{Ki}
\end{equation}
By a convenient choice of the functions $x_I(z)$ one can obtain $K_i=0$,
e.g.\ taking the special coordinates
\begin{equation}
  x_0=1\,,\qquad x_i=M_P^{-1}\phi _i\,.
\label{specCoor}
\end{equation}
However, this is not essential for what follows. On the other hand, the
constant $K_0$ can be set to zero by a real \Ka\ transformation $\Lambda
_\Yrho =\Lambda _\Yrho ^*=-\ft12 K_0$. That does not affect the
quantities in (\ref{tablefieldswk}), although, for example, $W$
separately (not in the combination $m$) is transformed. However, the
relevant quantities below transform as in (\ref{Qwk}), and are thus not
affected by this real \Ka\ transformation. The essential part of the \Ka\
potential ${\cal K}$ is thus the last term of (\ref{expK}).

Note that the \Ka\ metric is
\begin{equation}
  g^i{}_j =\frac{\partial }{\partial z_i}\frac{\partial }{\partial z^j}{\cal
  K}=\frac{\partial }{\partial \phi _i}\frac{\partial }{\partial \phi ^j} K\,,
\label{gijisgij}
\end{equation}
i.e.\ it does not change under the reparametrizations. Therefore, the
kinetic term for the scalars in (\ref{phenomL}) will lose its dependence
on $M_P$ by the reparametrization as chosen in (\ref{zepsphi}). This is,
in fact, the motivation for the proportionality factor $M_P^{-1}$ in
(\ref{zepsphi}). This also has the consequence that the fields $\phi _i$
again have the same dimension as the conformal fields $X_I$. The relation
can be made more direct in the special coordinates and in the \Ka\ gauge
$K_0=0$. Then, $\Yrho = M_P \exp [K/(6M_P^{2})]$ and
\begin{equation} X_0=\Yrho =M_P +{\cal O}(M_P^{-1})\,,\qquad
  X_i=\Yrho x_i= \Yrho M_P^{-1}\phi _i= \phi _i\exp [K/(6M_P^{2})] = \phi
  _i +{\cal O}(M_P^{-2})\,.
\label{Xispecial}
\end{equation}
Therefore, in the lowest order of $M_P^{-1}$, the fields $\phi _i$ are
equal to the conformal fields that we started from. Similarly, in this
case the conformal fermions $\Omega _I$ are
\begin{equation}
  \Omega _0={\cal O}(M_P^{-1})\,,\qquad \Omega _i=\chi _i+{\cal O}(M_P^{-2})\,.
\label{Omegaspecial}
\end{equation}
Therefore, this natural parametrization again makes close contact with
the conformal fields.

{}From now on, we will thus use the fields $\phi _i$ and complex
conjugates $\phi ^i$ rather than $z_i$ and $z^i$ to indicate the scalar
fields. Therefore, \emph{derivatives $\partial ^i$ will denote
derivatives with respect to $\phi_i$ rather than with respect to $z_i$}.
The difference is thus a factor $M_P$; e.g.\ from now on,
\begin{equation}
  f^i_{\alpha \beta }=\frac{\partial }{\partial \phi _i}f_{\alpha \beta
  }=M_P^{-1}\frac{\partial }{\partial z_i}f_{\alpha \beta }\,.
\label{fialphabetaNew}
\end{equation}
The rule for \Ka\ covariant derivatives in (\ref{Kacovder}) does not have
to be changed as all derivatives are now with respect to $\phi $. One can
also check that $ {\cal P}_\alpha $ has a finite rigid limit. Indeed,
considering its value (\ref{bosD}), one checks that
\begin{eqnarray}
\xi _{\alpha i} & = & \delta _\alpha z_i=M_P^{-1}\delta _\alpha \phi _i\,,
 \nonumber\\
\frac{\partial }{\partial z_i}{\cal K} & = & M_P^{-1} \frac{\partial
}{\partial \phi_i} K\,, \label{rigidlimitP}
\end{eqnarray}
where we assume, as argued above, that the linear terms in (\ref{expK})
are removed. This shows the finite limit for the first terms in
(\ref{bosD}). For the second terms, we have shown in
section~\ref{ss:mechD} how  $r_\alpha M_P^{-2}$ can also have a finite
rigid limit.

\section{Equations for the gravitino}\label{ss:eqsGravitino}
\subsection{Simplified action}\label{ss:simplAction}
In this section we {\it  omit the 4-fermion interactions} and use the
gauge $\Yrho=\Yrho^*= M_P \rme ^{ {\cal K}/6}$. We will also put the
gauge part of the theory  in the end of the action, since we plan to
focus mainly on the gravitino part of the theory,
\begin{eqnarray}
e^{-1}{\cal L}&=&-\ft12M_P^2 R
-g_i{}^j(\hat{\partial }_\mu \phi ^i)(\hat{\partial }^\mu \phi _j)
-V\nonumber\\
&-&\ft12 M_P^2\bar \psi_\mu R^\mu+
\ft12 m\,\bar \psi _{\mu R} \gamma ^{\mu \nu }\psi _{\nu R}
+\ft12 m^*\bar \psi _{\mu L} \gamma ^{\mu \nu }\psi _{\nu L}
\nonumber\\
& -&g_i{}^j \left[\bar \chi _j  \not\!\! {\cal D} \chi^i+ \bar \chi^i
 \not\!\! {\cal D} \chi_j \right]
  - m^{ij}\bar \chi _i\chi _j
 -  m_{ij}\bar \chi ^i\chi ^j
+e^{-1}{\cal L}  _\mathit{mix} \nonumber\\
&-& 2 m_{i\alpha} \bar \chi ^i\lambda^\alpha -
2 m^i{}_\alpha \bar \chi _i\lambda^\alpha
- m_{R,\alpha \beta } \bar \lambda_R ^\alpha \lambda_R ^\beta
- m_{L,\alpha \beta } \bar \lambda_L ^\alpha \lambda_L ^\beta
 \nonumber\\
&+&(\Re f_{\alpha \beta})\left[ -\ft14 F_{\mu \nu }^\alpha F^{\mu \nu
\,\beta } -\ft12 \bar \lambda ^\alpha \not\!\!{\cal D}\lambda ^\beta
\right] +\ft 14\rmi(\Im f_{\alpha \beta})\left[
 F_{\mu \nu }^\alpha \tilde F^{\mu \nu \,\beta }- \hat{\partial }_\mu\left(
\bar \lambda ^\alpha \gamma _5\gamma ^\mu \lambda
^\beta\right)\right] \nonumber\\ &+&\ft14\left\{ (\Re f_{\alpha
\beta})\bar \psi _\mu \gamma ^{\nu \rho } F_{\nu \rho }^\alpha \gamma
^\mu \lambda ^\beta
 - \left[ f^i_{\alpha \beta}\bar \chi _i\gamma ^{\mu \nu }
 F_{\mu \nu }^{-\alpha } \lambda _L^\beta +{\rm h.c.} \right] \right\} \,.
 \label{shortphenL}
\end{eqnarray}
The potential $V$ is given in (\ref{Vtotal}). The gravitino mass\footnote{Our
symbol $m$ does not represent a quantity with mass dimension~1. This choice is inspired
by the rigid limit explained in section~\ref{ss:rigidlimit}. The phase of $m$
is also a \Ka\ gauge degree of freedom, as can be seen from (\ref{Qwk}) and
(\ref{tablefieldswk}). The physical mass of the gravitino is $m_{3/2}=M_P^{-2}|m|$.}
is given in (\ref{defm})
and the mass matrix for the other fermions contains, using the
notation (\ref{defmima}),
\begin{eqnarray}
m^{ij} & = &  {\cal D}^i{\cal D}^j  m
= \left(\partial ^i+\ft12 (\partial ^i {\cal K})\right)
m^j-\Gamma ^{ij}_k m^k\,,
\nonumber\\
m_{i\alpha }&=&-\rmi\left[ \partial _i{\cal P}_\alpha -\ft 14 (\Re
f)^{-1\, \beta\gamma} {\cal P}_\beta f_{\gamma\alpha\,i  }\right]\,,
 \nonumber\\
m_{R,\alpha \beta }&=&-\ft14 f_{\alpha \beta i}  g^{-1}{}^i{}_j\, m^j \,,
\label{fermionmasses}
\end{eqnarray}
where the connection $\Gamma ^{ij}_k  $ is given in (\ref{Kaconnection})
interpreting in that formula again $\partial ^j$ as a derivative with
respect to $\phi _j$. Note that the fermion masses in this gauge are, in
principle, complex. However, in the applications we will mostly consider
real scalars, and hence this mass will be real.
 ${\cal L}_\mathit{mix}$ can be written in different ways as mentioned
 before:
\begin{eqnarray}
e^{-1}{\cal L}  _\mathit{mix}&=&
g _j{}^i \bar \psi_{\mu L}(\hat{\not\! \partial } \phi^j)
 \gamma^\mu \chi_i+ \bar  \psi_R  \cdot \gamma\upsilon^1_L+{\rm h.c.}
 \nonumber\\
&=& 2g_j{}^i\bar \psi_{\mu R}\gamma ^{\nu \mu}\chi ^j\hat{\partial }_\nu
\phi_i+\bar  \psi_R  \cdot \gamma\upsilon_L+{\rm h.c.}  \,,
\label{Lmixresc}
\end{eqnarray}
where
\begin{eqnarray}
  \upsilon_L&=& \upsilon^1_L+\upsilon^2_L\,,\nonumber\\
\upsilon^1_L&=& \ft12\rmi {\cal P} _\alpha \lambda _L^\alpha +m^i
\chi_i\,,\qquad \upsilon^2_L=
 (\not\!\hat{\partial }\phi_i) \chi ^jg_j{}^i  \,. \label{Goldstino}
\end{eqnarray}
The covariant derivatives on the scalar fields still contain a gauge
connection, while that on the fermions $\chi _i$ also contain Lorentz,
gauge and \Ka\ connections:
\begin{eqnarray}
  {\cal D}_\mu \chi _i&=&\left( \partial _\mu +\ft14\omega _\mu
  {}^{ab}(e)\gamma _{ab}\right) \chi _i -W_\mu ^\alpha \chi _j \partial
  ^j\xi _{\alpha i}   -\frac{\rmi}{2M_P^2}W_\mu ^\alpha {\cal P}_\alpha\chi _i
  \nonumber\\ &&+\ft14\left[ (\partial _j{\cal K}){\partial}_\mu \phi ^j -
  (\partial ^j{\cal K}){\partial}_\mu \phi _j\right] \chi _i
  +\Gamma _i^{jk}\chi
  _j \hat{\partial }_\mu \phi_k \,,
\label{Dmuchi}
\end{eqnarray}
where
\begin{equation}
  \partial   ^j\xi _{\alpha i}=\frac{\partial }{\partial z_j}\delta _\alpha
  z_i=\frac{\partial }{\partial \phi_j}\delta _\alpha\phi_i\,.
\label{deljchii}
\end{equation}

The parts of the supersymmetry transformation laws of the fermions
where they transform to bosons, and boson transformations
linear in fermions, are determined by
(\ref{confsusytransf}):
\begin{eqnarray}
\delta e_\mu ^a&=&\ft12\bar \epsilon\gamma ^a\psi _\mu \,,\qquad
\delta \phi _i=\bar \epsilon _L\chi _i\,,\qquad
\delta W_\mu ^\alpha =-\ft12\bar \epsilon \gamma _\mu \lambda
^\alpha \nonumber\\
\delta \psi _{\mu L}  & = &
\left( \partial _\mu  +\ft14 \omega _\mu {}^{ab}(e)\gamma _{ab}
+\ft 12\rmi A_\mu^B \right)\epsilon_L +\ft12 M_P^{-2}m\gamma _\mu
\epsilon _R \nonumber\\
\delta \chi _i& = & \ft12\not\! \hat{\partial }\phi_i \epsilon _R-
\ft12 g^{-1}{} _i^j m_j\epsilon _L
 \nonumber\\
\delta \lambda^\alpha  &=&\ft14\gamma ^{\mu \nu }
F_{\mu \nu }^\alpha+\ft12\rmi \gamma _5
(\Re f)^{-1\,\alpha \beta}{\cal P} _\beta
\epsilon  \,.
\label{gfsusyphenom}
\end{eqnarray}

\subsection{Field equations}\label{ss:fieldEqsGr}
For further analysis, we will restrict ourselves to backgrounds with
\emph{vanishing vector fields}. To analyse the gravitino propagation, we
first write down the field equations. We thus delete the vectors and
terms quadratic and cubic in spinors. To keep the flexibility in the
choice of gauge of the local supersymmetry at a later stage, we present the
equations of motion without choosing any gauge. Then the field equations
are the vanishing of
\begin{eqnarray}
S_{\mu \nu }&=&M_P^{2}G_{\mu \nu }+\partial _\mu \phi^ig_i{}^j\partial _\nu \phi_j
+\partial _\nu \phi^ig_i{}^j\partial _\mu \phi_j-
g_{\mu \nu }\left(\partial _\rho \phi^ig_i{}^j\partial ^\rho  \phi_j+V \right)
\nonumber\\
S_i&=&g_i{}^j{\cal D}_\mu \partial ^\mu \phi_j-\partial _iV
\nonumber\\
\Sigma ^\mu_R&= &M_P^{2}R^\mu _R - \gamma ^{\mu \nu }
\left[ m\psi _{\nu R}- 2\chi ^jg_j{}^i\partial _\nu \phi_i \right]
-\gamma ^\mu \upsilon_L \nonumber\\
\Sigma ^i&=&g^i{}_j\not\!\!{\cal D}\chi ^j+m^{ij}\chi _j
+m^i{}_\alpha \lambda ^\alpha _L
-\ft12\gamma^\mu \not\!\partial \phi^jg_j{}^i  \psi _{\mu L}
+\ft12\mxi^i\,\gamma \cdot \psi _R
\label{fe1}\\
\Sigma _{\alpha \,R}&=&(\Re f_{\alpha \beta })\not\!\!{\cal D}\lambda _L^\beta +2m_{i\alpha }\chi
^i+2m_{R\alpha \beta }\lambda _R^\beta -\ft14\left( f^i_{\alpha \beta }
\not\!\partial \phi_i-f_{\alpha \beta \,i}\not\!\partial \phi^i\right)\lambda _L^\beta
-\ft12\rmi {\cal P}_\alpha \gamma \cdot \psi _L\nonumber\,.
\end{eqnarray}
The definition of $G_{\mu \nu }$ is given in appendix~\ref{app:notations}.

In the scalar field equation\footnote{Observe that the covariant
derivative contains Christoffel connection and \Ka\ connection: ${\cal
D}_\mu
\partial ^\mu \phi _i=\partial _\mu \partial ^\mu \phi _i+ \Gamma ^\mu
_{\mu \nu }\partial ^\nu \phi _i+\Gamma ^{jk}_i\partial _\mu \phi _j
\partial _\mu \phi _k$.}, $\partial _iV$
is the variation of the potential, which can be written as
\begin{eqnarray}
V&=&-3M_P^{-2}|m|^2+m_ig^{-1}{} ^i_jm^j
+\ft12{\cal P}_\alpha (\Re f)^{-1\,\alpha \beta}{\cal P}
_\beta=-3M_P^{-2}|m|^2+V_+
\nonumber\\
\partial _i V&=&-2M_P^{-2}m\,\mxi_i+m_{ij}g^{-1\,j}{}_k\,\mxi^k+\rmi  m_{i\alpha }
(\Re f)^{-1\,\alpha \beta }{\cal P}_\beta \,.
\label{diV}
\end{eqnarray}
The price for not fixing the supersymmetry gauge is that the
equations are not independent. We have the identity
\begin{eqnarray}
&&\ft12S_{\mu \nu }\gamma ^\mu \psi ^\nu_L + S_i\chi ^i
+{\cal D}_\mu \Sigma ^\mu _R +\ft12M_P^{-2} m\gamma _\mu \Sigma ^\mu _L
\nonumber\\ &&
+ (\not\!\partial \phi_i)\Sigma ^i+ \mxi^j g^{-1}{} ^i_j\Sigma _i
+\ft12\rmi(\Re f_{\alpha \beta })^{-1} {\cal P}_\beta\Sigma _{\alpha
\,R}=0\,.
\label{susyinv}
\end{eqnarray}
The coefficients of the field equations in (\ref{susyinv}) are the supersymmetry
transformation laws in (\ref{gfsusyphenom}).

It may be useful to present the following equations which follow from the
definition of $R_\mu $:
\begin{eqnarray}
&&\gamma ^\mu R_\mu =2\gamma  ^{\mu \nu } {\cal D}_\mu \psi _\nu\,,
\nonumber\\
&&{\cal D}_\mu R^\mu =-\ft{1}{2}G_{\mu\nu}\gamma^\mu\psi^\nu+\rmi \tilde
  F^\mathit{quant}_{\mu \nu }\gamma ^\mu \psi ^\nu \,,\nonumber\\
  && R_\mu-\ft{1}{2}\gamma_\mu\gamma\cdot
  R  = \not\!\!  {\cal D} \psi_\mu -
 {\cal D}_\mu    \gamma \cdot  \psi\,.
\label{DR}
\end{eqnarray}
$F_{\mu \nu }^\mathit{quant}$ is the $U(1)$ curvature given in
(\ref{Fmunuquant}). The covariant derivatives indeed contain a $U(1)$
connection, and we also have to add Christoffels in order for the
covariant derivative of the vierbein to be zero:
\begin{equation}
{\cal D}_\mu\psi_\nu = \left(\partial_\mu
+ \ft14 \omega_\mu^{ab}(e) \gamma _{ab}
+\ft12\rmi A_\mu ^B\gamma _5
\right)\psi _\nu - \Gamma_{\mu\nu}^\lambda\psi_\lambda\,. \label{PvNRRel3}
\end{equation}

\subsection{Master gravitino field equation and its constraints}
\label{ss:MasterGrav} Considering the gravitino field equation, $\Sigma
^\mu =0$, we find that there is a combination of fermions and derivatives
of the bosons for which we will introduce a notation:
\begin{equation}
  \Upsilon_\mu \equiv g_j{}^i\left(  \chi_i\partial _\mu
  \phi^j +\chi ^j\partial _\mu \phi_i \right) \,.
\label{defUpsilon}
\end{equation}
The first term is its left-chiral part, and the second term
is the right-chiral part. Note that
\begin{equation}
  \upsilon^2= \gamma ^\mu\Upsilon_\mu\,,
\label{ups2Ups}
\end{equation}
and that in cosmological applications only $\Upsilon_0$ is non-zero.

Another useful notation in order to go to the Majorana form of the
fermions, is to introduce the complex mass as a matrix:
\begin{equation}
\begin{array}{rclrcl}
\mathbf{m}&=&\Re m -\rmi\gamma _5\, \Im m\,,\qquad &
\mathbf{m}^\dagger&=&\Re m +\rmi\gamma _5\, \Im m \\
    & = & P_R m + P_Lm^* \,,\qquad  &   & = &
   P_R m^*  + P_Lm \\[2mm]
  m & = & P_R \mathbf{m} + P_L\mathbf{m}^\dagger  &
  m^* & = &P_R \mathbf{m}^\dagger + P_L\mathbf{m}\,.
\end{array}
\label{complexm}
\end{equation}

We first write equations for acting with
a covariant ${\cal D}_\mu $ and with a $\gamma _\mu $ on the gravitino
field equation:
\begin{eqnarray}
0&=&{\cal D}_\mu\Sigma ^\mu = M_P^{2}{\cal D}_\mu R^\mu -\gamma ^{\mu \nu
}{\cal D}_\mu
 \left(  \mathbf{m}\psi_\nu  - 2 \Upsilon_\nu\right)-\not\!\!{\cal D}\upsilon
\,, \nonumber\\
0&=&\gamma _\mu\Sigma ^\mu = M_P^{2}  \gamma _\mu R^\mu -   3 \gamma^\mu
  \left(  \mathbf{m}\psi_\mu - 2 \Upsilon_\mu\right) -4\upsilon  \,.
\label{gammapsi}
\end{eqnarray}
Note that\footnote{\label{fn:Kacovnc}Under the K{\"a}hler transformations,
$\mathbf{m}$ behaves as a non-chiral quantity. Indeed, it is invariant
under the charge conjugation (see the end of
appendix~\ref{app:notations}). Non-chiral quantities do not transform as
in (\ref{Qwk}), but rather with a factor $\exp \rmi w_K\gamma _5\Lambda
$, where $w_K$ is the weight of its left component, which here is $m^*$.
Thus, one has, for example, the following K{\"a}hler transformations:
\begin{equation}
\mathbf{m}'=\exp (\rmi \gamma _5\Lambda)\mathbf{m}\,,\qquad
\psi _\mu '=\exp (-\ft12\rmi \gamma _5\Lambda)\psi _\mu\,,\qquad
\Upsilon_\mu '=\exp (\ft12\rmi \gamma _5\Lambda)\Upsilon_\mu \,.
\label{nonchiralKatr}
\end{equation}
}
\begin{eqnarray}
  {\cal D}_\mu \mathbf{m}\phantom{^\dagger}&\equiv&
  \left( \partial _\mu -\rmi\gamma _5 A_\mu ^B\right)\mathbf{m}\phantom{^\dagger} =
 P_Rm^i\partial _\mu
  \phi_i+P_Lm_i\partial _\mu \phi^i
  -\rmi\gamma _5M_P^{-2}W_\mu ^\alpha {\cal P}_\alpha\mathbf{m}\nonumber\\
   {\cal D}_\mu \mathbf{m}^\dagger &\equiv&
  \left( \partial _\mu +\rmi\gamma _5 A_\mu ^B\right)\mathbf{m}^\dagger =
 P_Rm_i\partial _\mu
  \phi^i+P_Lm^i\partial _\mu \phi_i+
  \rmi\gamma _5M_P^{-2}W_\mu ^\alpha {\cal P}_\alpha\mathbf{m}^\dagger
  \,.
\label{Dmpsi}
\end{eqnarray}
Therefore, we can combine these equations as they
appear in the supersymmetry rule (\ref{susyinv}), using the first two
equations of (\ref{DR}):
\begin{eqnarray}
0= {\cal D}_\mu \Sigma ^\mu  + \ft12 M_P^{-2}\mathbf{m}\gamma _\mu
\Sigma^\mu = -\ft{1}{2}M_P^{2}\mathbf{G}_{\mu\nu}\gamma^\mu\psi^\nu
  - \gamma ^{\mu \nu } ({\cal D}_\mu \mathbf{m})\psi _\nu
  -\ft32 M_P^{-2}|m|^2\gamma ^\mu \psi _\mu\nonumber\\
  + 2 \gamma ^{\mu \nu }{\cal D}_\mu \Upsilon_\nu
   +3M_P^{-2}\mathbf{m} \gamma^\mu \Upsilon_\mu -
   \not\!\!{\cal D}\upsilon -2M_P^{-2}\mathbf{m}\upsilon  \,,
\label{cons}
\end{eqnarray}
where we also use the complex matrix
\begin{equation}
\mathbf{G}_{\mu\nu}=G_{\mu \nu }-2\rmi \tilde F^\mathit{quant}_{\mu \nu }
 \,.
  \label{bfG}
\end{equation}

Using the combination as in the last equation of (\ref{DR}), we obtain
the `\emph{master equation}'
\begin{equation}
0=\Sigma _\mu -\ft12\gamma _\mu \gamma ^\nu \Sigma _\nu =
 M_P^{2}\not\!\! {\cal D} \psi _{\mu }+\mathbf{m}\psi _\mu   -
  \left( M_P^{2}{\cal D}_\mu -\ft12 \mathbf{m}\gamma_\mu\right) \gamma^\nu
  \psi_\nu- 2\Upsilon_\mu -\gamma _\mu \gamma \cdot \Upsilon+\gamma _\mu \upsilon  \,.
\label{master}
\end{equation}

So far, we have derived the master equation and the constraints for the
gravitino without specifying the gauge-fixing of the local supersymmetry.
Now we may decide how to fix the gauge. One possibility is to use the
gauge where the `goldstino' vanishes (see section~\ref{ss:Goldstino})
\begin{equation}
  \mbox{possible }Q\mbox{-gauge:}\quad  \upsilon=0\,.
\label{possibleQgauge}
\end{equation}
This choice thus eliminates one more spin-$\ft12$ fermion, as was the
case with the $S$-gauge (\ref{Sgauge}). The eliminated fermion is the
mode which is eaten by the gravitino to make it massive. Note that
$\upsilon ^1$ is the goldstino which was already considered in
\cite{general}. In the application considered there, the background was
stationary and everywhere constant.
 The backgrounds in cosmology may have
some non-vanishing time derivatives on scalar fields and therefore in the
context of cosmology it is natural to change the choice of the goldstino
such that also the derivative mixing\footnote{ In
section~\ref{ss:confInspBG} we will assume that there are no vector
fields in the background. Then the gaugino and the gravitino are mixed
only in the term (\ref{Lmixresc}).} terms between the gravitino and the
fermions are taken into account. Note, however, that in (\ref{Lmixresc})
we can never eliminate all the derivative terms by a choice of the
goldstino. The sum which we consider here is preferred because of the
supersymmetry transformation property (\ref{delvarpi2}). The unitary
gauge is a gauge where the massive gravitino has both $\pm \ft{3}{2}$ as
well as $\pm \ft12$ helicity states.

Another possibility to gauge-fix the local supersymmetry is to use the
condition $\gamma^\mu \psi_\mu=0$. We will consider such gauge-fixing in
the context
 of the equivalence theorem in section~\ref{ss:equivTh}.

\section{The cosmological background}\label{ss:confInspBG}
\subsection{Assumptions}\label{ss:Assumptions}
We now go to the situation which is important for cosmology. We will
consider a flat Friedmann universe produced by inflation, with
\begin{equation}
\rmd s^2= -\rmd t^2+ a^2(t)\,\rmd {\bf x}^2\,. \label{Friedmetric}
\end{equation}
The change of variables $\rmd t= \rmd\eta/a$ brings the \emph{metric} to
 the `conformal' form
\begin{equation}
  g_{\mu \nu }=a^2(\eta )\eta _{\mu \nu }\,,
\label{confmetric}
\end{equation}
where $\eta=x^0$ is the time coordinate (conformal time). The explicit
forms for curvatures, covariant derivatives,  etc are discussed in
appendix~\ref{app:conf_metric}. In this context, it will be necessary to
distinguish the flat gamma matrices from those with curved indices. We
will denote by $\gamma ^0$ and $\vec\gamma $ the components of flat
matrices, i.e.\ $\gamma ^a$. To make a bridge between the curved and flat
indices, it is useful to introduce the notation
\begin{equation}
  \overline {\not\!  \partial}=\gamma ^a \delta _a^\mu
\partial _\mu\,,\qquad \not\!  \partial=\gamma^\mu \partial _\mu =
a^{-1}\overline {\not\!  \partial}\,,\qquad \overline{\gamma }_\mu =\delta
_\mu ^a\gamma_a \,,\qquad \overline{\gamma }^\mu =\delta ^\mu
_a\gamma^a\,. \label{dslash}
\end{equation}
Thus, for example, we have
\begin{equation}
  \gamma ^\mu \psi _\mu= a^{-1}\bar \gamma^\mu \psi _\mu =a^{-1}
 \left(  \gamma ^0\psi _0 +\vec \gamma \cdot \vec \psi \right) \,.
\label{exampleflatcurved}
\end{equation}
Observe that $\psi _0$ and $\vec \psi $ are components of $\psi _\mu$.

A second assumption is that the \emph{vectors are zero} (because they are
scaled down by inflation) and the \emph{scalars depend only on time}
(because inflation makes everything almost exactly homogeneous). This has
immediate consequences:
\begin{eqnarray}
&& \vec \Upsilon=0\,,\qquad \upsilon ^2=a^{-1}\gamma
 ^0\Upsilon_0\,,\nonumber\\
&& A_0^B=\ft12\rmi\left( (\partial _i{\cal K})\partial _0\phi ^i
-(\partial ^i{\cal K})\partial _0\phi _i\right)  \,,\qquad
 \vec A^B=0\,,\qquad F_{\mu \nu }^\mathit{quant}=0 \,,
\label{immedScT}
\end{eqnarray}
and we will denote the remaining component $\Upsilon_0$ as $\Upsilon$ for
short. The graviton field equation $S_{\mu \nu }=0$ in (\ref{fe1}) then
implies that $G_{\mu \nu }=\mathbf{G}_{\mu \nu }$ is diagonal. Its
components are the background energy density and pressure, which we denote
as $\rho$ and $p$, such that, with curved indices, (see (\ref{Gconf}))
\begin{equation}
G^0_0= M^{-2}_P\rho\, ,   \qquad G=-\unity_3\, M^{-2}_P\,p \,,
\end{equation}
where $G$ denotes the $3\times 3$ matrix of spacelike components of
$G^\mu {}_\nu $. We obtain explicitly
\begin{eqnarray}
&&\rho = |\dot \phi|^2 +V\,, \qquad p = |\dot \phi|^2 -V\, ,\nonumber\\
&&\dot \phi\equiv a^{-1}\partial _0\phi\,,\qquad |\dot \phi|^2\equiv
g_i{}^j \dot \phi_j \dot \phi^i\,, \label{Gscalar}
\end{eqnarray}
(see appendix~\ref{app:conf_metric} concerning the dot notation). Using
the definition of the Hubble parameter (\ref{defHubble}), the scalar
field equation is
\begin{equation}
-S_i=  g_i{}^j(\ddot \phi_j +3H\dot \phi_j)+g_i^{jk}\dot \phi_j\dot \phi_k
+\partial _i V=0 \,,
\label{scFE}
\end{equation}
where (\ref{diV}) gives a useful expression for $\partial _i V$.
This leads to
\begin{equation}
  \dot \rho =-6H|\dot \phi|^2=-3H(\rho +p)\,.
\label{dotrho}
\end{equation}
The $S_{\mu \nu }$ field equations are (see (\ref{Gconf})),
\begin{equation}
  \rho =3M_P^2 H^2\,, \qquad  p=-M_P^2(3H^2+2\dot H)\,,
\label{rhoinH}
\end{equation}
consistent with (\ref{dotrho}).

When the scalars are constant, depending on the sign of the cosmological
constant (the sign of the constant value of the scalar potential $V$),
the gravitational background is either \Poin\ $(V=0$) or de~Sitter
($V>0$) or anti-de~Sitter ($V<0$). This scalar potential is $V=V_F+V_D$
as in (\ref{Vtotal}), and gives the contribution $H^2=\ft13 VM_P^{-2}$ to
the Hubble constant during inflation, see (\ref{curvdS}).

The third assumption is that for the fermions we use the \emph{plane-wave
ansatz for the space-dependent part} ($\sim \rme ^{\rmi \vec k \cdot\vec
x}$). We thus replace all $\vec \partial $ by $\rmi\vec k$.

\subsection{Super-Higgs effect in cosmology}
\label{ss:superHiggsCosm}
Having these simplifications, we can rewrite the expression for the
goldstino (\ref{Goldstino}) and the
supersymmetry transformations (\ref{gfsusyphenom}), as
\begin{eqnarray}
\upsilon^1 _L & = & \phantom{-}\ft12 \rmi{\cal P} _\alpha \lambda _L^\alpha +\mxi ^i\chi_i
\,,\qquad \upsilon^2 _L = \gamma ^0\nxi_i\chi ^i \nonumber\\
 \upsilon^1 _R  &=&  -\ft12\rmi {\cal P}_\alpha \lambda _R^\alpha +\mxi _i\chi^i
\,,\qquad \upsilon^2 _R= \gamma ^0\nxi ^i\chi _i \nonumber\\
\delta \chi _i & = & -\ft12 \gL ij \xi _j\epsilon \,,\qquad
\delta \lambda^\alpha  =\ft12\rmi \gamma _5(\Re f)^{-1\,\alpha \beta}{\cal P} _\beta
\epsilon \,.
\label{susyrepeat}
\end{eqnarray}
We introduced here a new notation. With
\begin{eqnarray}
\nxi_i=  g_i{}^j\dot \phi_j\,,&\qquad&
\nxi^i=  g_j{}^i\dot \phi^j\,.
 \label{xisimple}
\end{eqnarray}
we have
\begin{eqnarray}
\xi ^i  \equiv   m^i+\gamma _0n^i\,,&\qquad &
\xi ^{\dagger i}\equiv m^i-\gamma _0n^i
\nonumber\\
\xi _i  \equiv   m_i+\gamma _0n_i\,,&\qquad &
\xi ^\dagger_{ i}\equiv m_i-\gamma _0n_i\,.
\label{defxi}
\end{eqnarray}
Note that  the Hermitian conjugate of $\xi ^i$ is $\xi ^\dagger_{ i}$,
while its charge conjugate is $\xi _i$. In this notation, the goldstino is
\begin{equation}
 \upsilon =\upsilon ^1+\upsilon ^2=\xi ^{\dagger i}\chi _i+\xi ^\dagger_{ i}\chi
  ^i+\ft12\rmi\gamma _5{\cal P} _\alpha \lambda ^\alpha\,.
\label{goldstino_in_xi}
\end{equation}
As we wrote already at the end of section~\ref{ss:Goldstino}, this has a
non-zero transformation. We can make this now explicit:
\begin{equation}
-2\delta \upsilon = \left( \xi ^{\dagger i}\gL ij \xi _j +\xi ^\dagger_ j\gR ij\xi ^i
+\ft12{\cal P} _\alpha (\Re f)^{-1\,\alpha \beta}{\cal P} _\beta\right) \epsilon
 = \alpha\, \epsilon \,.
\label{posrho}
\end{equation}
We may write several expressions for $\alpha $ that will be useful
below (using $V_D$ as in (\ref{Vtotal}))
\begin{eqnarray}
\alpha  & = & \xi ^{\dagger i}\gL ij \xi _j +\xi ^\dagger_ j\gR ij\xi ^i
+V_D \nonumber\\
  & = &\ft12\left(  \xi ^{\dagger i}\gI ij \xi _j +\xi ^i\gI ij\xi^\dagger _j\right)
+V_D
\nonumber\\
&=& \mxi^i\gI ij \mxi_j +\nxi^i\gI ij\nxi_j +V_D
\nonumber\\
&=&|\dot \phi|^2+ \mxi^i\gI ij \mxi_j +V_D=|\dot \phi|^2 +V_+
 \nonumber\\
&=&\rho+ 3M_P^{-2}| m|^2=3(M_P^2H^2+M_P^{-2}|m|^2) \nonumber\\ &=&
3M_P^2(H^2+ m_{3/2}^2)\,.
\label{alpha}
\end{eqnarray}
The last line has several important implications. First of all, it shows
that in a flat universe  (\ref{Friedmetric}) the parameter $\alpha $ is
strictly positive\footnote{To avoid misunderstandings, we should note
that, in general, one may consider  the situations when the energy density
$\rho$ is negative, The famous example is anti-de~Sitter space with a
negative cosmological constant. However, in the context of inflationary
cosmology, the {\it energy density can never become negative}, so anti-de
Sitter space cannot appear. The reason is that inflation makes the
universe almost exactly flat. As a result, the term $k\over a^2$ drops
out from  the Einstein equation for the scale factor independently of
whether the universe is closed, open or flat. The resulting equation
acquires the form that we use in this paper: $ \left({\dot a\over a
}\right)^2 = H^2 = {\rho\over 3 M_P^2}$. In the early universe, according
to inflationary theory, $\rho > 0$. Then gradually the energy density
decreases, but it can never become negative even if a negative
cosmological constant is present, as in anti-de~Sitter space. Indeed, the
equation  $ \left({\dot a\over a }\right)^2   = {\rho\over 3 M_P^2}$
implies that as soon as the energy density becomes zero, expansion stops.
Then the universe recollapses, and the energy density becomes positive
again.}. This implies that supersymmetry is {\it always broken}. The
symmetry breaking is associated to an equal extent with the expansion of
the universe and with the non-vanishing gravitino mass (the term
$3M_P^2(H^2+ m_{3/2}^2)$). This is a rather interesting result because
usually supersymmetry breaking is associated with the existence of
gravitino mass. Here we see that in an expanding universe the Hubble
parameter $H$ plays an equally important role.

Once we have such a fermion and $\alpha >0 $, we can split the
spin-$\ft12$ fermions into parts which are invariant under supersymmetry
and a part proportional to the goldstino. In fact, the general rule for a
set of spinors $\varpi_\Lambda$ transforming under the symmetry as $\delta
\varpi_\Lambda (\epsilon )$ is that the combinations
\begin{equation}
  \varpi_\Lambda +\delta \varpi_\Lambda \left(\frac 2{\alpha }\upsilon
\right)
\label{invariantFermions}
\end{equation}
are invariant. For our case, we can write
\begin{eqnarray}
g_i^j\chi _j & = & \hat\Pi{}_i{}^j\chi _j +\hat\Pi{}_{ij}\chi ^j+\hat\Pi{}_{i\alpha }\lambda ^\alpha
+\frac 1\alpha P_L \xi _i\, \upsilon\,,
 \nonumber\\
g^i_j\chi ^j & = & \hat\Pi{}^{ij}\chi _j +\hat\Pi{}^i{}_{j}\chi ^j+\hat\Pi{}^i{}_{\alpha }\lambda ^\alpha
+\frac 1\alpha P_R  \xi ^i\, \upsilon\,,
 \nonumber\\
(\Re f_{\alpha \beta })\lambda ^\beta   & = & \hat\Pi{}_\alpha{}^ j\chi _j +\hat\Pi{}_{\alpha j}\chi ^j+\hat\Pi{}_{\alpha \beta }
\lambda ^\beta-\frac \rmi\alpha \gamma _5 {\cal P} _\alpha
\,\upsilon\,,
\label{projectedFermions}
\end{eqnarray}
where
\begin{eqnarray}
&&  \hat\Pi{}_i{}^j= P_L\left( g _i{}^j-\frac 1\alpha \xi _i \xi ^{\dagger j}\right)
P_L  \,, \qquad \hat\Pi{} _{ij}=-\frac 1\alpha P_L  \xi _i \xi ^\dagger _jP_R\,,
  \qquad
  \hat\Pi{} _{i\alpha }=-\frac{\rmi}{2\alpha }P_L \xi ^\dagger _i{\cal P}_\alpha\,,
  \nonumber\\
&&  \hat\Pi{} ^{ij}=-\frac 1\alpha P_R  \xi ^i \xi ^{\dagger j}P_L
  \,, \qquad \hat\Pi{}^i{}_j= P_R\left( g ^i{}_j-\frac 1\alpha \xi ^i \xi ^\dagger _j
  \right) P_R
\,,  \qquad
  \hat\Pi{} ^i{}_{\alpha }=\frac{\rmi}{2\alpha }P_R \xi ^{\dagger i}  {\cal P}_\alpha\,,
  \nonumber\\
&& \hat\Pi{} _\alpha {}^j=\frac{\rmi}{\alpha }
{\cal P} _\alpha  \xi ^jP_L\,,\qquad\hspace{7mm}
\hat\Pi{} _{\alpha j}=-\frac{\rmi}{\alpha }{\cal P} _\alpha  \xi _jP_R\,,\qquad
\hspace{2mm}\hat\Pi{} _{\alpha\beta} =\Re f_{\alpha \beta }
- \frac{1}{2\alpha }{\cal P} _\alpha  {\cal P}_\beta   \,.
\label{valuePi}
\end{eqnarray}
One can check that the transformations of the fermions on the left-hand
sides of (\ref{projectedFermions}) are provided only by the term
proportional to the goldstino.

The projectors $\hat\Pi{}_i{}^j$ and $\hat\Pi{}_{ij}$ have simple form in
terms of $m$ and $n$
\begin{eqnarray}
\hat\Pi{}_i{}^j  & = & P_L\Pi _i{}^j\qquad \mbox{with}\qquad
\Pi _i{}^j= g_i {}^j-\frac 1\alpha\left( m_im^j+n_in^j\right)\,, \nonumber\\
\hat\Pi{}_{ij}  & = & P_L\gamma _0 \Pi _{ij}\qquad \mbox{with}\qquad
\Pi _{ij}=\frac{1}{\alpha }\left(m_i n_j-n_im_j\right)\,.
\label{PiPure}
\end{eqnarray}
Note the properties
\begin{equation}
  \Pi _i{}^j \gI ji =\frac{1}{\alpha }V_D\,,\qquad  \Pi _{ij}=-\Pi
  _{ji}\,,
\label{propPi}
\end{equation}
reflecting, for example, that the $\Pi $ vanish for only one chiral
multiplet and no vector multiplets.

Observe that in the Higgs effect as discussed usually, where the scalar
background is constant, the $\Pi _{ij}$ and complex conjugates $\Pi
^{ij}$ vanish. Indeed, then $n_i=0$. The mixing between the left and right
chiralities of the $\chi $ fields is thus a new feature of this
super-Higgs effect in a background with time-dependent
scalars\footnote{The definition of the goldstino  in global supersymmetry
in a time-dependent scalar-field background with real scalars and without
vector multiplets was given in \cite{GTR}. It is similar to our
definition for this particular case. However, since no distinction was
made in \cite{GTR} between left and right chiral fermions, their
projector operators are different.}.

{}From now on, we adopt the unitary gauge (\ref{possibleQgauge}).
Then the fermion $\Upsilon$ has the expression
\begin{equation}
  \Upsilon= a(n_i\chi ^i+n^i\chi _i)=-\ft12a\gamma _0(\xi _i\chi ^i+\xi
  ^i\chi _i)\,.
\label{UpsilonInGauge}
\end{equation}
Another useful expression that we will use below, can be derived by
this, using $P_R \xi ^{\dagger i}\xi _jP_R=P_R\xi ^i\xi ^\dagger
_jP_R$, and the vanishing of the goldstino:
\begin{equation}
  P_L\xi ^{\dagger i}\Upsilon= \alpha a \Pi ^{ij}\chi _j\,.
\label{Usef2}
\end{equation}

\subsection{Constraints in the unitary gauge $\upsilon =0$.}
\label{ss:constraintsGr}
The first constraint, (\ref{cons}), is an algebraic relation between
$ \gamma^0\psi_0$ and $\vec\gamma\cdot \vec \psi$. This can now be
made explicit as
\begin{eqnarray}
0&=&  -\alpha \gamma ^0\psi _0+(\alpha _1+\gamma
  _0 \alpha _2) \theta +4\left(a^{-1} \rmi\vec\gamma \cdot \vec
k +\ft32M_P^{-2}\widehat{m}\right)\gamma ^0\Upsilon
\nonumber\\
 && \alpha _1\equiv  p- 3 M_P^{-2}|m|^2  \,,\qquad
  \alpha _2\equiv 2 \dot{\mathbf{m}}{}^\dagger  \nonumber\\
 && \theta \equiv \vec\gamma\cdot \vec \psi\,,\qquad \mbox{and}\qquad
  \widehat{m}\equiv  \mathbf{m} +M_P^2H\gamma _0\,,\qquad
  \widehat{m}^\dagger \equiv  \mathbf{m}^\dagger-M_P^2H\gamma _0\,,
\label{constralpha}
\end{eqnarray}
where we have used that
\begin{equation}
a^2  \gamma ^{\mu \nu }{\cal D}_\mu \Upsilon_\nu =\left( \rmi\vec\gamma \cdot \vec
k \gamma ^0+\ft32\dot a\right)\Upsilon_0\,.
\label{derU}
\end{equation}
On quantities of non-zero K{\"a}hler weight, the dot denotes a K{\"a}hler
covariant derivative, scaled as in (\ref{Gscalar}), e.g.
\begin{equation}
  \dot{\mathbf{m}}{}^\dagger=a^{-1}{\cal D}_0\mathbf{m}{}^\dagger=
  a^{-1}\left(\partial _0 +\rmi \gamma _5
  A_0^B\right)\mathbf{m}{}^\dagger\,,
\label{dotm}
\end{equation}
for which one may use the expression (\ref{Dmpsi}).

The explicit form of the scalar potential $V$, (\ref{Vtotal}), is
used to relate these quantities to inner products of the vectors
$m_i$, $n_i$ or $\xi $ and $\xi ^\dagger $, as we did for $\alpha $ in (\ref{alpha}):
\begin{eqnarray}
\alpha_1& =&-\mxi^i\gI ij \mxi_j +\nxi^i\gI ij\nxi_j
-V_D=-\ft12\left(\xi ^{\dagger i}\gI ij \xi _j +\xi ^i\gI ij \xi ^\dagger
_j\right)-V_D\,,
\nonumber\\
\alpha_2& =&2\mxi^ig^{-1}{} _i^j\nxi_j P_L +2\nxi^ig^{-1}{}
_i^jm_jP_R=\gamma _0\left( \xi ^{\dagger i}\gR ij \xi^\dagger _j -\xi
^i\gR ij \xi_j \right)\,,
 \nonumber\\
\alpha^\dagger _2 & =&2\mxi^ig^{-1}{} _i^j\nxi_j P_R
+2\nxi^ig^{-1}{} _i^j\mxi_jP_L= \gamma ^0\alpha _2 \gamma _0\,.
\label{xialpha}
\end{eqnarray}
Note that $\alpha $, $\alpha _1$ and $\alpha _2$ are invariant under the
charge conjugation, but $\alpha _2$ is not invariant under Hermitian
conjugation\footnote{See appendix~\ref{app:notations}, to see that
$(P_L)^\dagger=P_L$, but $(P_L)^C=P_R$.}. It is convenient to write these
equations and (\ref{xialpha}) in the following matrix form:
\begin{equation}
  \pmatrix{\mxi^i \cr\nxi^i}g^{-1}{} _i^j \pmatrix{m_j
  & \nxi_j}+\pmatrix{{\cal P}_\alpha \cr 0}\frac12(\Re f)^{-1\,\alpha \beta}
  \pmatrix{{\cal P} _\beta& 0}=\frac{1}{2}
  \pmatrix{\alpha -\alpha _1&\alpha _2P_R + \alpha _2^\dagger P_L\cr
 \alpha _2P_L + \alpha _2^\dagger P_R  &\alpha
 +\alpha_1 }\,.
\label{matrixeqnal}
\end{equation}
The determinant of the last matrix is
\begin{eqnarray}
\ft14\alpha ^2\Delta ^2&\equiv&
 \ft14( \alpha ^2-\alpha _1^2-|\alpha _2|^2) \nonumber\\
&=&  \nxi^i\nxi_j  \left[\mxi _k\mxi ^\ell
\left( g^{-1}{}_i{}^j g^{-1}{}_\ell{}^k -g^{-1}{}_i{}^k
g^{-1}{}_\ell {}^j\right)  +\ft12g^{-1}{}_i{}^j {\cal P} _\alpha
(\Re f)^{-1\,\alpha \beta }
{\cal P}_\beta \right]\nonumber\\
&=& \dot \phi^i\dot \phi_j \mxi_k\mxi^\ell
  \left( g^{-1}{}^k_\ell g_i{}^j -\delta _i{}^k\delta_\ell {}^j\right)
+\ft12|\dot \phi|^2 {\cal P}_\alpha (\Re f)^{-1\,\alpha \beta }
  {\cal P}_\beta  \geq 0  \,,
\label{B2isnul}
\end{eqnarray}
(one can also use the explicit form of $V$ and ${\cal D}_0\mathbf{m}$,
(\ref{Dmpsi}), to derive this relation). This expression vanishes in the
case of one chiral multiplet only. $\Delta ^2$ is also related to the
modulus squared of $\Pi _{ij}$ in (\ref{PiPure}):
\begin{equation}
  \Pi ^{ij}\gI ik\gI j\ell \Pi _{k\ell }=\frac12\Delta
  ^2-\frac 2{\alpha ^2}V_Dn^i\gI ijn_j\,.
\label{Pisquare}
\end{equation}

The constraint equation~(\ref{constralpha}) is then
\begin{equation}
  \gamma ^0\psi _0=\hat{A}\theta+ \hat C \Upsilon\,,
\label{constraint1A}
\end{equation}
where
\begin{eqnarray}
\hat{A}&\equiv &\frac{1}{\alpha }\left( \alpha _1+\gamma _0\alpha _2\right)
  \,,\qquad
\hat{A}^\dagger \equiv \frac{1}{\alpha }\left( \alpha _1-\gamma _0\alpha _2\right)\,,
 \nonumber\\
\hat C &\equiv &\frac{4}{\alpha }\left(a^{-1} \rmi\vec\gamma \cdot \vec k
+\ft32M_P^{-2}\widehat{m}\right)\gamma^0  \,. \label{defA} \label{C}
\end{eqnarray}
The matrix $\hat{A}$ can be written as
\begin{equation}
  \hat{A}=-\frac{\xi ^{\dagger i}\gR ij \xi^\dagger  _j
  +\xi ^\dagger_{ j}\gL ij\xi^{\dagger i}+V_D}
  {\xi ^{\dagger i}\gL ij \xi _j +\xi ^\dagger_{ j}\gR ij\xi
  ^i+V_D}\,.
\label{A_in_xi}
\end{equation}
It satisfies the equation
\begin{equation}\label{A1}1-|\hat A|^2=\Delta ^2 \geq  0\,,\qquad \mbox{and }\
|\hat A|^2 = 1 \quad \mbox{in the case of 1 chiral multiplet only.}
\end{equation}

A second constraint is obtained from the field equation $\Sigma
^0=0$, which does not involve a time derivative. That equation leads
to an algebraic expression for ${\vec k \cdot \vec\psi}$ in terms of
$\theta $:
\begin{equation}
\rmi \vec k\cdot \vec \psi   =  \left( \igk
-M_P^{-2}a\mathbf{m}^\dagger-\gamma
  _0\dot a\right) \theta \,.
\label{long}
\end{equation}

These constraints determine the independent components of $\psi _\mu $.
The first constraint (\ref{constraint1A}) can be solved for $\psi _0$. We
thus remain with $\vec \psi $. In general, the 12 components of $\vec
\psi$ can be decomposed into
 4 components of its  transverse part
$\vec \psi^T$, 4 components of the trace $\theta$ and 4
components of the trace ${\vec  k \cdot \vec \psi}$:
\begin{equation}
\vec \psi=\vec \psi^T+\left(\frac{1}{2}\vec \gamma-\frac{1}{2\vec k^2}{\vec k}
 ({\vec k} \cdot \vec \gamma) \right) \theta +
\left(\frac{3}{2\vec k^2}\vec k-\frac{1}{2\vec k^2} \vec \gamma ({\vec k} \cdot
\vec \gamma) \right) { \vec k} \cdot \vec \psi \,, \label{split}
\end{equation}
 so that $\vec \gamma\cdot \vec \psi^T={\vec k}\cdot \vec \psi^T=0$.
 The transverse part $\vec \psi^T$ can be
obtained from $\vec \psi$ with the
 projector operator $\vec \psi^T=\mathbf{P}\vec  \psi$:
\begin{equation}
\mathbf{P}=\unity _3-\left(\frac{1}{2}\vec \gamma-\frac{1}{2\vec k^2}{\vec k}
 (\vec k \cdot \vec \gamma) \right) \vec \gamma^t -
\left(\frac{3}{2\vec k^2}{\vec k}-\frac{1}{2\vec k^2} \vec \gamma (\vec  k \cdot
\vec \gamma) \right) \vec  k^t \,, \label{proj}
\end{equation}
where the $t$ denotes transpose vectors (no transpose of the gamma
matrices). The projection operator satisfies the properties
\begin{equation}
  \mathbf{P}\vec  \gamma=\mathbf{P}\vec  k=
  \vec \gamma ^t\mathbf{P}=\vec k^t\mathbf{P}=0\,,
  \qquad
\mathbf{P} \gamma_0=\gamma _0\mathbf{P}\,,\qquad \vec \gamma \cdot\vec
k\mathbf{P} =\mathbf{P}\vec \gamma \cdot \vec k\,. \label{propP}
\end{equation}

After the constraint (\ref{long})
we have  the on-shell decomposition for the longitudinal part
\begin{equation}
\vec   \psi =\vec \psi ^T+\frac{1}{\vec k^2}\left[ \vec k\,(\vec \gamma \cdot \vec  k)
+\frac{1}{2}\rmi
  \left(3\vec k-\vec \gamma (\vec k\cdot \vec \gamma )\right) \left(\dot a\gamma _0
  +M_P^{-2}a\mathbf{m}^\dagger \right)\right]\theta \,.
\label{decomppsi}
\end{equation}
Thus,  essentially there are two degrees  of freedom associated with the
transverse part $\vec \psi^T$, which correspond to helicity
$\pm\ft{3}{2}$, and two degree of freedom associated with
 $\theta$ (or $\psi_0$)
which correspond to helicity $\pm\ft12$. For vanishing mass, the
transverse part $\vec \psi^T$ is conformal with weight $+\ft12$.
Meanwhile, two remaining degrees of freedom imprinted in $\theta$ are not
conformal; see the discussion after (\ref{gravitinoconf}).

Equations~(\ref{constraint1A}), (\ref{trace1}) and (\ref{decomppsi}) for
gravitinos, which we derived in this section, are applicable for an
arbitrary  FRW metric. They are also applicable for vanishing gravitino
mass $m$. This takes place, for instance, in $D$-term inflation, where the
superpotential $W=0$ during and after inflation.

\subsection{Dynamical equations in the unitary gauge}
\label{ss:dynEqGr} We now derive  dynamical equations for the transverse
part, $\vec \psi $, and the longitudinal part, $\theta $, of the
gravitino. The latter part couples to the spin-$\ft{1}{2}$ fermions, and
we thus have to also consider their dynamical equations. For the
dynamical equations, it will be convenient to use the
 master equation, (\ref{master}), using the constraint equations
in the form (\ref{constraint1A}) and (\ref{long}).

We will first derive the equation for $\vec \psi^T$. For this, consider
the spacelike components of the master equation~(\ref{master}). Useful
intermediate steps are obtained from (\ref{Dmupsimu}), complemented with
the K{\"a}hler connection:
\begin{eqnarray}
a\not\!\! {\cal D} \vec \psi&=& \left(\overline{\not\!
{\partial}} +\ft12\dot a\gamma^0 +\ft12\gamma ^0
\rmi\gamma _5A_0^B\right)\vec \psi-\ft12\dot a
\vec \gamma \psi_0\,, \nonumber\\
a\vec {\cal D} {\gamma }^{\mu} \psi_{\mu}&=&\rmi\vec k\overline{\gamma }^{\mu}
 \psi_{\mu}-\ft12\dot a\vec \gamma \left( \psi_0-\gamma _0\theta
 \right)\,.
\end{eqnarray}

Applying the projector operator $\mathbf{P}$ to the spacelike part of the
master equation~(\ref{master}), we obtain
\begin{equation}
\left(\overline{\not\! {\partial}} + \ft12\dot a\gamma^0+\ft12\gamma ^0
\rmi\gamma _5A_0^B +
M_P^{-2}\mathbf{ m}a \right)\vec \psi^T=0 \,.
\label{trans}
\end{equation}
The transformation $\vec\psi^T=a^{-1/2}\vec\Psi^T$ reduces
 the equation for the transverse part to the
free Dirac equation.
This is the massive Dirac equation in an expanding universe.

Consider now the $\mu=0$ component of (\ref{master}). Using again
(\ref{Dmupsimu}) complemented with the K{\"a}hler connection,
\begin{eqnarray}
&&a\not\!\! {\cal D} \psi_{0}= \left(\overline{\not\!
{\partial}} +\rmi\ft12\gamma ^0\gamma _5A_0^B+\ft12\dot a\gamma^0 \right)\psi_{0}-
\dot a\theta \,,\nonumber\\
&&a {\cal D}_{0} {\gamma }^{\mu} \psi_{\mu}=\overline{\gamma }^{\mu}
a\dot {\psi}_{\mu}-\dot a\overline{\gamma }^{\mu}\psi_{\mu}
\,,
\end{eqnarray}
we have
\begin{equation}
\left(\ft{3}{2}(M_P^{-2}\mathbf{m}a -\dot a\gamma _0)+\igk\right) \psi_{0}=
\dot \theta-\ft12M_P^{-2}\mathbf{m}a\gamma _0
\theta +3M_P^{-2}a\Upsilon_0\,. \label{0}
\end{equation}
For the following equations we introduce the notation
$\hat{\partial}_0$ for the K{\"a}hler-covariant derivative\footnote{The
K{\"a}hler transformations of some quantities appeared in
footnote~\ref{fn:Kacovnc}. Those that we encounter here have
 $\theta '= \rme ^{\rmi\gamma _5\Lambda/2
}\theta $, $\hat{A}'= \rme ^{\rmi\gamma _5\Lambda /2}\hat{A}\rme
^{-\rmi\gamma _5\Lambda /2}$, $\hat{B}'= \rme ^{\rmi\gamma _5\Lambda/2
}\hat{B}\rme ^{-\rmi\gamma _5\Lambda/2 }$. For the covariant derivatives,
this thus means, for example, $\hat{\partial}_0\theta =\partial _0\theta
-\rmi\gamma _5A_0^B/2$, or using the explicit form of $A_0^B$ in
(\ref{immedScT}), $\dot \theta =a^{-1}\partial _0 \theta +\ft14\gamma _5
(\dot \phi ^i\partial _i {\cal K}-\dot \phi _i\partial ^i{\cal K})$.}. We
thus have that $a\dot \theta =\hat{\partial}_0\theta $. On bosonic
quantities $\hat{\partial}_0$ is the same as ${\cal D}_0$. With real
backgrounds, the K{\"a}hler connection vanishes and $\hat{\partial }_0$ is
just $\partial _0$. Substituting $\psi_{0}$ from  (\ref{constraint1A})
into (\ref{0}), we obtain an equation for $\theta$:
\begin{equation}
\left[\hat{\partial}_0- \ft12 M_P^{-2}\mathbf{m}a\gamma _0 -\gamma _0
\left(\ft{3}{2}M_P^{-2}a\widehat{m}^\dagger  +\igk\right) \hat{A} \right]
\theta -\frac{4}{\alpha a}\vec k^2 \Upsilon=0 \,, \label{trace1}
\end{equation}
where we used (see (\ref{alpha}))
\begin{equation}
  \widehat{m}^\dagger \widehat{m}=  |m|^2+M_P^4H^2=\ft13M_P^2\alpha \,.
\label{hatmmdagger}
\end{equation}
Let us define
\begin{eqnarray}
  \hat{B}&=&-\ft32\dot a\hat{A}-\ft12 M_P^{-2}\mathbf{m}a\gamma
_0(1+3\hat{A})=-\ft12 M_P^{-2}\mathbf{m}a\gamma _0
-\ft{3}{2}M_P^{-2}\gamma _0 a\widehat{m}^\dagger    \hat{A}
\,,\nonumber\\
\hat{B}^\dagger  & = & -\ft32\dot a\hat{A}{}^\dagger
+\ft12M_P^{-2}(1+3\hat{A}{}^\dagger )a \mathbf{m} \gamma   _0\,,
\label{defB}
\end{eqnarray}
to write (\ref{trace1}) as
\begin{equation}
\left(\hat{\partial}_0+ \hat{B}-\igk \gamma _0 \hat{A} \right)
 \theta-\frac{4}{\alpha a}\vec k^2 \Upsilon=0 \,.
\label{trace5B}
\end{equation}
This equation is not conformally invariant. In general, the parameters
$\hat{A}$ and $m$ are time-dependent parameters during the fast rolling
down and oscillations  of the inflaton field. The effects of the
background metric and the scalar field variation cannot be eliminated by
the conformal transformation.

We now consider the equation of motion for the fermion $\Upsilon$.
To do so, consider
\begin{eqnarray}
  \gamma ^0\dot \phi_i \Sigma ^i&=&
  -a^{-2}  \left(\hat{\partial }_0+\igk\gamma ^0+\ft12\dot a\right)    \Upsilon_{R}
  +\chi ^j\left( g_j{}^i\ddot \phi _i+g_j^{ik}\dot \phi _i\dot \phi _k\right)
   \nonumber\\
&&+\gamma ^0\dot \phi _i  \left( m^{ij}\chi _j
  + m^i{}_\alpha \lambda ^\alpha _L\right) -
  \ft12 |\dot \phi |^2 a^{-1}(-\gamma ^0\psi _{0L}+\theta _R)
+\ft12 \gamma ^0\dot \phi _i \mxi^i\,\gamma \cdot \psi
_R\,.
\label{feUps0}
\end{eqnarray}
Here again the $\dot\Upsilon$ contains \Ka\ connection, while the Lorentz
connection has been extracted using (\ref{Dmupsimu}). We now use the
scalar field equation~(\ref{scFE}) to rewrite this as
\begin{eqnarray}
\gamma ^0\dot \phi _i \Sigma ^i&=&
 -a^{-2}\left(\hat{\partial }_0+\igk \gamma ^0 +\ft72\dot a\right) \Upsilon_R
  -(\partial _i V) \chi ^i+
\gamma ^0 \dot \phi _i  m^{ij}\chi _j
  + \gamma ^0 \dot \phi _i m^i{}_\alpha \lambda ^\alpha _L \nonumber\\
&&+ \ft14 a^{-1} \left[(\alpha +\alpha _1)(\gamma ^0\psi _{0L}-\theta
  _R)+\gamma ^0  \alpha _2(\gamma ^0\psi
  _{0R}+\theta _L)  \right]\nonumber\\
  &=&-a^{-2}
\left(\hat{\partial }_0+\igk \gamma ^0 +\ft72\dot a\right) \Upsilon_R-2M_P^{-2}a^{-1}m\gamma^0
\Upsilon_L+\Xi_R  \nonumber\\
&& + \ft14 a^{-1} P_R\alpha \left[(1+\hat{A}^\dagger
  )\gamma ^0\psi _0 -(1+\hat{A})\theta \right]  \,.
\label{feUps}
\end{eqnarray}
where
\begin{equation}
  \Xi _R=   -\mxi ^kg^{-1}{}_k{}^jm_{ji}\chi ^i
-\rmi {\cal P}_\alpha  (\Re f)^{-1\,\alpha \beta}m_{\beta i}\chi ^i
   +\gamma ^0\dot \phi _j\left( m^{ji}\chi _i+m^j{}_\alpha \lambda ^\alpha _L\right)
   +\rmi M_P^{-2} m\, {\cal P}_\alpha \lambda _R^\alpha \,.
\label{defOmega}
\end{equation}
For the second expression of (\ref{feUps}) we used (\ref{diV}), whose
first term is absorbed into a new $\Upsilon$ term using (\ref{susyrepeat})
and  $\upsilon ^2=-\upsilon ^1$. Finally, using the constraint
(\ref{constraint1A}), the expression in square brackets in the last line
of (\ref{feUps}) is $(\hat{A}^\dagger\hat{A}-1)\theta +(1+\hat{A}^\dagger
)\hat{C}\Upsilon$. We thus obtain from the field equation $\Sigma ^i=0$,
\begin{equation}
 \left[\hat{\partial }_0+\igk \gamma ^0 +\ft72\dot a
 -\ft14 a \alpha(1+\hat{A}^\dagger )\hat{C}+2\mathbf{m}a\gamma ^0\right] \Upsilon
-a^2\Xi +\ft14 a \alpha
\Delta ^2\theta=0   \,.
\label{eqUps1}
\end{equation}
Using the explicit form of $\hat{C}$, this is
\begin{equation}
  \left[\hat{\partial }_0+\igk \gamma _0\hat{A} +\dot a (2-\ft32\hat{A}^\dagger
  )-\ft12M_P^{-2}(1-3\hat{A}^\dagger ) a\mathbf{m}\gamma _0\right] \Upsilon
-a^2\Xi +\ft14 a \alpha \Delta ^2\theta=0\,.
\label{eqUps}
\end{equation}

We thus conclude that the transverse mode obeys a simple massive
Dirac equation, (\ref{trans}), while in the longitudinal mode
we have a system of coupled equations.
Using the quantities
\begin{equation}
   \hat{A}=\frac{p-3M_P^{-2}|m|^2 +2\gamma _0\dot \mathbf{m}^\dagger }{\rho
   +3M_P^{-2}|m|^2}\,,
\label{Ain_p_and_rho}
\end{equation}
and $\hat{B}$ defined in (\ref{defB}) we have as dynamical equations
\begin{eqnarray}
&& \left(\hat{\partial}_0+ \hat{B}-\igk \gamma _0 \hat{A} \right)
 \theta-\frac{4}{\alpha a}\vec k^2 \Upsilon=0 \,,\nonumber\\
&&\left[\hat{\partial }_0+\igk \gamma _0\hat{A} +\hat{B}^\dagger
 +2\dot a - M_P^{-2}a\mathbf{m}\gamma _0\right] \Upsilon
-a^2\Xi +\ft14 a \alpha \Delta ^2\theta=0\,.
\label{systemLinEqs}
\end{eqnarray}
They couple to the spinor $\Xi $, (\ref{defOmega}), which is, in general,
another independent spinor, and which, neglecting contributions of the
vector multiplets,  can be written as
\begin{equation}
  \Xi =-\xi ^k\gI kj m_{ji}\chi ^i-\xi _k\gI jk m^{ji}\chi _i+\mbox{ vector mult.
  contr.}\,.
\label{XiNOVM}
\end{equation}
We can again
extract its dynamical equation from (\ref{fe1}), but in section~\ref{ss:2mult} we will
consider two cases in which this is not necessary.
\subsection{Higher-order equation for the longitudinal part of the gravitino}
\label{ss:longGr} We can obtain a second-order equation for the
longitudinal part of the gravitino, by applying the operator that acts on
$\Upsilon$ in the second line of (\ref{systemLinEqs}) to the first line.
First, note the following property of $\hat{B}$, defined in (\ref{defB}):
\begin{equation}
  2B_1\equiv \hat{B}+\hat{B}^\dagger=-3\dot a\frac{\alpha _1}{\alpha }
   +{3 a \over 2\alpha }M_P^{-2}\left(
   \mathbf{m}\alpha _2 +\alpha _2^\dagger \mathbf{m}^\dagger \right) =
   a\frac{\dot \alpha }{\alpha }+3\dot a \,.
\label{eqnB1}
\end{equation}
Then we obtain,
\begin{eqnarray}0 & = & \frac{1}{\alpha a} \left[\hat{\partial }_0+\igk \gamma _0\hat{A} +\hat{B}^\dagger
 +2\dot a -M_P^{-2} a\mathbf{m}\gamma _0\right]\alpha a
   \left[ \hat{\partial }_0 +\hat{B}-\igk \gamma _0 \hat{A} \right]
 \theta \nonumber\\
  &   &-\frac{4\vec k^2}{\alpha a}\left[ a^2\Xi -\ft14 a \alpha \Delta ^2\theta
\right]   \nonumber\\
  &=&\left[\hat{\partial }_0\hat{\partial }_0 +\vec k^2+|\hat{B}|^2+2B_1\hat{\partial
  }_0  +a\dot{\hat{B}}-\igk\gamma
  _0a\dot{\hat{A}} \right] \theta \nonumber\\
  &&-\frac{4a\vec k^2}{\alpha }\Xi +\left( 2B_1-M_P^{-2}a\mathbf{m}\gamma
  _0\right) \left[\hat{\partial }_0 +\hat{B}-\igk \gamma _0 \hat{A} \right]
 \theta\,.
\label{quadreqn}
\end{eqnarray}
As all the fermions are involved in this equation, due to the mixing with
$\Xi $, this equation will lead to higher-order equations, in general.
\section{Gravitino--goldstino equivalence theorem at large $M_P$}
\label{ss:equivTh} The proof of the high-energy equivalence theorem
relating $S$-matrix elements for helicity-$\pm \ft12$ gravitinos to the
corresponding goldstinos was given in \cite{casalbuoni}, clarifying and
generalizing the original proposal given in \cite{fayet}. The basic idea
was to use the analogue of the $R_\xi$ gauges as in the non-Abelian gauge
theories. In the cosmological setting, when scalars are time dependent,
in the case of one chiral multiplet the relevant $R_\xi$ gauge was
introduced in \cite{Maroto:1999vd}. In the limit when $\xi \rightarrow
\infty$ it is reduced to the analogue of the renormalizable gauge
$-\ft12\gamma^\mu \not\!\partial \phi^1 g_1{}^1
 \, \psi _{\mu L}
+\ft12\mxi^1\,\gamma \cdot \psi _R =0$. In the opposite limit, when $\xi
\rightarrow 0$, it becomes a unitary gauge in which there is no goldstino,
$\chi_1=0$. {\it In the presence of many chiral and vector multiplets and
time-dependent  scalars, the analogue of the renormalizable gauge in which
the chiral spinors would decouple from gravitino, is not available}.   The
equations for chiral fermions are
\begin{eqnarray}
g^i{}_j\not\!\!{\cal D}\chi ^j+m^{ij}\chi _j
+m^i{}_\alpha \lambda ^\alpha _L
-\ft12\gamma^\mu \not\!\partial \phi^jg_j{}^i  \psi _{\mu L}
+\ft12\mxi^i\,\gamma \cdot \psi _R=0\,.
\end{eqnarray}
By a choice of a single fermionic gravitino-dependent function, in
general, one cannot remove both gravitino-dependent terms from the
equations for $\chi_i$. Still the nearest analogue of the
gravitino--goldstino equivalence theorem can be established in the
cosmological background when working in the gauge where $\gamma \cdot
\psi=0$.

Consider the set of fermionic field equations in the cosmological
background in the gauge where $\gamma \cdot \psi=0$,
\begin{eqnarray}
&&g^i{}_j\not\!\!{\cal D}\chi ^j+m^{ij}\chi _j +m^i{}_\alpha \lambda
^\alpha _L -\dot  \phi^jg_j{}^i \, \psi _{0 L} =0\,, \nonumber\\
&&(\Re f_{\alpha \beta })\not\!\!{\cal D}\lambda _L^\beta
 +2m_{i\alpha }\chi
^i+2m_{R\alpha \beta }\lambda _R^\beta -\ft14\left( f^i_{\alpha \beta }
\not\!\partial \phi_i-f_{\alpha \beta \,i}\not\!\partial
\phi^i\right)\lambda _L^\beta=0 \,. \label{fe2}
\end{eqnarray}
The gaugino has already decoupled from the gravitino. However, the
presence of the $\psi_0$ term in the equations for the chiral fermions
shows that, in general, the chiral fermions $\chi_i$ are mixed with the
gravitino. The gravitino master equation~(\ref{master}) in the gauge
$\gamma \cdot \psi=0$ is
\begin{equation}
 M_P^{2}\not\!\! {\cal D} \psi _{\mu }+\mathbf{m}\psi _\mu
 - 2\Upsilon_\mu -\gamma _\mu \gamma \cdot \Upsilon +\gamma _\mu \upsilon =0   \,.
\label{masterR}
\end{equation}
Thus equations for the gravitino and the chiral fermions are mixed in
this gauge and only the gaugino is decoupled. At this stage it is not
clear why starting with the global supersymmetry equations for $\chi$,
decoupled from the gravitino, one can hope to find the same result as in
supergravity in the spirit of the equivalence theorem. However, now we
may consider an approximation to this system of equations where $M_P$ is
large and the geometry tends to a flat one. The $0$-component of the
gravitino equation becomes
\begin{equation}
\not\! \partial \psi _{0 }+  M_P^{-2} (\mathbf{m}\psi _0
 - 3\Upsilon +a\gamma _0 \upsilon) =0
 \,,
\label{masterR0}
\end{equation}
and the equations for the other fermions in this approximation reduce to
\begin{eqnarray}
&&g^i{}_j\not\!\partial \chi ^j+m^{ij}\chi _j +m^i{}_\alpha \lambda
^\alpha _L -\dot  \phi^jg_j{}^i \, \psi _{0 L} =0\,, \nonumber\\
&&(\Re f_{\alpha \beta })\not\!\partial \lambda _L^\beta
 +2m_{i\alpha }\chi
^i+2m_{R\alpha \beta }\lambda _R^\beta -\ft14\left( f^i_{\alpha \beta }
\not\!\partial \phi_i-f_{\alpha \beta \,i}\not\!\partial
\phi^i\right)\lambda _L^\beta=0 \,. \label{fe3}
\end{eqnarray}
In the large-$M_P$ limit, the gravitino equation~(\ref{masterR0}) reduces
to  $\not\! \partial \psi _{0 }=0$ and it has a consistent solution with
$\psi_0=0$. Note that without such an approximation,  $ \psi _{0 }=0$ is
not consistent with (\ref{masterR0}) and thus the chiral fermions are not
decoupled when the scalars depend on time, $\dot \phi\neq 0$.

Thus in the gauge $\gamma \cdot \psi=0$ in the limit of large $M_P$ the
equations for matter fermions  as obtained in supergravity (with the
rescalings corresponding to the fields of the original superconformal
theory) are
\begin{eqnarray}
&&g^i{}_j\not\!\partial \chi ^j+m^{ij}\chi _j +m^i{}_\alpha \lambda
^\alpha _L =0\,,\nonumber\\
&& (\Re f_{\alpha \beta })\not\!\partial \lambda _L^\beta
 +2m_{i\alpha }\chi
^i+2m_{R\alpha \beta }\lambda _R^\beta -\ft14\left( f^i_{\alpha \beta }
\not\!\partial \phi_i-f_{\alpha \beta \,i}\not\!\partial
\phi^i\right)\lambda _L^\beta=0 \,. \label{fe4}
\end{eqnarray}
Those are precisely the equations of global supersymmetry.

The fact that in the gauge $\gamma \cdot \psi=0$ one has $\psi_0 = 0$,
implies that in this gauge one does not have gravitinos with helicity
$\ft12$. One can derive an equation for the goldstino $\upsilon =\xi
^{\dagger i}\chi _i+\xi ^\dagger_{ i}\chi ^i+\ft12\rmi\gamma _5{\cal P}
_\alpha \lambda ^\alpha$ directly from the system of equations for the
fields $\phi_i$, $\chi_i$ and $\lambda$ of the globally supersymmetric
theory. This can be done ignoring gravitinos, which decouple from
$\phi_i$, $\chi_i$ and $\lambda$ in the limit $M_P \to \infty$ in the
gauge $\gamma \cdot \psi=0$ with $\psi_0 = 0$. Then one should find a
solution of the goldstino equation and make a transformation from the
gauge $\gamma \cdot \psi=0$ to the unitary gauge, in which the goldstino
becomes the helicity-$\ft12$ component of the gravitino.

That is why the component of the gravitino with helicity $\ft12$ in the
limit $M_p \to \infty$ must satisfy the same equation as the goldstino in
the underlying globally supersymmetric theory. This conclusion should be
valid for the theories including any number of scalar and vector
multiplets. In the particular case with an arbitrary number of scalar
multiplets but without vector multiplets, our conclusion can be verified
by a direct comparison of our equations for the gravitino with the
equations for the goldstino derived in \cite{GTR}.

In the next section we will present the system of equations for the
gravitino and other fermions in the unitary gauge in the limit of large
$M_P$ and compare it with the equations for the goldstino derived in
global supersymmetry in \cite{GTR}. In the case with two chiral
multiplets, this will be done in detail. The equivalence theorem
presented above, explains the reason why the equation for the goldstino
derived from decoupled equations for the chiral fermions, gives a correct
equation for the helicity-$\pm \ft12$ gravitino at large $M_P$.

This result is most important for the theory of gravitino production
because it shows that the rate of their production can be as large as the
rate of production of usual matter fields of the globally supersymmetric
theory, i.e.\ it is not suppressed by the small gravitational coupling. In
other words, one can expect that the density of gravitinos soon after
inflation can be as large as the density of other fermions. In some
theories their density can be much smaller than the density of bosons
after inflation. Also, gravitinos can be diluted or can be converted to
other particles. However, the gravitino--goldstino equivalence theorem
implies that gravitino production in the early universe can potentially be
quite significant.

\section{Gravitino equations with one and two chiral multiplets}
\label{ss:2mult}
In order to have a closed set of equations, we consider
the coupling of the supergravity multiplet to just one and then two chiral
multiplets.
\subsection{One chiral multiplet \label{ss:one}}
Let us show how the gravitino equations for 1 chiral multiplet
\cite{GravProd}--\cite{Lyth:1999ph}, follow from the general case studied
above.

First of all, consider (\ref{trans}) for the transverse gravitino
component. This equation, in the simplest case of real scalars, when
$A_0^B$ vanishes, can be represented as follows:
\begin{equation}
\left(\gamma ^0\partial _0 +\igk+ \Omega _T\right)\vec\Psi^T=0\,, \qquad
\vec\Psi{}^T\equiv a^{1/2}\vec\psi^T\,. \label{trans2}
\end{equation}
where the effective mass is $\Omega_T=m_{3/2}(\eta)a(\eta)$. In the limit
of vanishing gravitino mass, the transverse part $\vec \Psi{}^T$ is
conformal with a weight $+\ft12$. Equation~(\ref{trans2}) is the free
Dirac equation with a time-varying mass term. It is well known how to
treat this type of equations (see, e.g., \cite{GK}). Acting on the
equation with the Hermitian conjugate operation $\left(-\gamma ^0\partial
_0 -\igk+ \Omega _T\right)$ gives rise to the second-order equation
\begin{equation}
\left(\partial _0^2+k^2+\Omega_T^2-\gamma ^0\Omega_T'\right)\vec\Psi _T=0
\,.
 \label{dirac}
\end{equation}
The matrix $\gamma ^0$ can be diagonalized to $\pm\rmi$.

As usual, the situation with the longitudinal component is more
complicated. If there is only one chiral multiplet, and there are no
vector multiplets, then there is only one spin-$\ft12$ fermion, which is
thus the goldstino. Therefore, $\Upsilon$ and $\Xi $ are proportional to
the goldstino and vanish in the unitary gauge. Also, as written in
(\ref{A1}), the quantity $\Delta ^2= 1- |\hat A|^2$ vanishes, and thus
the second line of (\ref{systemLinEqs}) is satisfied in a trivial way.
Therefore,  in this case there remains only one first-order
 equation for the longitudinal part of the gravitino,  $\theta$,
\begin{equation}
   \left( \hat{\partial  }_0+\hat{B}-\igk \gamma _0 \hat{A} \right)  \theta=0
   \,,
\label{1mult}
\end{equation}
and $\hat{A} ^2=1$. The relevant constraint for the longitudinal part is
(\ref{constraint1A}), reducing to
\begin{equation}
  \gamma ^0\psi _0=\hat{A}\theta\,.
\label{constr1m}
\end{equation}
 This is the key equation explaining why gravitinos are efficiently
produced in an expanding universe. The constraint equation in Minkowski
space would be $\gamma ^0\psi _0=-\theta = -\vec\gamma\cdot\vec \psi$.
Meanwhile, as we will see soon, the matrix $\hat A$ in an expanding
universe dominated by the oscillating field $\phi$ rotates with a
frequency twice as great as the frequency of oscillation of the scalar
field $\phi$ (not suppressed by $M_P^{-1}$). This may lead to
non-adiabatic changes in the wavefunction of the gravitino component with
helicity $\ft12$, which results in gravitino production \cite{GravProd}.

The differential equation~(\ref{1mult}) is an unusual equation, which at
first glance indicates the non-permissible strongest effect for the
largest $k$. In fact, it is not so, and equation~(\ref{1mult}) can be
reduced to a physically more transparent equation. The second-order
equation in (\ref{quadreqn}) reduces to the first line, and, inserting
$\hat{A}^\dagger \hat{A}$ in the last term, it can be written as
\begin{equation}
  \left[\hat{\partial }_0\hat{\partial }_0 +\vec k^2+|\hat{B}|^2+2B_1\hat{\partial
  }_0  +a\dot{\hat{B}}-a\dot{\hat{A}}^\dagger \hat{A}
  \left( \hat{\partial  }_0+\hat{B}\right)  \right] \theta=0\,.
\label{2ndorder1m}
\end{equation}

We now restrict ourselves to \emph{real scalars}\footnote{Here and below,
when we consider real scalars, we also assume that $W$ is real.} and use
the relation $m_{3/2} = |m|M_P^{-2}$. Then the expression (\ref{defB})
gives
\begin{eqnarray}
\hat{B}&=&  B_1+\gamma _0 B_2=-\ft32\dot a\hat{A} -\ft12 \gamma
_0m_{3/2}\, a(1+3\hat{A})\,,
  \nonumber\\
  \hat{B}{}^\dagger&=&B_1-\gamma _0 B_2=-\ft32\dot a\hat{A}{}^\dagger
   +\ft12 m_{3/2}\,a\,\gamma_0(1+3\hat{A}{}^\dagger)\,,\nonumber\\
  B_1&=&\frac3{2\alpha }\left(-\dot a\alpha _1 + m_{3/2}a\alpha _2 \right) \,,\quad
  B_2=-\frac{1}{2\alpha }\left( 3\dot a\alpha _2 + m_{3/2} a(\alpha  +3\alpha _1)\right)   \,,
\label{defB1}
\end{eqnarray}
(also see (\ref{eqnB1}) for $B_1$). We can split the spinors into
eigenvectors of $\gamma _0$
\begin{eqnarray}
  \theta &=&\theta _++\theta _-\,,\qquad \theta _\pm =\ft12
  (1\mp \rmi\gamma _0)\theta\, , \nonumber\\
  \theta _\pm(\vec k) ^*&=&\mp {\cal C}\theta _\mp(-\vec k) \,,
\label{thetapm}
\end{eqnarray}
where the last line follows from the Majorana condition, with ${\cal C}$
the charge conjugation matrix. In a representation with diagonal $\gamma
_0$, the components $\theta _\pm $ correspond to the $\gamma
_0$-eigenvalues $\pm \rmi$. Choosing for each $k$ a spinor basis
$u_{1,2}(k)$ for the two components of $\theta _+$, and two independent
solutions of the differential equations $f_{1,2}(k,\eta )$, the general
solution is thus in the variables of (\ref{thetapm})
\begin{equation}
\theta _+=
  \sum_{\alpha ,\beta =1}^2a^{\alpha \beta }(k)f_\alpha(k,\eta )
  u_\beta(k)\,, \qquad
\theta _-= -{\cal C}^{-1}
  \sum_{\alpha ,\beta =1}^2a^{*\alpha \beta }(-k)f^*_\alpha(-k,\eta )
  u^*_\beta(-k)\,.
\label{gensolL}
\end{equation}

On the $\theta _+$ components the
differential equation is then
\begin{equation}\label{x1}
\left[ \partial ^2_0 +2\left(B_1+\rmi a\mu\right)\partial _0 + \left(\vec
k^2+ \vert B\vert^2+ (\partial _0B)+ 2\rmi Ba\mu \right)\right] f(k,\eta
)=0 \,,
\end{equation}
where $B$ now stands for $B_1+\rmi B_2$,
\begin{equation}
A=\frac{1}{\alpha }(\alpha _1+\rmi\alpha _2)\,,\qquad
  \mu \equiv \frac\rmi 2\dot A^*A=\frac{\rmi}{2}\frac{\dot
  A^*}{A^*}\,,
\label{defmu}
\end{equation}
and $\mu $ is real due to $|A|^2=1$. An explicit expression for $\mu$ is
derived in  appendix~\ref{app:calcmu}:
\begin{equation}
  \mu =\left(m_{11}+m_{3/2}\right)
  +3(H\dot \phi - m_{3/2}m_1)\frac{m_1}{\dot \phi ^2 +m_1^2}\,,
\label{muism2a}
\end{equation}

Now we have to find the properties of the function $f(k,\eta)$.
Equation~(\ref{x1}) can be simplified significantly further. We can
reduce this equation to an oscillator-like equation by the substitution
\begin{equation}
f(k,\eta)= E(\eta) y(k,\eta)\,\qquad  E(\eta)=(-A^*)^{1/2} \exp\left(-
 \int^\eta \,\rmd \eta\, B_1(\eta)\right) \,.
 \label{sub}
\end{equation}
 Then equation~(\ref{x1}) is reduced to the final equation
\begin{equation}
\left(\partial _0^2+k^2+\Omega^2-\rmi(\partial _0\Omega)\right)y=0 \,,
 \label{final}
\end{equation}
where $\Omega=-B_2+a\mu$. Taking $A=1$ at $t=-\infty $, and using
(\ref{defmu}), we can write ($\rmd t= a\,\rmd \eta $)
\begin{equation}
  A^*=-\exp \left(- 2\rmi\int_{-\infty }^t \rmd t \,\mu (\eta)\right)\,.
\label{Astarmu}
\end{equation}

Thus, the character of the solution of the longitudinal gravitino
component is ultimately defined by the matrix $A$ in terms of the phase
function $\mu(\eta)$. The effective frequency $\Omega$ and   mass $\tilde
m$ of the gravitino with helicity $\pm \ft12$  is given in terms of $\mu$:
\begin{eqnarray}
 \tilde m \equiv \frac{\Omega }{a}&=&
  \mu+\frac3{2\alpha }H \alpha _2+ \frac1{2\alpha }
m_{3/2}  \left(\alpha  +3 \alpha _1\right) \nonumber\\ &=&  \mu-\ft32 H
\sin 2{\int \rmd t \, \mu }+ \ft12  m_{3/2}   \left(1 -3 \cos{2\int \rmd
t \, \mu }\right) \,.
 \label{eff2}
\end{eqnarray}
 These results     \cite{GravProd}
coincide with the results   obtained later in \cite{Giudice:1999yt}, up
to the   difference in notation\footnote{A detailed comparison of our
results with those of \cite{Giudice:1999yt,GTR} requires some work
because of differences in notation, which are not always explicitly
defined in \cite{Giudice:1999yt,GTR}. }.

\subsection{Two chiral multiplets \label{ss:sol2mult}}
Let us consider a slightly more complicated case, when  the supergravity
 multiplet couples to  two chiral multiplets, and we again allow \emph{complex
 scalars}.
Then there are at first 2 spin-$\ft12$ fields. However, one is the
goldstino mode, and one remains. This is thus the $\Upsilon$ spinor, but,
as there is no other spin-$\ft{1}{2}$ left, also $\Xi$ should be
proportional to it,
\begin{equation}
  \Xi =-a^{-1} \hat{F} \Upsilon\,,
\label{OmegaFUps}
\end{equation}
where $\hat{F}$ is a matrix that we will determine. Therefore, in this
case we can still suffice with the dynamical
equations~(\ref{systemLinEqs}), and do not need to extract a further
dynamical equation for $\Xi $.

If there are only two chiral multiplets, then $\Pi_{ij}$ has only one
non-vanishing component, $\Pi _{12}=-\Pi_{21}$. Also the factors $g^{-1}$
in (\ref{Pisquare}) then reduce to the determinant of the K{\"a}hler metric,
so that this equation reduces to
\begin{equation}
  |\Pi _{12}|^2=\ft14 \Delta^2 \det g\,.
\label{Pi12square}
\end{equation}
Therefore, equation~(\ref{Usef2}) leads to an expression of $\chi _i$ in
terms of $\Upsilon$:
\begin{equation}
  \Pi _{ij}P_L \xi ^{\dagger j}\Upsilon= -\ft14 a\alpha \chi _i\Delta ^2
  \det g\,.
\label{xiinverse}
\end{equation}
Therefore, we have in this case (\ref{OmegaFUps}) with
\begin{equation}
\hat{F}=-\,\frac{4}{\alpha \Delta ^2\det g} \left[\xi ^k
\gR k\ell\, m_{\ell i}\Pi ^{ij}
\xi^\dagger _j +\xi _k\gL\ell k
m^{\ell i}\Pi _{ij}\xi^{\dagger j}
\right] \,.
\label{Omega2}
\end{equation}

We may thus now use (\ref{OmegaFUps}) in either the first-order
equations~(\ref{systemLinEqs}) or in the second-order
equation~(\ref{quadreqn}). This gives for the case of 2 chiral multiplets
either
\begin{eqnarray}
&&\left(\hat{\partial}_0+ \hat{B}-\igk \gamma _0 \hat{A} \right)
 \theta-\frac{4}{\alpha a}\vec k^2 \Upsilon=0 \,, \nonumber\\
 & &\left(\hat{\partial }_0+\igk \gamma _0\hat{A} +\hat{B}^\dagger
+a\hat{F} +2\dot a - M_P^{-2}a\mathbf{m}\gamma _0\right) \Upsilon
 +\ft14 a \alpha \Delta ^2\theta=0\,,
\label{summ2mult}
\end{eqnarray}
or
\begin{eqnarray}
  0 & =&
  \left[\hat{\partial }_0\hat{\partial }_0 +\vec k^2+|\hat{B}|^2+2B_1\hat{\partial
  }_0  +a\dot{\hat{B}}-\igk\gamma
  _0a\dot{\hat{A}} \right] \theta \nonumber\\
  &&+\left( 2B_1+a\hat{F}-M_P^{-2}a\mathbf{m}\gamma
  _0\right) \left[\hat{\partial }_0 +\hat{B}-\igk \gamma _0 \hat{A} \right]
 \theta\,.
\label{result2m}
\end{eqnarray}
In the case of real backgrounds, one may neglect the hat on the
$\partial _0$ derivative, and $\mathbf{m}$ is then just $\mathbf{m}=m=M_P^2
m_{3/2}$.
\subsection{The limit of large $M_P$\label{ss:LargeMP}}
Now let us consider the limit $M_P \to \infty$. In this limit one can
neglect expansion of the universe and put $a(t) = 1$. This removes the
difference between the usual time $t$ and the conformal time $\eta$. One
can omit all time derivatives of the scale factor, all $B$-terms, as well
as the terms containing $M_P^{-2}$ in the dynamical equations. We will
consider here the theory  with no contributions of vector multiplets and a
minimal K{\"a}hler potential, ${\cal K}=M_P^{-2}\phi_i \phi^i$ so that in the
limit $M_P \to \infty$ one can take
\begin{equation}
g_i^j = \delta_i^j\,, \qquad
m_i= W_i = {\partial W \over
\partial \phi^i} \,,\qquad  n_i=\dot \phi _i\,,\qquad
m_{ij}=W_{ij} = {\partial^2 W \over \partial
\phi^i\partial\phi^j}\,.
\label{mW}
\end{equation}
This implies that
\begin{equation}
  \xi_k = W_k +\gamma_0 \dot\phi_k\,, \qquad \xi ^\dagger _k = W_k -\gamma_0
\dot\phi_k\, , \label{xiMinimal}
\end{equation}
with derivative
\begin{equation}
\gamma_0 \dot \xi_i= W_{i j}\xi^j\,, \qquad
\gamma_0  \dot \xi_i^{\dagger  }=- W_{i j}\xi^{\dagger j}\,.
\label{calM2aa}
\end{equation}
Furthermore, in this limit, we have a simpler expression for $\hat A$.
One has
\begin{eqnarray}
\alpha =\rho \,, \qquad \alpha _1\!\!\!&=\!\!\!&p\,,\qquad
\alpha _2=\dot W+\dot W^*+\gamma _5(\dot W-\dot W^*)\,,\nonumber\\ \nonumber\\
&&  \hat A = {{ p} \over { \rho}} +\gamma_0{\alpha _2
 \over { \rho}}\,.
\label{A3}
\end{eqnarray}
For a real background $\alpha _2=2\dot W$, and
\begin{equation}
\hat A =  {{ p} \over { \rho}} +{2\gamma_0\dot W \over { \rho}}\ , \qquad
\Delta^2\equiv 1- |\hat A |^2 \,. \label{realA}
\end{equation}
The linear equations reduce to
\begin{eqnarray}
&& \left({\partial}_0 -\igk \gamma _0 \hat{A} \right)
 \theta-\frac{4}{\rho }\vec k^2 \Upsilon=0 \,,\nonumber\\
&&\left[{\partial }_0+\igk \gamma _0\hat{A}\right] \Upsilon
-\Xi +\ft14\rho \Delta ^2\theta=0\,. \label{systemLinEqsinLim}
\end{eqnarray}
Here
\begin{equation}
  \Upsilon= -\ft12 \gamma _0(\xi _i\chi ^i+\xi^i\chi _i)\,,
  \qquad
  \Xi =-\xi ^j W_{ji}\chi ^i-\xi _j W^{ji}\chi _i\, .
\label{XiLimit}
\end{equation}
The chiral fermions satisfy the constraint (see (\ref{goldstino_in_xi}))
\begin{equation}
\upsilon =  \xi ^{\dagger i}\chi _i+\xi
^\dagger_{ i}\chi ^i=0 \,. \label{orthogonal}
\end{equation}
One can compare the large-$M_P$ gravitino equation coupled to the chiral
fermions, constrained as shown in (\ref{orthogonal}), with the relevant
system of equations presented for the goldstino in global supersymmetry in
\cite{GTR}. Taking into account the difference in notation and the
treatment of right and left chiral fermions, one can identify
equations~(34)--(37) of \cite{GTR} with our equations presented above and
derived from the large-$M_P$ limit of supergravity.

We may also present our equations as a quadratic equation with $\Upsilon$
excluded.
\begin{equation}
  \ddot\theta +\left[\vec k^2-\igk\gamma
  _0 \dot{\hat{A}} \right] \theta-\frac{4\vec k^2}{\rho }\Xi  =0\,.
\label{quadrInLim}
\end{equation}
Here again taking account of the difference in notation one can recognize
equation~(38) in \cite{GTR}. Thus we have shown that the equations for a
helicity-$\pm \ft12$ gravitino in the limit of large $M_P$ coincide with
equations for the goldstino that can be derived within the framework of
global supersymmetry from equations for chiral fermions and scalars. The
large-$M_P$ equivalence theorem proved in section~\ref{ss:equivTh} has
explained the reason for this.

\subsection{One  chiral multiplet in the limit of large $M_P$
\label{ss:1mulLimit}}

It is instructive to consider a weak-coupling limit $M_P \to \infty$ of
the gravitino equation~(\ref{final}) for a single chiral multiplet. As we
just mentioned, in this limit one has $a(t)=1$, and the conformal time
$\eta$ is equal to the coordinate time $t$. A lengthy expression for the
effective gravitino mass (\ref{eff2}) with $\mu$ given by (\ref{muism2})
is reduced to the very simple form
\begin{equation}
\Omega=\tilde m = \mu= m_{11}= \partial^2_\phi W\,.
\end{equation}
This is the mass of the chiral fermion field $\chi$ in the weak-coupling
limit (i.e.\ at $M_P\to \infty$). Also, in this limit we have from
(\ref{sub}) and (\ref{Astarmu}) that $E(\eta )= \rme ^{- \rmi \int \rmd t
\, \mu}$. Therefore, \begin{equation}
 f(k,t) = \rme ^{- \rmi \int \rmd t \, \mu}\, y(k,t)\, ,
\end{equation}
 see (\ref{sub}). Removing the overall factor $\rme ^{- \rmi \int \rmd t \, \mu}$
 into the field redefinition, one can investigate the gravitino production by
 solving equation~(\ref{final}) with time-dependent effective mass $\Omega$.

Let us now compare this result with the properties of the goldstino. We
have (\ref{orthogonal}), and, using (\ref{calM2aa}) in the case of a
single real scalar field, we obtain
\begin{equation}
\xi ^{\dagger 1 }= \xi ^\dagger_{1}=\rho^{1/2}\rme ^{ \gamma_0 \int dt
\mu},
\end{equation}
where $\rho$ is the energy density of the background scalar
 field; it is constant in the weak-coupling limit.
 Finally, we obtain
\begin{equation}\label{goldmany}
 \upsilon =  \rho^{1/2}\rme ^{ \gamma_0 \int \rmd t \, \mu}\, (\chi_1 + \chi^1) .
\end{equation}
Here chiral fermions $\chi_1$ and $\chi^1$ obey  the Dirac equation with
mass $\mu$ and without gravity ($a(t) = 1 $).

According to the gravitino--goldstino equivalence theorem,  in the limit
of weak gravitational coupling the longitudinal gravitino $\theta$ has
the same properties of the goldstino $ \upsilon$, which is proportional
to the symmetric combination $(\chi_1 + \chi^1)$ of the right and left
components of the chiral fermion $\chi$ (`inflatino' in the context of
inflation). It can be shown directly that (\ref{goldmany}) is a solution
of the gravitino equation~(\ref{1mult}), in the limit $M_P \to \infty$.

As was demonstrated in \cite{GravProd,Giudice:1999yt},
equation~(\ref{final}) describes the creation of gravitinos in
cosmological backgrounds of varying $a(t)$ and $\phi(t)$. The most
important result, however, is that even in the weak-coupling limit
 $M_P \to \infty$, the effect does not disappear. This is because the
longitudinal gravitino couples to the  time-dependent scalar background
in the same way  as the goldstino.

\subsection{Two chiral multiplets in the limit of large $M_P$
\label{ss:2mulLimit}} Let us consider  the case with 2 scalar fields. In
this case one can use (\ref{quadrInLim}) and relation
$\Xi=-\hat{F}\Upsilon$, to obtain the following second-order equation
for $\theta$:
\begin{eqnarray}
  \ddot \theta +\left[ \vec k^2 -\igk\,\gamma
  _0\, \dot{\hat A}   +\hat{F} (\hat{\partial }_0-\igk\, \gamma _0\, \hat{A})\right]
 \theta = 0\,.
\label{result2ma}
\end{eqnarray}
If, for simplicity, one considers   two \emph{ real } scalar fields,
there is no difference between upper and
lower indices. Then one has
\begin{equation}
\Delta = 2\Pi _{12}=\frac{2}{\rho }\left(W_1\dot\phi_2 -W_2\dot \phi_1\right)
\,.
\label{calM2a}
\end{equation}
Therefore, the expression for $\hat F$ reads (with
$\varepsilon^{12}=-\varepsilon ^{21}=1$)
\begin{eqnarray}
 \hat F &=&-\frac{2}{\rho \Delta }\xi^k\, W_{k i}\varepsilon ^{ij}
  \xi^\dagger _j=- \frac{\gamma_0 (\dot \xi_1 \xi^\dagger _2 -\dot \xi_2
\xi^\dagger _1)}{W_1\dot\phi_2 -W_2\dot \phi_1 } \,. \label{Omega2aa}
\end{eqnarray}

Up to the difference in notation, our equation for the helicity-$\ft12$
gravitino in the limit $M_p \to \infty$,  (\ref{result2ma}), coincides
with the equation for the goldstino   obtained in \cite{GTR} in the
context of a globally supersymmetric theory. This is exactly what one
could  expect  in accordance with   the gravitino--goldstino equivalence
theorem, which should be valid for any number of chiral and vector
multiplets in the limit $M_P \to \infty$ (see section~\ref{ss:equivTh}).

This implies that the goldstino $ \upsilon =  \xi ^{\dagger i }  \chi_i
+\xi ^\dagger_{i}  \chi^i $, with all terms taken in the limit $M_P \to
\infty$, satisfies the gravitino equation~(\ref{result2ma}). Thus,
instead of trying to solve  (\ref{result2ma}) directly, which can be
rather complicated, one may try to find time-dependent solutions for
$\phi_i$ and $\chi_i$, which can be much simpler. Then one should build
from them the combination $\xi ^{\dagger i }  \chi_i +\xi ^\dagger_{i}
\chi^i$. This combination should solve the gravitino
equation~(\ref{result2ma}).

This method is not quite sufficient for the investigation of the problem
of gravitino production because the initial conditions for the
wavefunction of gravitinos should be formulated in terms of gravitinos
(because they correspond to physical degrees of freedom) rather than in
terms of $\chi_i$ and $\phi_i$. Moreover, in some important cases which
we are going to discuss later, the limit $M_P \to \infty$ is inadequate
at  intermediate stages  of the gravitino production. Still the method
described above  can be  very useful   if one wants to increase one's
intuition by working with extremely complicated equations for the
gravitino in theories with several chiral multiplets.

For example, let us consider a model describing several chiral multiplets.
In simple cases when the mass matrix $m^{ij}$ is diagonal, $m^{ij}=\delta
_{ij}\mu _i$, the combination $\upsilon  =   \xi ^{\dagger i }  \chi_i
+\xi ^\dagger_{i} \chi^i $ has a transparent form
\begin{equation}
 \upsilon  = \sum_i \rho_i^{1/2} \rme ^{ \gamma_0\int \rmd t \, \mu_i}(\chi_i+
\chi^i) \ , \label{form}
\end{equation}
where $\rho_i$ is a fraction of total energy in the scalar field
$\phi_i$:
\begin{equation}
  \rho _i=\xi _i^\dagger \xi _i\,,\qquad \rho =\sum_i \rho _i\,.
 \label{rhoi}
\end{equation}

It would be rather difficult to find this solution by solving
(\ref{result2ma}) directly.  For example, in the qualitative analysis of
this equation performed  in  \cite{GTR} it was assumed that $\Delta$
never vanishes and the matrix $\hat F$ is non-singular. Let us consider,
however,  the simplest model of two non-interacting chiral fields with $W
= {\mu\over 2}\phi_1^2 + \zeta\phi_2 + \beta$. This model mimics the
theory involving the chiral field $\phi_1$ with mass $\mu$ and the
Pol\'{o}nyi \cite{Polonyi:1977pj} field $\phi_2$ in the hidden sector.
Equations for the scalar fields in this theory have a solution $\phi_1 =
C \cos \mu t$ (with $C>0$) and $\phi_2 = 0$, which gives
\begin{eqnarray}
&& \xi _1=\mu  C \rme ^{-\gamma _0\mu t}\,,\qquad \rho _1=(\mu C) ^2\,,
\qquad  \xi _2=\zeta\,,\qquad \rho _2=\zeta ^2\,,\nonumber\\
 && \Delta  =  {2 (\rho_1\rho_2)^{1/2} \sin  \mu t \over
\rho }\,, \qquad \hat F =  - {\mu  \, \rme ^{ \gamma_0 \mu  t }\over  \sin
\mu t}\,   . \label{deltaF1}
\end{eqnarray}
 Thus we see that $\Delta$ vanishes and the  matrix $\hat
F$ blows up each time when $\sin mt = 0$.

A similar result appears if one considers a theory of two non-interacting
oscillating fields $\phi_i$ with masses $\mu _i$ and superpotentials
${\mu _i\over 2} \phi^2_i$. In this case, for the solutions $\phi_1 = C_i
\cos \mu _it$, one has
\begin{equation}\label{deltaF2}
\Delta =  {2 (\rho_1\rho_2)^{1/2} \sin  (\mu _1-\mu _2)t \over \rho }\,,
\qquad \hat F =  - { \mu _1\rme ^{ \gamma_0 (\mu _2-\mu _1) t } -\mu
_2\rme ^{ \gamma_0 (\mu _1-\mu _2) t } \over  \sin  (\mu _1-\mu _2)t}\,  .
\end{equation}

Thus the matrix $\hat F$  is generically singular, so one should be very
careful in investigating the gravitino equation in the form
(\ref{result2ma}). One of the sources of this problem is obvious: we
wanted to write a single equation for the gravitino (or goldstino), which
is a complicated time-dependent combination of several other fields. Even
if  each of these fields changes in a simple way, the evolution of their
nonlinear combination can be pretty complicated. Therefore, one should try
to find the best way to investigate this equation and to study the
gravitino production, depending on the choice of a particular model.

As a useful step in this direction, one may try to establish some
relation between the cases of one and two chiral multiplets. With this
purpose, we will return to the original system of
equations~(\ref{summ2mult}) for 2 chiral multiplets, and present them in
the large-$M_P$ limit:
\begin{eqnarray}
&&\left({\partial}_0-\igk \gamma _0 \hat{A} \right)
 \theta-\frac{4}{\rho  }\vec k^2 \Upsilon=0 \,, \nonumber\\
 & &\left({\partial }_0+\igk \gamma _0\hat{A}
+ \hat{F}\right) \Upsilon
 +\ft14  \rho \Delta ^2\theta=0\,.
\label{summ2multreal}
\end{eqnarray}
Now let us consider, as an example, a generic model of two real
non-interacting fields $\phi_i$ and take into account that $\Delta =
\frac{2}{\rho }\left(W_1\dot\phi_2 -W_2\dot \phi_1\right)$ and $|W_i|,
|\dot\phi_i|\leq \sqrt\rho_i$. This implies that
\begin{equation}
\Delta^2 \leq \frac{16 \rho_1\rho_2}{\rho^2 }
 \,. \label{calM2a2}
\end{equation}
This inequality agrees with  (\ref{deltaF1}) and (\ref{deltaF2}).

In the early universe one often encounters situations when the energy
densities $\rho_i$ of different fields differ from each other by many
orders of magnitude. Suppose  that in the early universe $\rho  \approx
\rho_1 \gg \rho_2$. Then one finds that $\Delta^2 \leq \frac{16
\rho_2}{\rho_1} \ll 1$.

Let us substitute $\Upsilon$ from the first of
equations~(\ref{summ2multreal}) to the second one. This gives
\begin{equation}
\left({\partial }_0+\igk \gamma _0\hat{A} + \hat{F}\right)
\left({\partial}_0-\igk \gamma _0 \hat{A} \right)\theta
 + \vec k^2 \Delta ^2\theta=0
 \,. \label{calM2a2a}
\end{equation}
In the limit $\Delta ^2 \ll 1$ the last term in this equation vanishes as
compared with $\vec k^2 A^2\theta$. Omitting  $\vec k^2 \Delta ^2\theta$
leads to the equation $({\partial }_0+\igk \gamma _0\hat{A} + \hat{F})
({\partial}_0-\igk \gamma _0 \hat{A})\theta = 0$ which is solved by the
solutions of the one multiplet equation $({\partial}_0-\igk \gamma _0
\hat{A})\theta=0$.

One could come to a similar conclusion in a different way. From the
expression for the goldstino (\ref{form}) (which was obtained in the limit
$M_P \to \infty$, assuming that the matrix $m^{ij}$ is diagonal) one may
conclude that if, say, $\rho_1$ is much greater than all other $\rho_i$,
the expression for the goldstino will be the same as in the theory of a
single chiral field: $\upsilon  = \rho_1^{1/2} \rme ^{ \gamma_0\int \rmd
t\, \mu_1}(\chi_1+ \chi^1)$. Therefore, in this case  the equation for the
goldstino (and, accordingly, the equation for the gravitino in the
large-$M_P$ limit) will be the same as in the theory of a single chiral
multiplet.

Thus, in the situations when the energy density of the universe is
dominated by one particular scalar field, which does not interact with
other scalars, or if $\Delta^2 \ll 1$ for any other reason, the gravitino
equation can be reduced to the equation in the theory with one multiplet,
which is much easier to solve. This may help to find the solution to the
gravitino equations at some particular stages of the process; the main
complication appears when the energy density of various fields become of
the same order.  We will discuss this issue later.

Another lesson is that it may be useful to rearrange (\ref{result2ma}) in
an equivalent form that does not contain singular functions:
\begin{eqnarray}
   (\hat{\partial }_0-\igk\, \gamma _0\, \hat{A}) \theta +
   \hat{F}^{-1}\left[\ddot \theta +(\vec k^2 -\igk\,\gamma
  _0\, \dot{\hat A} )\theta\right] = 0\,.
\label{result2ma2}
\end{eqnarray}
The term proportional to $\hat{F}^{-1}$ in this equation appears because
of the existence of two multiplets. In those cases when the solution of
the one-multiplet equation $(\hat{\partial }_0-\igk\, \gamma _0\,
\hat{A})) \theta=0$ is a good approximation to the exact solution, one
should be able to verify that the term proportional to $\hat{F}^{-1}$ is
small.

Finally, in this section we shall note that $(\theta, \Upsilon)$ is not
the only possible basis when we work with two chiral multiplets in the
unitary gauge. On can try, for example, to work with $\theta$ and $\Xi$,
or with another combination of $\chi_1$, $\chi_2$ instead of $\Upsilon$
and $\Xi$.

\section{Towards the theory of the gravitino production
in the early universe \label{ss:creation}}

\begin{figure}
\centerline{\leavevmode\epsfxsize=0.7\columnwidth \epsfbox{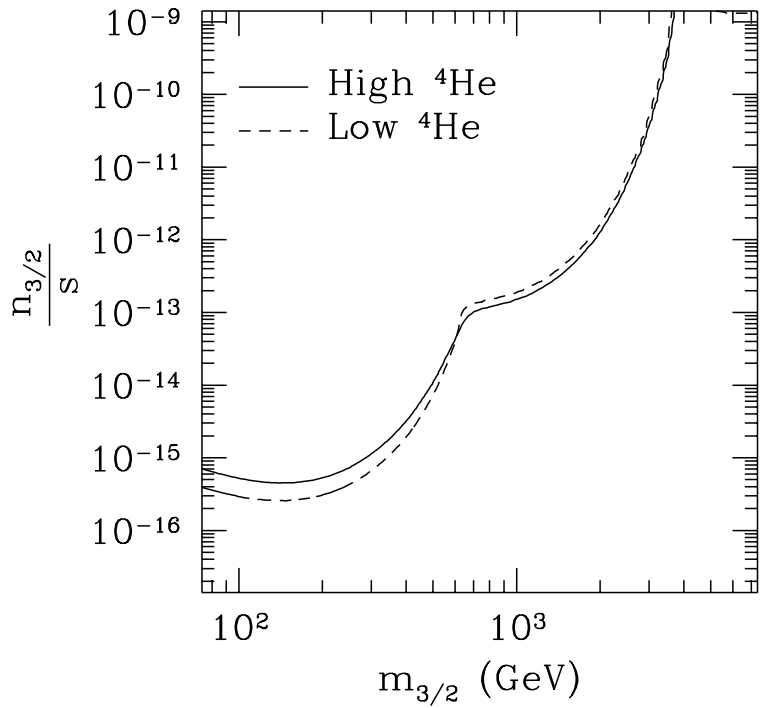}}

\caption{The constraint on the ratio $n_{3/2}/s$ that follows from the
theory of nucleosynthesis, according to M. Kawasaki and K. Kohri. The
allowed values for $n_{3/2}/s$ are below the full (or broken) curve. The
two curves correspond to the two slightly different observational results
concerning the cosmological abundance of $\rm{}^4He$. We are grateful to
Kawasaki and Kohri for permission to present their results in our paper.}
\label{fig0}
\end{figure}

There are many different versions of phenomenological supergravity, each
giving its own predictions for the mass of the gravitino. In the standard
textbook version  \cite{Nilles,BW,BL,gravitinos,Moroi} the gravitino mass
is supposed to be of the same order as the electroweak scale, $m_{3/2}
\sim 10^2 $--$ 10^3$~GeV. Such particles decay very late, and lead to
disasterous cosmological consequences unless the ratio of the number
density of gravitinos $n_{3/2}$ to the entropy density $s$ is extremely
small. For example, the ratio of the number density of gravitinos
$n_{3/2}$ to the entropy density $s$ should be smaller than ${\rm
O}(10^{-15})$ for gravitinos  with mass ${\rm O}(100)$~GeV
\cite{gravitinos,Moroi}. This constraint depends on the value of the
gravitino mass, as shown in figure~\ref{fig0}.

It would be very interesting to apply the results of our investigation to
the problem of gravitino production in the early universe in realistic
cosmological models based on supergravity. However, this complicated
problem comes in the same package with the cosmological moduli problem.
Also, to obtain a completely consistent scenario, one should
simultaneously find a natural realization of the inflationary scenario in
supergravity.  In this paper we will only briefly discuss certain
features of the theory of gravitino production after inflation. We hope
that our comments will be helpful for future studies of this issue.

\subsection{Initial conditions \label{se:deSitter}}

One could expect that the gravitino production may begin already at the
stage of inflation, due to the breaking of conformal invariance. However,
there is no massive particle production in de~Sitter space. Indeed,
expansion in de~Sitter space is in a sense fictitious; one can always use
coordinates in which it is collapsing or even static. An internal observer
living in de~Sitter space would not see any time dependence of his
surroundings caused by particle production; he would only notice that he
is surrounded by particle excitations at the Hawking temperature $H/2\pi$.

Still it may be important to incorporate an investigation of the gravitino
during inflation in order to set initial conditions for the subsequent
stage of gravitino production. During inflation we cannot assume that the
gravitational coupling is weak. In this situation, there is no simple
interpretation of the effect with the gravitino in terms of the components
of the  goldstino.

For simplicity, consider the de~Sitter background, where all time
derivatives of the scalar fields vanish, $\dot \phi_i=0$. In this case
$\dot m=0$, $p=-\rho$, thus the parameter $A=-1$. We can also choose the
VEVs of scalars in real directions, so that their imaginary parts vanish.
Equation~(\ref{trace1}) in this case is reduced to ($\Upsilon=0$ as
scalars are constant, and therefore we can use below the equations for
one multiplet)
\begin{equation}
\left( \partial_{0} \theta+\vec  \gamma\cdot \vec \partial +\ft32\dot a+
m_{3/2} a \gamma_0\right) \theta=0 \,. \label{DS}
\end{equation}
Also, in this case $\theta=- \gamma^0 \psi_0$. Then, equation~(\ref{DS})
can be rewritten as
\begin{equation}
\left(  \not\! {\partial}+\ft32\dot a\gamma_0   + m_{3/2}a\right)\psi_0=0
\,. \label{Dir}
\end{equation}
A simple conformal transformation $\psi_0=a^{-3/2}\Psi_0$ reduces this
equation to the Dirac equation in flat spacetime,
\begin{equation}
\left(  \not\! {\partial}+  m_{3/2} a\right)\Psi_0=0 \,. \label{Dirflat1}
\end{equation}
Note that the degrees of freedom associated with the helicity $\pm\ft12$
in de~Sitter space have conformal properties of fermions with spin
$\ft12$.

The transversal gravitino component $\psi^T_i$ can after the
transformation $\psi^T_i=a^{-1/2}\Psi^T_i$  be reduced to the same
equation~(\ref{Dirflat1}). Therefore, further results will be valid for
both helicities. In terms of  (\ref{final}) we have $B_1=-\ft32\dot a$,
$\mu =0$, $\Omega=-B_2=- m_{3/2}\, a$. The factor of $E(\eta)$ in
(\ref{sub}) is equal to $a^{3/2}$. Therefore, the factor $y(k,\eta)$ in
the expression for $\Psi_0$ (or $\theta$) and $\Psi^T$   satisfies the
equation
\begin{equation}\label{sp39}
\partial _0^2 y  +    \left( k^2 + m_{3/2}^2a^2
+\rmi   m_{3/2} a\dot a \right) y = 0\,.
\end{equation}

For $a(\eta)=-{1\over {H \eta}}$, \ $-\infty < \eta < 0$ (which is $C=0$
in (\ref{curvdS})), the exact solution of  (\ref{sp39}) with the
vacuum-like initial condition $y(k,\eta)={1 \over \sqrt{2}}\rme ^{-\rmi
k\eta}$ for $\eta \to - \infty$ is given by
\begin{equation}\label{solut}
y(k,\eta)={1 \over 2} \sqrt{\pi k \vert \eta \vert} \,  \exp\Bigl( {{\pi
m_{3/2}} \over 2H}\Bigr) \,
 {\cal H}^{(1)}_{{1 \over 2} -\rmi {m_{3/2} / H}}(|k\eta|)\,,
\end{equation}
where $ {\cal H}^{(1)}$ are Hankel functions \cite{KLSu}.

Although equations for the gravitino at the de~Sitter stage in the
unitary gauge turn out to be decoupled  from the  chiral fermions and
gauginos, equations for chiral fermions and gauginos at this stage are not
decoupled from the gravitino.

Gravitino production may occur at the stage of inflation due to the
(slow) motion of the scalar field, but the most interesting effects occur
at the end of inflation, when the scalar field $\phi$  rolls rapidly down
toward the minimum of its effective potential $V(\phi)$ and oscillates
there. During this stage the vacuum fluctuations of the gravitino field
are amplified, which corresponds to gravitino production.

This effect depends on the detailed structure of the theory. First we
will comment on the theory of gravitino production in the models with one
chiral multiplet, and then we will make some comments on this process in
more realistic models with several multiplets.

\subsection{Theories with one chiral multiplet}
Production of  gravitinos with helicity $\ft{3}{2}$  in the theory with
one chiral multiplet is described in terms of the mode function
$y_T(\eta)$. This function  obeys equation~(\ref{dirac}) with
$\Omega_T=m_{3/2}a$, which is suppressed by $M_P^{-2}$. Non-adiabaticity
of the effective mass $\Omega_T(\eta)$ results in the departure of
$y_T(\eta)$ from its positive-frequency initial condition $\rme ^{\rmi
k\eta}$, which can be interpreted as particle production. The theory of
this effect was investigated in
\cite{Maroto:2000ch,Lemoine:1999sc,GravProd,Giudice:1999yt,GTR}; it is
completely analogous to the theory of production of usual fermions of
spin $\ft12$ and mass $m_{3/2}$ \cite{GK}. Indeed,  (\ref{dirac})
coincides with the basic equation that was used in \cite{GK} for the
investigation of production of Dirac fermions during preheating.

The description of production of gravitinos with helicity $\ft12$ is
similar but somewhat more involved. The wavefunction of the
helicity-$\ft12$ gravitino is a product of the  factor $E(\eta)$ and the
function $y(\eta)$. The factor $E(\eta)$ does not depend on momenta and
controls only the overall scaling of the solution. It is the function
$y(\eta)$ that controls particle production which occurs because of the
non-adiabatic variations of the effective mass parameter $\tilde m =
\Omega(\eta)/a(\eta)$. The function $y(\eta)$ obeys
equation~(\ref{final}) with effective mass $\tilde m$, which is given by
the superposition (\ref{eff2}) of all three mass scales in the problem:
$\mu$, $H$ and $m_{3/2}$.

In different models of the inflation, different terms of $\tilde m$ will
have a different impact on the helicity-$\ft12$ gravitino production. The
strongest effect usually comes from the largest mass scale $\mu$, if it
is varying with time. This makes the production of gravitinos of helicity
$\ft12$ especially important.

To fully appreciate this fact, we will  consider several toy models where
the effective potential at the end of inflation has a simple shape such as
$V \sim \phi^n$. We will not discuss here the problem of finding
superpotentials which lead to such potentials (and inflation) at $\phi >
M_P$ \cite{book}, because we are only interested in what happens after
the end of inflation, which occurs at $\phi \sim M_P$.

First, consider the superpotential $W = \ft12m_\phi \phi^2$.  Here we will
consider real fields $\phi$ and switch to the minimally normalized field,
$\phi \to \phi/\sqrt 2$. At $\phi \ll M_P$ it leads to the simple
quadratic potential $V ={ m_\phi^2\over 2} \phi^2$. The parameter $\mu$
in this case coincides with the inflaton mass $m_\phi$. In a realistic
inflationary model one should take $m_\phi \sim 10^{13}$~GeV, which is
equal to $5\times 10^{-6} M_P$ \cite{book}. The Hubble constant during the
field oscillations is given by ${m_\phi \phi_0\over \sqrt{6} M_P}$, where
$\phi_0(t)$ here is the amplitude of the field oscillations, which
decreases during the expansion of the universe. The gravitino mass is
given by $m_{3/2} = {m_\phi \phi^2\over 4 M_P^2}$.

Thus, at the end of inflation in this model, which occurs at $\phi \sim
M_P$, all parameters determining the behaviour of the gravitino
wavefunction are of the same order, $\mu \sim m_\phi \sim H \sim m_{3/2}$.
However, later the amplitude of $\phi$ decreases as $\phi_0 \sim {1.5
M_P\over m_\phi t} \sim {M_P\over 4 N}$, where $N$ is the number of
oscillations of the field $\phi$ after the end of inflation \cite{KLS97}.
Thus already after a single oscillation there emerges a hierarchy of
scales, $\mu \sim m_\phi \gg H \gg m_{3/2}$.

Since $m_\phi =\rm constant$, after the first oscillation the parameter
$\mu$ becomes nearly constant,  the parameters $H$ and $m_{3/2}$ become
very small, and their contribution to the gravitino production becomes
strongly suppressed. As a result, the dominant contribution to the
gravitino production in this model occurs within the first oscillation of
the scalar field after the end of inflation. Each of the parameters
$\mu$, $H$ and $m_{3/2}$ at the end of inflation changes by $O(m_\phi)$
within the time ${\rm O}(m_\phi^{-1})$. This means that (because of the
uncertainty relation) gravitinos of both helicities will be produced,
they will have physical momenta $k = {\rm O}(m_\phi)$ and their
occupation numbers $n_k$ will be not much smaller than ${\rm O}(1)$. This
leads to the following conservative estimate of the number density of
gravitinos produced: $n_{3/2} \sim 10^{-2} m^3_\phi$.

Now let us assume for a moment that all the energy of the oscillating
field $\phi$ transfers to thermal energy $\sim T^4$ within one
oscillation of the field $\phi$. This produces a gas with entropy density
$s \sim T^3 \sim \left({m_\phi^2 M_P^2\over 2}\right)^{3/4}$. As a
result, the ratio of $n_{3/2}$ to the entropy density becomes
\begin{equation}
{n_{3/2}\over s} \sim 10^{-2} \left({m_\phi\over M_P}\right)^{3/2} \sim
10^{-10} \,. \label{final1}
\end{equation}
This violates the bound ${n_{3/2}\over s} < 10^{-15}$ for the gravitino
with $m_{3/2} \sim 10^2$~GeV by about 5 orders of magnitude. Thus one may
encounter the gravitino problem even if one neglects their thermal
production.

In this particular model one can overcome the gravitino problem if
reheating and thermalization occur sufficiently late. Indeed, during the
post-inflationary expansion the number  density of gravitinos decreases as
$a^{-3}$. The energy density of the oscillating massive scalar field $\rho
= m_\phi^2 \phi_0^2(t)/2$ also decreases as $a^{-3}$. However, the entropy
produced at the moment of reheating is proportional to $\rho^{3/4}$, so it
depends on the scale factor at the moment of reheating as $a^{-9/4}$. If
reheating occurs late enough (which is necessary anyway to avoid thermal
production of gravitinos), the ratio ${n_{3/2}\over s} \sim 10^{-10}
a^{-3/4}$ becomes smaller, and the gravitino problem can be relaxed.
 For a more detailed numerical investigation of this model see \cite{GTR}.

However, this simple resolution is not possible in some other models. As
an example, consider the model with the superpotential $W=
\sqrt{\lambda}\phi^3/3$. Again,  we will consider real fields $\phi$ and
switch to the minimally normalized field, $\phi \to \phi/\sqrt 2$. The
effective potential in this theory at  $\phi \ll M_P$ is
$\lambda\phi^4/4$. The oscillations of the scalar field near the minimum
of this potential are described by an elliptic cosine, $\phi(\eta)={\phi_0
\over a}\mathop{\rm cn}(\sqrt{\lambda}\phi_0, {1 \over \sqrt{2}})$. The
frequency of oscillations is $0.8472 \sqrt{\lambda}\phi_0$ and the initial
amplitude $\phi_0\simeq M_P$ \cite{KLS97}.

The parameter $\mu$  for this model is given by $\mu=\sqrt{2\lambda}
\phi$. It changes rapidly in the interval between $0$ and
$\sqrt{2\lambda} \phi_0$ within each oscillation of the inflaton field
$\phi$. Initially it is of the same order as $H$ and $m_{3/2}$, but then
$H$ and $m_{3/2}$ decrease rapidly compared with $\mu$, and therefore the
oscillations of $\mu$ remain the main source of gravitino production. In
this case production of gravitinos with helicity $\ft12$ is much more
efficient than that of helicity $\ft{3}{2}$.

The theory of the production of gravitinos with helicity $\ft12$ in this
model is similar to the theory of production of spin-$\ft{1}{2}$ fermions
with mass $ \sqrt{2\lambda}\phi$ by the coherently oscillating scalar
field in the theory $\lambda\phi^4/4$. This theory has been investigated
in \cite{GK}. The result can be formulated as follows. Even though the
expression for $\Omega$ contains a small factor $\sqrt{2\lambda}$, one
cannot use the perturbation expansion in ${\lambda}$. This is because the
frequency of the background field oscillations is also proportional to
$\sqrt{\lambda}$. Growth of fermionic modes (\ref{final}) occurs in the
non-perturbative regime of parametric excitation. The modes are fully
excited with occupation numbers $n_k \simeq \ft12$ within about ten
oscillations of the field $\phi$, and the width of the parametric
excitation of fermions in momentum space is about $\sqrt\lambda\phi_0$.
This leads to the following estimate for the energy density of created
gravitinos,
\begin{equation}\label{dens}
\rho_{3/2} \sim (\sqrt\lambda \phi_0)^4 \sim \lambda V(\phi_0)\,,
\end{equation}
and the number density of gravitinos
\begin{equation}\label{numberdens}
n_{3/2} \sim   \lambda^{3/4} V^{3/4}(\phi_0)  \,.
\end{equation}

Now let us suppose that at some later moment reheating occurs and the
energy density $V(\phi_0)$ becomes transferred to the energy density of a
hot gas of relativistic particles with temperature $T \sim V^{1/4}$. Then
the total entropy of such particles will be $s \sim T^3 \sim V^{3/4}$, so
that
\begin{equation}\label{numberdens2}
{n_{3/2}\over s} \sim   \lambda^{3/4}  \sim 10^{-10}  \,.
\end{equation}
This result violates the cosmological constraints on the abundance of
gravitinos with mass $\sim 10^2$~GeV by 5 orders of magnitude. In this
model the ratio ${n_{3/2}\over s}$ does not depend on the time of
thermalization, because both $n_{3/2}$ and $V^{3/4}(\phi_0)$ decrease as
$a^{-3}$. To avoid this problem one may, for example, change the shape of
$V(\phi)$ at small $\phi$, making it quadratic.

These estimates have been obtained in   \cite{GravProd}. A more detailed
numerical investigation of the  models with quadratic and quartic
effective potentials was performed in \cite{GTR}. However, even though
these toy models correctly illustrate some basic features of the new
mechanism of gravitino production, they can be somewhat misleading.

First of all, the strongest constraint  on the gravitino abundance
$n_{3/2}/s < 10^{-15}$ was derived for realistic models with more than
one chiral multiplet. More importantly, this constraint was derived under
the condition that at the end of the process the gravitino mass becomes
$m_{3/2} = \rme ^{{\cal K}/2}\, W M_P^{-2} \sim 10^2$~GeV. Meanwhile in
both models $W= 0$ in the minimum of the effective potential, so that
supersymmetry eventually becomes restored and $m_{3/2}$ vanishes.

In this case at the end of the process the super-Higgs mechanism does not
work, and instead of the massive gravitinos with spin $\ft32$ and helicity
$\ft12$ we  have chiral fermions with spin $\ft12$.

One could expect that this problem could be easily cured by adding small
terms such as $\zeta \phi + \beta$ to the superpotential. These terms
could shift a position of the minimum of the effective potential in such
a way as to ensure supersymmetry breaking.

However, the situation is more complicated. All of our attempts to achieve
this goal by a small modification of the superpotentials $\phi^2$ and
$\phi^3$ in the theories with one multiplet have been unsuccessful so
far, for a rather non-trivial reason. It was necessary to satisfy two
conditions: to have a vanishing vacuum energy density at the minimum of
the effective potential, and to have $m_{3/2} \sim 10^2$~GeV there. In
certain cases it was possible to achieve these two conditions for real
$\phi$, but the true minima of the effective potential in such cases
appeared at imaginary $\phi$. The effective potential in these minima was
large  and negative: $V \sim - M_P^2 m_{3/2}^2 \sim -10^{-32} M_P^4$,
which may seem small, but, in fact, it is 90 orders of magnitude greater
than the observable value of the vacuum energy, $\rho_{vac} \sim
10^{-122} M_P^4$.

One can see the problem especially clearly using the theory  $W =
\ft12m_\phi \phi^2$ with $m_\phi \sim 10^{13}$~GeV as an example. Suppose
one can add some small corrections to this superpotential that will shift
the position of the minimum and ensure that the energy density vanishes at
this minimum and   $m_{3/2}   \sim 10^2$~GeV. In the original minimum
$W_{\phi\phi} = m_\phi$. We suppose that the corrections are indeed
small, so that they do not appreciably change $W_{\phi\phi}$ at $\phi \ll
M_P$. Therefore, neglecting higher-order corrections in $M_P^{-1}$, which
is always possible if the additional terms are small, and keeping $\phi
\ll M_P$ in the minimum of the effective potential, one has $\mu \approx
m_\phi \gg m_{3/2}$.

Equation~(\ref{Ain_p_and_rho})  implies that if $m_{3/2}$ becomes constant
in the end of the process and  if  $|\rho|  \ll M_P^2m_{3/2}^2$ in the
minimum of the effective potential, then at the moment when the energy
density of the oscillations becomes smaller than $M_P^2m_{3/2}^2$, the
matrix $\hat A$ becomes constant, $\hat A = -1$ \cite{GravProd}. This
could be possible only if $\mu$ rapidly changes its sign during each
oscillation, which could imply the existence of an additional stage of
strong non-adiabaticity at the very end of reheating \cite{Lyth:2000yc}.
However, this is impossible because  $\mu \approx  m_\phi  = \rm
constant$ in our model. Thus the condition that   the absolute value of
energy density in the minimum of the effective potential is much smaller
than $M_P^2m_{3/2}^2$ cannot be satisfied in our model with any minor
modifications of the superpotential $W = \ft12m_\phi \phi^2$.

This does not mean, of course, that one cannot construct any consistent
cosmological models of one chiral field with a simple superpotential. The
simplest example is provided by the Pol\'{o}nyi model, with $W = \zeta( \phi
+ 2-\sqrt3)$, see, e.g., \cite{BL}. To find $m_{3/2} \sim 10^2$~GeV in
this model one should take $\zeta \sim 5\times 10^{-17} M_P^2$. The
minimum of the effective potential in this model occurs at $\phi =
M_P(\sqrt 3 - 1)$. The vacuum energy density in the minimum vanishes, $V
= 0$, but this happens only because the minimum occurs at $\phi \sim
M_P$, where one cannot neglect corrections suppressed by $M_P^{-1}$.

\begin{figure}
\centerline{\leavevmode\epsfxsize=0.7\columnwidth \epsfbox{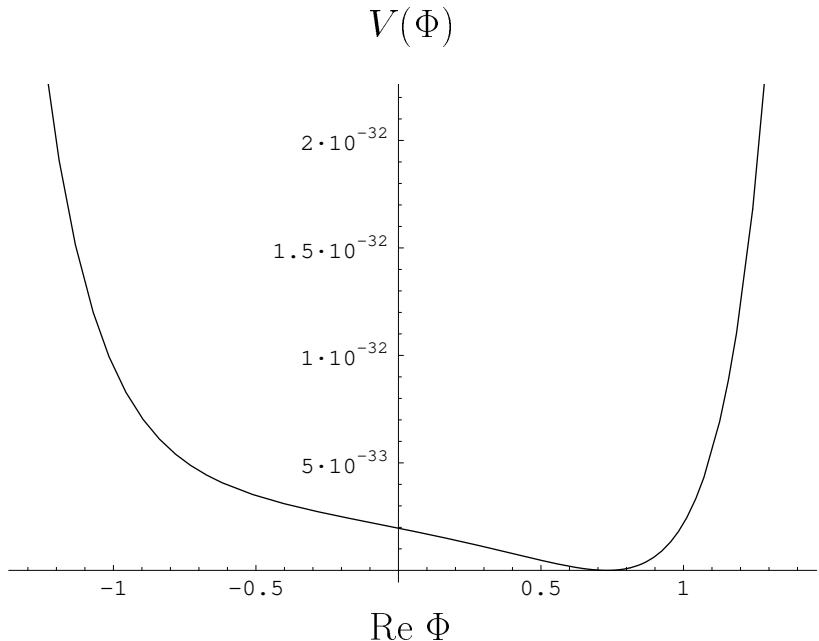}}

\

\caption{Effective potential in the Pol\'{o}nyi model, $W= \zeta (\phi +
2-\sqrt3)$, as a function of $\Re \phi$.} \label{fig1}
\end{figure}

The effective potential of the Pol\'{o}nyi model (for real $\phi$) is shown
in figure~\ref{fig1}. It is very flat at small positive $\phi$, but
becomes extremely curved at large $|\phi|$ due to the {\Ka} term $ \rme
^{\cal K} = \exp(\phi^2/M_P^2)$. As a result, the Pol\'{o}nyi potential
rapidly approaches $M_P^4$ at large $|\phi|$, and the corresponding
parameter $\mu$ approaches $M_P$. If the oscillations of the field $\phi$
begin from the point when the effective potential approaches $M_P^4$, the
parameter $\mu$ changes by ${\rm O}(M_P)$ during each oscillation, which
takes time ${\rm O}(M_P^{-1})$. This leads to extremely efficient
gravitino production, and these gravitinos remain massive, with $m_{3/2}
\sim 10^2$~GeV, at the end of the process.

One may wonder, whether one can obtain a complete scenario, including
inflation and gravitino production, in the context of a simple model of
one chiral field. Usually those who study inflation in supergravity
consider only the high-energy scale of inflation and neglect  details of
the effective potential required to give the gravitino small mass
$m_{3/2} \sim 10^2$~GeV. Even this programme is extremely complicated, and
having both inflation and supersymmetry breaking in a model of a single
chiral field is even more difficult.

However, thanks to the functional freedom in the choice of $W(\phi)$,
this problem is not unsolvable. One of the possible tricks is to keep the
superpotential equal to $\zeta( \phi + 2-\sqrt3)$ near  the minimum of
the effective potential at $\phi = M_P(\sqrt 3 - 1)$, and use the
functional freedom in the choice of $W$ to modify the effective potential
far away from its minimum.

\begin{figure}
\centerline{\leavevmode\epsfxsize=0.7\columnwidth \epsfbox{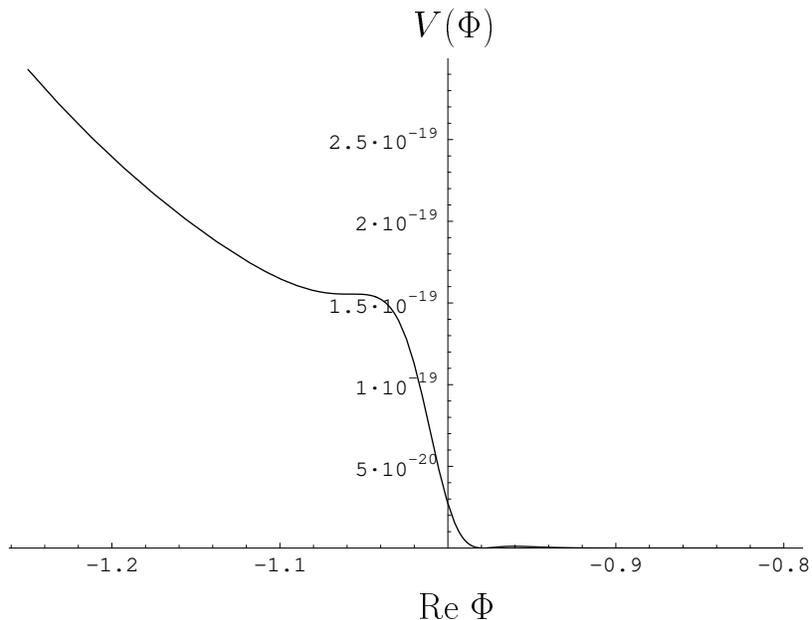}}

\

\caption{Effective potential in the theory $W= \zeta (\phi + 2-\sqrt3) +
C_1 (\phi+C_2)(1-\tanh(C_3(\phi+ C_2)))$ as a function of $\Re \phi$.}
\label{fig2}
\end{figure}

\begin{figure}
\leavevmode\epsfxsize=1\columnwidth \epsfbox{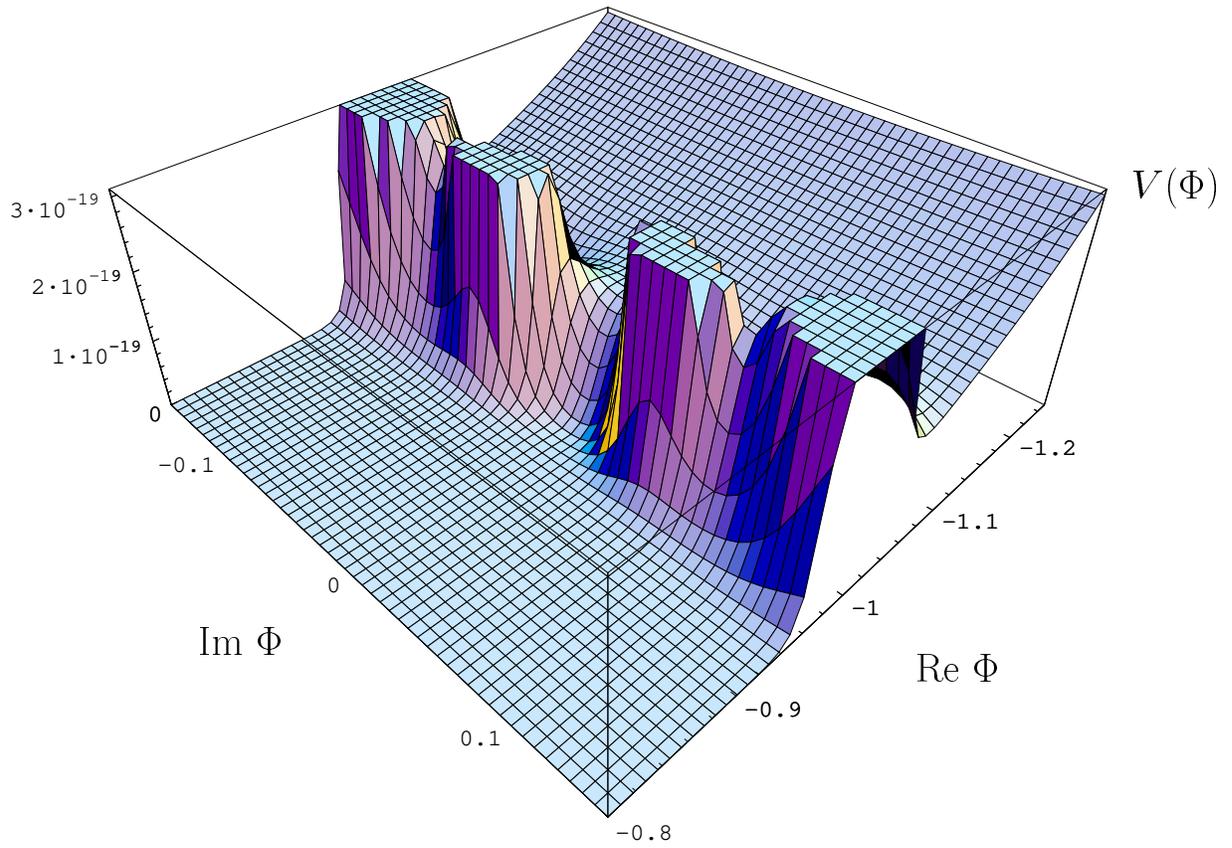}

\

\caption{Effective potential in the theory $W= \zeta (\phi + 2-\sqrt3) +
C_1 (\phi+C_2)(1-\tanh(C_3(\phi+ C_2)))$ as a function of $\Re \phi$ and
$\Im \phi$.} \label{fig3}
\end{figure}

To give an example of such a potential, let us consider $W= \zeta (\phi +
2-\sqrt3) + C_1 (\phi+C_2)(1-\tanh(C_3(\phi+ C_2)))$. The first part of
$W$ is the Pol\'{o}nyi superpotential. The second part is chosen in such a way
as to become exponentially small at the minimum of the Pol\'{o}nyi potential
at $\phi = M_P(\sqrt 3 - 1)$, and to rise sharply at $\phi < -C_2$.
Figure~\ref{fig2} shows the effective potential in this model for real
$\phi$, for $C_1 = 10^{-10}$, $C_2 = 1$ and $C_3 = 30$. As we see, the
potential has a plateau at $\phi \sim -1.05$. By a slight change of the
parameters one can change the height and the length of this plateau. This
suggests a possibility of an inflationary regime, which ends when the
field $\phi$ falls from the plateau, rapidly oscillates and produces the
gravitino. A complete investigation of this possibility is beyond the
scope of this paper. We would like to point out, however, that the
dynamical behaviour of the field $\phi$ in such models can be amazingly
rich. To understand it, it is sufficient to take a look at the effective
potential of this model  in the complex plane (see figure~\ref{fig3}). In
other models of a similar type, where instead of $\tanh(C_3(\phi+ C_2))$
one uses, for example, $\tanh(C_3(\phi+ C_2)^n)$, the effective potential
looks even more interesting and complicated.

 The main goal of presenting these models was to demonstrate that with
a proper choice of the superpotential one can obtain a broad variety of
potentials, even in the theories with a single chiral multiplet with a
minimal {\Ka} potential. In particular, one can have models with such
potentials where in the early universe one has efficient gravitino
production, whereas at the end of the process one has $m_{3/2} \sim
10^2$~GeV, and a nearly vanishing vacuum energy density $V \sim
10^{-29}$~g~cm$^{-3}$, consistent with the observational data.

Before one becomes too excited, one should remember that near the minimum
of the effective potential all these models look like the simple Pol\'{o}nyi
model with $W= \zeta (\phi + 2-\sqrt3) $.  This is good, because the
vacuum energy in the minimum of the effective potential in this model
vanishes, but $m_{3/2}$ does not, so we definitely have production of
gravitinos with helicity $\ft12$ in this model. However, it is also bad,
because the gravitino production in the Pol\'{o}nyi model only exacerbates
the moduli problem \cite{Polonyi1} which exists in this theory. The
energy density stored in the oscillating Pol\'{o}nyi field $\phi$ typically
leads to a much greater conflict with the observational data than the
decay of gravitinos produced during the oscillations of the field $\phi$.
Note, however, that the  physical origin of the gravitino problem differs
from that of  the moduli problem, so, in general, the problem of
gravitinos  produced after inflation may appear even in the models where
the moduli problem can be solved.

\subsection{Theories with two chiral multiplets}
To examine more realistic models of gravitino production, one should
consider theories with more than one chiral multiplet. Here the situation
becomes even more complicated. It was extremely difficult to derive
equations for the gravitino in such models. Solving them presents an even
greater challenge and requires more effort. In this paper we will only
give several comments on this subject.

First of all, one may try, whenever possible, to reduce the problem
involving two chiral multiplets to the theory of one multiplet. We have
shown in section~\ref{ss:2mulLimit} that this can be done in some cases
when one of the scalar fields has a much greater energy density than all
the other fields.

However, in other cases this method  may not work. For example, the
simplest supersymmetric version of the hybrid inflation scenario studied
in \cite{Bastero-Gil:2000je} has a very interesting property. After the
end of inflation, two scalar fields fall down to the minimum of the
effective potential simultaneously and oscillate synchronously, as a
single field. For this reason it was assumed in \cite{Bastero-Gil:2000je}
that one can use (\ref{1mult}) to study the gravitino production, just as
in the theory with one chiral multiplet. However, according to
\cite{Bastero-Gil:2000je}, $\Delta^2 = 1-|\hat A|^2$ can be ${\rm O}(1)$
in this model. However, in this case the second equation in the system of
equations~(\ref{summ2mult}) derived in our paper implies that $\Upsilon
\not = 0$. Therefore,  instead of solving (\ref{1mult}) one should solve
the system of two equations~(\ref{summ2mult}). This is a much more
complicated problem; it was not addressed in \cite{Bastero-Gil:2000je}.

Even more importantly, the existence of gravitino production in the very
early universe is not a guarantee of their survival until the late stages
of the evolution of the universe. This issue requires a special
investigation, and the answer may be model-dependent.

Indeed, suppose for a moment that the gravitino--goldstino correspondence
is valid at all stages of the process, and instead of solving equations
for the gravitino in supergravity one can solve equations for the chiral
fermions $\chi_i$ in the underlying globally supersymmetric theory. Then
one may wonder whether the particles produced after inflation are
gravitinos, or usual chiral fermions, superpartners of the inflaton,
which can be called an inflatino \cite{Nilles99}.

As we have already mentioned, supersymmetry is always broken during the
expansion of the universe supported by the energy density of scalar
fields. Suppose, however, that the relatively small supersymmetry
breaking remaining when the universe slows down involves chiral fermions
belonging to the hidden sector, such as the Pol\'{o}nyi field. Suppose also
that the inflaton field $\phi_1$ does not interact with the Pol\'{o}nyi field
$\phi_2$ in the limit $M_P \to \infty$. As an example of such a theory one
may consider the theory with the superpotential $W= m_\phi\phi_1^2/2 +
\zeta (\phi_2 + 2-\sqrt3)$ or $W= \sqrt{\lambda}\phi_1^3/3 + \zeta
(\phi_2 + 2-\sqrt3)$.

Soon after inflation, the energy density is dominated by the energy of the
inflaton field,  i.e.\  $\rho \approx \rho_1$ is many orders of magnitude
greater than $\rho_2$. In the expression for the effective potential of
the Pol\'{o}nyi field there is a term $\exp(\phi_2^2/M_P^{2}) |W_1|^2$. This
term at $\phi_2 \ll M_p$ becomes $(1 + {\phi_2^2/M_P^{2}})\rho_1 \approx
(\rho_1 + 3\phi_2^2 H^2 )$, which implies that the effective potential for
the Pol\'{o}nyi field in the early universe has a minimum at $\phi_2 \approx
0$, and the mass squared of this field is $m_2^2 \approx 3 H^2$. (Note
that in order to go to the canonically normalized field one should divide
$\phi_2$ by $\sqrt 2$.) Therefore, the field $\phi_2$ rapidly oscillates
about $\phi_2 =0$. The amplitude of oscillations decreases as $a^{-1}$,
which eventually puts this field at the vicinity of the point $\phi_2 =
0$, where it stays until $H$ drops below ${\rm O}(m_{3/2})$. At that
stage the Pol\'{o}nyi field starts rolling towards the flat-space minimum of
its potential at $\phi_2 = M_P(\sqrt 3 - 1)$, and oscillates about it. The
energy density of this field decreases very slowly, which constitutes the
essence of the moduli problem as formulated in \cite{Polonyi2}. We will
not consider this problem here, but one should remember that in a
realistic cosmological model based on supergravity this problem should be
taken into account and resolved.

The solution for the goldstino in this model in the limit $M_P \to
\infty$ is given by (\ref{form}), so that at the early stages, when
$\rho_1$ dominates, one has $\upsilon = \rho_1^{1/2} \rme ^{ \gamma_0\int
\rmd t \, \mu_1}(\chi_1+ \chi^1)$. Therefore, one can use the results of
our previous investigation of the gravitino production
\cite{GravProd,Giudice:1999yt,GTR} to estimate the number of gravitinos
produced at that stage.

However, eventually the energy density of the oscillating field $\phi_1$
drops down, and the energy becomes dominated by the energy density of the
oscillating Pol\'{o}nyi field. At that stage one has $\upsilon = \rho_2^{1/2}
\rme ^{ \gamma_0\int \rmd t \, \mu_2}(\chi_2+ \chi^2)$ (except for at
that stage one should use a more accurate expression because the
effective potential of the Pol\'{o}nyi field cannot be correctly described in
the limit $M_P \to \infty$).

Thus one may argue that  the efficient production of the chiral fermions
$\chi_1$, which can be interpreted as the process of gravitino production
at the first stage of the process, does not imply anything about the
density of gravitinos at the second stage: the gravitino must first spit
the chiral fermion $\chi_1$, and then eat the chiral fermion $\chi_2$, if
there are any of them around. However, in this model one does not expect
efficient production of the fermions $\chi_2$ at the stage of the
oscillations of the Pol\'{o}nyi field, so at the end of the process one will
have many fermions $\chi_1$ (inflatino) and practically no gravitinos with
helicity $\ft12$.

This argument is very interesting but insufficient to give a definite
answer to the question concerning the number of gravitinos at the last
stages of the process. It is based on the assumption that the fermions
$\chi_i$ do not mix with each other, so that the fermions $\chi_1$
produced at the first stage of the process  cannot be converted into
$\chi_2$. This assumption is justified in the limit $M_P \to \infty$, but
in this limit there is no expansion of the universe, the ratio
$\rho_2/\rho_1$  is constant and the definition of the goldstino does not
change in time. The transition from the regime $\rho_1 \gg \rho_2$ to the
regime $\rho_2 \gg \rho_1$ in the expanding universe takes time ${\rm
O}(H^{-1}) \sim M_p /\sqrt\rho$. This time interval becomes infinitely
long in the limit $M_p \to \infty$.

Thus one may wonder whether some of the particles $\chi_1$ can convert to
$\chi_2$ at the intermediate stage of the process, when $\rho_1 \sim
\rho_2$ and the definition of the goldstino gradually changes from
$\rho_1^{1/2} \rme ^{ \gamma_0\int \rmd t \, \mu_1}(\chi_1+ \chi^1)$ to
$\rho_2^{1/2} \rme ^{ \gamma_0\int \rmd t \, \mu_2}(\chi_2+ \chi^2)$. In
order to study this question one should go beyond the approximation $M_P
\to \infty$, i.e.\ one cannot apply the gravitino--goldstino equivalence
theorem used in the argument given above.

One may try to approach this issue in a more direct and unambiguous way,
using the system of  equations~(\ref{summ2mult}) for $\theta$ and
$\Upsilon$ in the unitary gauge. This is a much more powerful method. As
we have emphasized, this method applies at {\it all} stages of the
process, because the only condition required for its validity, $H^2 +
m^2_{3/2}> 0$, is always satisfied in the FRW universe. It allows one to
avoid using imprecise language based on the gravitino--goldstino
equivalence and the limit $M_P \to \infty$. Also, it allows us to
formulate adequate initial conditions for the gravitino equations at the
end of inflation (see section~\ref{se:deSitter}).

In this formalism one could expect that the total number of massive
gravitinos with helicity $\ft12$ can be represented by some adiabatic
invariant, so the gravitinos produced at the first stage of the process
cannot simply disappear because of the change of the regime if this
change is slow enough (adiabatic). Moreover, one could even expect that
if this change is {\it not} adiabatic, it may lead to an additional
gravitino production similar to that discussed in \cite{Lyth:2000yc}.
Consider, for example, the model with $W= m_\phi\phi_1^2/2 + \zeta
(\phi_2 + 2-\sqrt3)$ with $m_\phi \sim 10^{13}$~GeV, as in the previous
section. In this model, at the first stage of the process one has $\mu_1
\sim m_\phi \sim 10^{13}$~GeV. Then, within the time $H^{-1} \sim
{m^{-1}_{3/2}}$ this parameter drops down to ${\rm O}(m_{3/2}) \sim
10^2$~GeV. The last stages of this process are non-adiabatic. An estimate
similar to that made in the previous section indicates the possibility
of  production of gravitinos with $n_{3/2} \sim 10^{-2}(m_\phi
m_{3/2})^{3/2}$. If one ignores for a moment the moduli problem and
assumes that the energy of the scalar field $\phi_2$ immediately
transfers to thermal energy as soon as it begins oscillating, one finds
$s \sim (M_P m_{3/2})^{3/2}$ and ${n_{3/2}\over s} \sim
10^{-2}\left({m_\phi\over M_P}\right)^{3/2}$, as in (\ref{final1}).

However, the situation is, in fact, more complicated, and the arguments
based on adiabaticity must be examined in a detailed way. They should be
applicable at the time before and after the change of the regime from
$\rho_1 \gg \rho_2$ to $\rho_1 \ll \rho_2$, but they may not apply at the
transitional stage when the energy densities of the scalar fields $\phi_1$
and $\phi_2$ become comparable. At that time the fields $\theta$ and
$\Upsilon$ start mixing because of the term $\vec k^2\Delta^2\theta$ in
(\ref{calM2a2a}), so at that intermediate epoch the number of gravitinos,
generally speaking, is not conserved. This resembles the neutrino
oscillations, when the neutrino registered in the detector may be
different from the neutrino emitted by the Sun. Without performing a
quantization of the system of coupled fields $\theta$ and $\Upsilon$  and
solving the corresponding equations in the time-dependent background, one
cannot tell whether the number of gravitinos becomes depleted at the
transitional stage, or, vice~versa, the existence of this stage creates a
new type of non-adiabaticity which may enhance gravitino production.

We hope that the results of our investigation create a proper framework
which may help to address these questions.

\subsection{Other mechanisms of gravitino production}
There exist some other mechanisms of gravitino production, which we did
not discuss in this paper, and which can be particularly interesting in
the context of hybrid inflation. We will briefly describe them here.

These mechanisms are related to the possibility that the scalar fields
are inhomogeneous, or that other oscillating fields, such as vector
fields, appear in the universe. We did not consider this possibility in
the main part of our paper because all fields become nearly homogeneous
during inflation, and vector fields become extremely small. However, in
some models inhomogeneous scalar fields and vector fields may rapidly
appear soon after the end of inflation.

The first mechanism is related to the formation of cosmic strings, which
is a typical property of the hybrid inflation scenario
\cite{hybrid,LythRiotto}. These cosmic strings, associated with the
topologically stable configuration of the complex scalar fields $z_i$,
interact with the gravitino field $\psi_{\mu}$. Dynamics of the strings
lead to emission of gravitinos. Indeed, the generic gravitino master
equation~(\ref{master}) is constructed with
 the long covariant derivative
${\cal D}_\mu\psi_\nu = \left(\partial_\mu + \ft14 \omega_\mu^{ab}(e)
\gamma _{ab} +\ft12\rmi A_\mu ^B\gamma _5 \right)\psi _\nu -
\Gamma_{\mu\nu}^\lambda\psi_\lambda$, which contains,  in particular, a
bosonic part of the $U(1)$ connection,
 $A_\mu ^B$. In the  gauge that is not singular at $W = 0$,
the form of  $A_\mu ^B$ is given by equation~(\ref{AB2gauge}), $  A_\mu
^B=\ft12\rmi\left[(\partial_i{\cal K})  {\partial }_\mu z^i
  -(\partial ^i{\cal K}){\partial}_\mu z_i\right] +
M_P^{-2}W_\mu ^\alpha {\cal P}_\alpha$. Consider for simplicity the
minimal  K{\"a}hler potential ${\cal K}=z_iz^i$. Then in the long covariant
derivative we will have a term $J_\mu=z_i {\partial }_\mu z^i-
z^i{\partial}_\mu z_i$, which does not vanish for the complex field
configuration around the string. Spacetime variations of the term
$J_\mu(t, {\vec x})$ due to the motion and excitations of the string will
lead to the radiation of gravitinos by the string, similar to the
gravitational radiation from the cosmic strings.

However, the  term  $J_\mu(t, {\vec x})$ may appear in the gravitino
equations even in the cases where strings are not generated. For example,
very strong spinodal (tachyonic) instability of scalar field fluctuations
at the `waterfall' stage in the hybrid inflation results in a very rapid
decay of the homogeneous inflaton field into inhomogeneous classical
scalar fields in the model. For the inhomogeneous complex scalar fields
we also have a non-vanishing term $J_\mu(t, {\vec x})$, which leads to
additional gravitino production.

In both cases the magnitude  of $J_\mu(t, {\vec x})$ is proportional to
the small quantity $\vert \phi \vert^2  M_P^{-4}$, where $\vert \phi
\vert$ is the typical amplitude of the scalar field. Therefore, the
gravitino production in this case can be studied using perturbation
theory. The net gravitino abundance will depend on the duration of the
gravitino emission. One may think about gravitino production in this case
as resulting from collisions of classical waves of various scalar fields.
One may also consider gravitino production due to the collisions of
non-thermalized particles produced at the first stages of preheating.
This and other mechanisms of gravitino production after inflation deserve
a separate investigation.

\section{Discussion}

The initial goal of our investigation was rather limited. We wanted to
study the gravitino production by an oscillating scalar field at the end
of inflation. However, soon we discovered that the existing tools that we
could use in our work were not quite adequate.

We have found that it is rather difficult to study the conformal
properties of gravitinos in the standard formulation of supergravity.
This forced us to reformulate phenomenological supergravity in a way
revealing its hidden superconformal symmetry $SU(2,2|1)$. Until now, the
superconformal version of phenomenological supergravity was  only used as
a tool for the derivation of phenomenological supergravity, We have found
that this formulation has several important advantages over the standard
one.

The superconformal formulation of supergravity may simplify the
investigation of the processes in the early universe. Indeed, the FRW
universe is conformally flat. Therefore, by making appropriate
redefinitions of the fields and the metric in a conformally invariant
theory, one can reduce the investigation of all processes in an expanding
FRW universe to an investigation of processes in Minkowski space. We did
not use this method in our work, but it may be extremely interesting and
rewarding to study the standard issues of big-bang cosmology from this
new perspective.

For us it was important that the $SU(2,2|1)$ formulation of supergravity
provides a flexible framework unifying various formulations of $N=1$
supergravity interacting with matter, depending on the choice of the
$R$-symmetry fixing. It explains the superconformal origin of the
Fayet--Iliopoulos terms. It also allows us to study the weak-coupling
limit of supergravity, $M_P\rightarrow \infty$, and to formulate the
gravitino--goldstino equivalence theorem which is valid in this limit.
Indeed, it is not very simple to make sense out of this limit in the
standard versions of $N = 1$ supergravity unless one makes a certain
field rescaling, which brings us back to the original field variables of
the underlying superconformal theory. This made it possible to prove the
equivalence theorem, which explains why gravitino production in the early
universe in some models is not suppressed in the limit of weak
gravitational coupling, $M_P \to \infty$.

The superconformal formulation helped us to study  the super-Higgs effect
in cosmology and  to derive the equations for the gravitino interacting
with any number of chiral and vector multiplets in the gravitational
background with varying scalar fields.

There are several other aspects related to the superconformal formulation
which require further study. One of them is the possible existence of
strings associated with the lines where the conformon field vanishes.
Such objects could not appear in the usual $N = 1$ supergravity where
$M_P$ is already fixed. Note that the conformon field does not have any
dynamical degrees of freedom associated with it; it can be set equal to
$M_P$ everywhere where it does not vanish. However, this does not include
the string, where this field vanishes and the vector field $A_\mu$
exists. It would be very interesting to study the question of whether the
strings which may appear in the $SU(2,2|1)$ formulation of supergravity
may have any interesting dynamics associated with them. In such a case
one would have stringy excitations as part of supergravity.

Returning to the problem of gravitino production, we should note that the
derivation of the gravitino equations was only part of the problem; it is
very difficult to find their solutions in the cosmological background. We
analysed them in some particular cases and studied their properties,
which can be helpful for a future investigation of gravitino production
in realistic models based on supergravity. The main conclusion is that
due to the gravitino--goldstino equivalence theorem, the production of
gravitinos in the very early universe can be as efficient as the
production of ordinary spin-$\ft{1}{2}$ fermions, i.e.\ it is suppressed
by small coupling constants such as $\lambda$ or $g^2$, but not by the
gravitational coupling.

Thus, despite naive expectations, gravitinos created by the oscillating
scalar fields soon after inflation  may provide a contribution to the
matter content of the universe as large as that from the usual particles.
This result can be quite important for our understanding of the physical
processes in the very early universe.

On the other hand, the definition of the gravitino (and goldstino) in the
early universe changes in time. Therefore, an additional investigation is
required to check whether the number of gravitinos produced soon after
the end of inflation is conserved until the late stages of the evolution
of the universe, or they become converted to other, less harmful,
particles. The answer to this question may be model dependent. Many
models based on supergravity are plagued by the moduli problem, which is
usually even more severe than the problem of the gravitino created after
inflation. During the last few years there have been many attempts to
solve the moduli problem. One of the possible solutions involves an
additional short stage of the late-time `thermal' inflation
\cite{LythStewart}. If this mechanism solves the moduli problem, it may
solve the gravitino problem as well.

However, in some models the gravitino problem may be as severe or even
more severe than the moduli problem. For example, in the supersymmetric
versions of the hybrid inflation scenario all coupling constants are
${\rm O}(10^{-1})$. A full investigation of gravitino production in such
models have not been performed as yet, but preliminary estimates indicate
that the ratio $n_{3/2}/n_s$ at the first stages of preheating can be as
large as ${\rm O}(1)$ \cite{GravProd,Giudice:1999yt,GTR}. This may lead to
drastic cosmological consequences, unless the gravitinos created soon
after inflation become converted into less harmful particles at the stage
when the energy density of the inflaton field becomes sub-dominant and the
definition of the gravitino changes. Therefore, it is very important to
study whether the gravitino conversion is actually possible. We hope that
this problem, as well as other issues related to the physical properties
of the gravitinos in the early universe, can be addressed using the
formalism developed in our paper.

\subsection*{Acknowledgments}
It is a pleasure to thank P. Bin{\'e}truy, A. Chamseddine,  G.F. Giudice, P.
Greene, H.P. Nilles, A. Riotto,  I. Tkachev, W. Troost and S. Vandoren
for useful discussions.  The work of R.K and A.L. was
supported by NSF grant PHY-9870115, L.K. was supported by NSERC and CIAR.
A.V.P. thanks the Department of Physics at Stanford and the CERN theory
division for the hospitality. L.K. and A.L. thank NATO Linkage Grant
975389 for support.
\newpage

\appendix

\section{Notation}\label{app:notations}

We mostly use the notation of \cite{PvNPR}, which agrees with
\cite{general}. However, rather than using the index range $\mu
=1,2,3,4$, for curved indices we use $\mu =1,2,3,0$, thus converting the
metric $(++++)$ in \cite{PvNPR} into the familiar $(+++-)$. In the same
token, the Levi--Civita tensor $\varepsilon^{abcd}$ (with flat indices
$a=1,2,3,4$) is taken to be imaginary. Indeed, we have
$\varepsilon^{0123}=\rmi$. The normalization for $\gamma
_5=\varepsilon^{0123}\gamma _0\gamma _1\gamma _2\gamma _3$
 is chosen such that $\ft12\varepsilon^{abcd}\gamma  _{cd}=-\gamma _5\gamma
^{ab}$. The comparison with \cite{PvNPR} is $\gamma _4=-\rmi\gamma _0$.
Then $\gamma _0$ is anti-Hermitian, while $\vec\gamma $ (where this
denotes the spacelike components) and $\gamma _5$ are Hermitian.
Furthermore, for antisymmetrization, we use $[ab]=\frac{1}{2}(ab-ba)$. An
explicit realization of $\gamma $-matrices is
\begin{equation}
\gamma_0=\pmatrix{\rmi\unity _2 &0\cr 0 & -\rmi\unity _2\cr}\ ;\qquad
\vec\gamma=\pmatrix{0&-\rmi\vec\sigma\cr \rmi\vec\sigma & 0\cr}\ ; \qquad
\gamma_5=\pmatrix{0&-\unity _2\cr -\unity _2 & 0\cr}\,.
\label{explgamma}
\end{equation}
We use also left and right projections
\begin{equation}
 P_L=\ft12(1+\gamma _5)\,,\qquad P_R=\ft12(1-\gamma _5)\,.
\label{LRProj}
\end{equation}
The Majorana condition, defined by $-\rmi\lambda ^\dagger \gamma _0=
\lambda ^T{\cal C}$, with ${\cal C}$ the charge conjugation $\gamma
_0\gamma _2$ in this representation, then amounts to
\begin{equation}
  \lambda ^*=\pmatrix{0&-\sigma _2\cr \sigma _2&0}\lambda \,.
\label{Majoranaexpl}
\end{equation}
The barred spinors are Majorana conjugates. For the chiral spinors, e.g.\
$\lambda _L\equiv P_L\lambda $, not being Majorana spinors, this implies
\begin{equation}
  \bar \lambda _L\equiv (\lambda _L)^T{\cal C}=
\bar \lambda P_L=-\rmi(\lambda _R)^\dagger\gamma _0\,.
\label{barlambdac}
\end{equation}
For more detailed explanations, see, e.g., \cite{Tools}.
\par
The Einstein tensor and energy--momentum tensor are
\begin{eqnarray}
R_{\mu \nu }{}^{ab}&=&2\partial _{[\mu }\omega _{\nu ]}{}^{ab}
+2\omega _{[\mu }{}^{ac}\omega _{\nu ]c}{}^{b}\nonumber\\
R_{\mu \nu }&=& R_{\mu \rho }{}^{ab}e^\rho _ae_\nu ^b\,,\qquad
R=g^{\mu \nu }R_{\mu \nu }\nonumber\\
  G_{\mu \nu }&=&e^{-1}\frac{\delta }{\delta g^{\mu \nu }}\int \rmd ^4x\, eR
  =R_{\mu \nu }-\ft12 g_{\mu \nu }R\nonumber\\
  T_{\mu \nu }&=& -e^{-1}e_\nu ^a\frac{\delta }{\delta e^\mu _a }M_P^{-2}
  \int \rmd ^4 x\,{\cal L}^{(m)}\,,
\label{defGmunu}
\end{eqnarray}
with ${\cal L}^{(m)}$ the matter action, i.e.\ all but the graviton
term. Due to  invariance under Lorentz rotations, $T_{\mu \nu }$ is
symmetric by use of field equations.

Complex conjugation by definition reverses the order of the spinors. The
action of complex conjugation on spinors is often complicated. A simpler
equivalent procedure is taking the charge conjugate. For pure bosonic
numbers this is just complex conjugation. For matrices in spinor space,
we have $\gamma _\mu ^C=\gamma _\mu $ and $\gamma _5^C=-\gamma _5$. For
this operation we do not have to interchange the spinors. Majorana spinors
are invariant under this operation, but the chiral spinors change
chirality. The same holds for the barred (Majorana conjugate) spinors. In
our notation for chiral multiplets, the operation thus interchanges upper
with lower $i$ indices, but does not change gauge indices $\alpha $. As a
practical summary of the rules, see the following equations:
\begin{eqnarray}
 &   & X_I^C=X^I\,,\qquad  Y^C=Y^*\,,\qquad
 z_i^C= z^i\,, \qquad \phi _i^C=\phi ^i\,,  \nonumber\\
 &&\gamma _\mu ^C=\gamma _\mu\,,\qquad \gamma _5^C=-\gamma _5\,,
\qquad  P_L^C=P_R \nonumber\\
  &   & \Omega _I^C= \Omega ^I\,, \qquad \chi _i^C=\chi ^i \,,
\qquad \bar \chi _i^C=\bar \chi ^i \,,  \nonumber\\
&&  \lambda^{\alpha \, C}=\lambda ^\alpha\,, \qquad \lambda ^\alpha
_L{}^C=\lambda ^\alpha _R\,, \qquad {\cal P}_\alpha {}^C= {\cal
P}_\alpha\,.
\label{sumC}
\end{eqnarray}
 An example of these rules can be seen by obtaining the second line
from the first line in (\ref{susyrepeat}), where at
the end one also checks the charge conjugation invariance of $\delta \lambda
^\alpha $ due to the charge conjugation invariance of $\rmi\gamma _5$.
For more intrinsic definitions of the charge conjugation
operation, see \cite{Tools}.
\section{Conformal metric}\label{app:conf_metric}
We take a conformal metric, i.e.\
\begin{equation} e^a_\mu = a(\eta)
\delta_\mu^a \,, \qquad g_{\mu\nu} =a^2(\eta) \eta_{\mu\nu}\,,
\end{equation}
where $\eta =x^0$ is a time coordinate. However, to start with, we will
still allow $a$ to be a general scalar function of spacetime. The
connections are then (from now on, indices are raised and lowered
with flat metric $\eta _{\mu \nu }$)
\begin{equation}
\omega ^{ab}_\mu = 2  \delta_\mu^{[a} \partial ^{b]} \ln a \,,\qquad
\Gamma ^{\rho}_{\mu \nu} = 2  \delta_{(\nu}^{\rho}
\partial_ {\mu )}
\ln a - \eta_{\mu\nu}\partial^\rho  \ln a\,.
\end{equation}
The curvature is then
\begin{eqnarray}
R^{\mu }{}_{\nu  \rho \sigma } & = &\eta _{\nu \nu '}\left[
 4\delta^{[\mu }_{ [\sigma }\partial _{\rho ]}\partial^{\nu']}\ln a
+4\delta _{[\rho }^{[\mu }(\partial ^{\nu ']}\ln a)( \partial
_{\sigma ]}\ln a) -2\delta _{[\rho }^{[\mu }\delta _{\sigma ]}^{\nu'
]}(\partial \ln a)^2\right]
 \nonumber\\
R_{\mu\nu}&=& 2 \partial_\mu   \partial _\nu   \ln a -2 (\partial_\mu
\ln a)(\partial _\nu   \ln a)
  + \eta_{\mu\nu} [ {\,\lower0.9pt\vbox{\hrule \hbox{\vrule height 0.2 cm
\hskip 0.2 cm \vrule height 0.2 cm}\hrule}\,}  \ln a+2(\partial  \ln
a)^2]
  \nonumber\\
a^2 R&=& 6 {\,\lower0.9pt\vbox{\hrule \hbox{\vrule height 0.2 cm \hskip
0.2 cm \vrule height 0.2 cm}\hrule}\,}  \ln a  +6 (\partial \ln a)^2
\nonumber\\ G_{\mu\nu}&=& 2 \partial_\mu   \partial _\nu \ln a -2
(\partial_\mu \ln a)( \partial _\nu   \ln a) - \eta_{\mu\nu} [ 2
{\,\lower0.9pt\vbox{\hrule \hbox{\vrule height 0.2 cm \hskip 0.2 cm
\vrule height 0.2 cm}\hrule}\,} \ln a+(\partial \ln a)^2]\,.
\label{curvatures}
\end{eqnarray}
The Lorentz connection term in covariant derivatives on spinors is
\begin{equation}
+\ft14 \omega ^{ab}_\mu \gamma_{ab}= +\ft12 \delta _\mu ^a
\gamma_a{}^b \delta _b^\nu  \partial_\nu  \ln a\,.
\end{equation}
Using these expressions we obtain useful equations for the covariant
derivatives of the gravitino (in this appendix the K{\"a}hler connections are
omitted), for example,
\begin{eqnarray}
a  \not\!\! {\cal D} \psi _\mu  \equiv  \overline{ \not\!
\partial} \psi_\mu
+\ft{1}{2} \left(\overline{ \not\!\partial}\ln a\right)
\psi_\mu +\overline{\gamma }_\mu(\partial _\sigma  \ln a ) \eta
^{\nu \sigma }\psi_\nu-\overline{\gamma}\cdot \psi \partial
_\mu \ln a \,.
 \end{eqnarray}
\par
Now we specify the case that $a$ depends only on the time coordinate,
which is called $\eta $. It is, however, easy to express derivatives with
respect to a time coordinate defined by $\rmd t= a(\eta)\,\rmd \eta$. We
thus introduce
\begin{equation}
  \dot a\equiv \frac{\partial _0 a}{a}\,,
\label{dota}
\end{equation}
and also the Hubble parameter
\begin{equation}
  H=\frac{\dot a}{a}=\frac{\partial _0 a}{a^2}\,.
\label{defHubble}
\end{equation}

Then we have the following expressions for covariant derivatives where we
use the notation introduced at the beginning of
section~\ref{ss:confInspBG},
\begin{eqnarray}
{\cal D}_\mu \chi &=& \partial _\mu \chi -\ft12 \overline{\gamma }_{\mu
0}\dot a\chi\nonumber\\
 a \not\!\!{\cal D}\chi &=& \overline{\not\!  \partial} \chi +\ft32 \gamma ^0 \dot
  a\chi \nonumber\\
 {\cal D}_{\mu} \psi^{\mu}&=&    \partial^\mu \psi_\mu -2a^{-2}\dot a\psi_0
 +\ft12a^{-2}\dot a\gamma_0 \vec\gamma\cdot \vec \psi\nonumber\\
  a\not\!\! {\cal D} \psi _\mu  &=&    \overline{\not\!  \partial} \tilde
\psi_\mu
  -\dot a \left(\overline{\gamma }_\mu
  \psi_0+\ft{1}{2} \gamma _0 \psi_\mu +\overline{\gamma }\cdot \psi
  \delta _\mu ^0  \right)  \,.
 \label{Dmupsimu}
\end{eqnarray}

Furthermore, we now have
\begin{eqnarray}
-M^{-2}_P\rho \equiv G^0{}_0 & = &  -3 a^{-2}(\partial _0 \ln
a)^2=-3H^2 \nonumber\\ -M^{-2}_P p\unity _3\equiv G & =a^{-2}
&\unity _3\left[ 2\partial _0 ^2\ln a + (\partial _0\ln
a)^2\right]=\unity _3(3H^2 +2\dot H) \,, \label{Gconf}
\end{eqnarray}
where $G$ denotes the $3\times 3$ part of $G^\mu {}_\nu $ in the spacelike
directions.

The condition for a de~Sitter metric is $p=-\rho $ which implies that $H$
is constant. We then have
\begin{eqnarray}
a^{-1} & = & H (C-\eta ) \nonumber\\ R_{\mu \nu }{}^{\rho \sigma } &
= & 2\delta _\mu ^{[\rho }\delta _\nu ^{\sigma ]}H^2 \nonumber\\ R
&=&-12 H^2=-4 VM_P^{-2}\nonumber\\ G_{\mu \nu }&=& -\ft14 g_{\mu \nu
}R=g_{\mu \nu }VM_P^{-2}\,, \label{curvdS}
\end{eqnarray}
where $C$ is also a constant, and we also gave the expression in terms of
the constant value of the potential $V$.

\section{\Ka\ geometry from the conformal formulation}
\label{app:elimA} In this section we make the transition from the scalar
action in terms of $\Nphi_I{}^J$ to a \Ka ian action\footnote{An earlier
treatment in terms of special coordinates, that already contains many of
the steps which we perform in this section can be found in
\cite{BernardTrieste84}.}. The proof that we present here can also be
applied to the $N=2$ theory as to $N=1$.\vspace{2mm}

\noindent \textbf{Theorem}.

We consider the action
\begin{equation}
   {\cal L}=-\Nphi_I{}^J{\cal D}_\mu X^I {\cal D}^\mu X_J
\label{LscInPhi}
\end{equation}
in terms of $n+1$ complex scalars $X^I$ (where $X_I$ are their
complex conjugates). In (\ref{LscInPhi}) the covariant derivatives
have a $U(1)$ connection, and we write them as
\begin{equation}
   {\cal D}_\mu X_I=\partial _\mu X_I + \ft{1}{3}\rmi A_\mu X_I
\,,\qquad
   {\cal D}_\mu X^I=\partial _\mu X^I - \ft{1}{3}\rmi A_\mu X^I\,.
\label{cDapp}
\end{equation}
The functions $\Nphi_I{}^J$ depend on $X^I$ and $X_I$, and satisfy
\begin{equation}
X^I \frac{\partial }{\partial X^K} \Nphi_I {}^J= X^K
 \frac{\partial }{\partial X^K} \Nphi_I {}^J= 0\,,
\label{relNphi}
\end{equation}
and their complex conjugates. Note that here we took $\Nphi_I {}^J$
as the basic object, rather than as the second derivative of a scalar
$\Nphi$. If it is the second derivative, as in the main text, then
the two equations are equivalent.

One can choose $n$ complex coordinates $z_i$ (and complex conjugates
$z^i$) by defining
\begin{equation}
  X_I =\Yrho x_I(z)\,,\qquad  X^I=\Yrho ^* x^I(z^*)\,,
\label{Xrhoapp}
\end{equation}
where $x_I(z)$ are $n+1$ non-degenerate\footnote{The matrix $\partial
^i x_I$ has to be of rank $n$ and the matrix $(x_I, \partial ^ix_I)$
has to be of rank $n+1$.} arbitrary holomorphic functions of
the $z_i$, and $\Yrho $ is the $(n+1)$th complex variable. The action
can then be written as
\begin{eqnarray}
  {\cal L}&=& -\frac14 \Nphi^{-1}\left( \partial _\mu \Nphi\right) ^2
 -\frac19\Nphi \left( A_\mu+ \frac{\rmi}{2}\left( \partial ^i{\cal K}\,\partial _\mu z_i -
  \partial _i{\cal K}\,\partial _\mu z^i\right) -\frac{3\rmi}{2}\partial _\mu
  \ln \frac{\Yrho }{\Yrho ^*}\right) ^2\nonumber\\ &&
  +\frac13\Nphi\left(\partial ^j\partial _i {\cal K}\right)
  \partial  _\mu z^i\, \partial ^\mu z_j\,,
\label{rewrittenL}
\end{eqnarray}
where
\begin{eqnarray}
\Nphi& \equiv &  X^I \Nphi_I{}^J X_J \nonumber\\
 {\cal K}(z,z^*)&\equiv &-3 \ln \left[-\ft13x^I(z^*) \Nphi_I{}^J(z,z^*) x_J(z)\right] \,.
\label{defNphiK}
\end{eqnarray}
This scalar function $\Nphi$ coincides with the one that we started with
in the main text.

\noindent \textbf{Interpretation}.

The metric defines a cone (using $r^2=-\Nphi$ one obtains the canonical
parametrization $\rmd s^2= \rmd r^2+ r^2\ldots $). When $U(1)$ is not
gauged ($A_\mu =0$), the base of the cone
 (the manifold with fixed $\Nphi$) is a Sasakian
manifold with a $U(1)$ invariance\footnote{This has been remarked first
in a similar situation with hypermultiplets in $N=2$ in \cite{dWKV}, and
has been looked at systematically in \cite{GibbonsR}. }. Here we gauge
$U(1)$, which implies that the auxiliary field $A_\mu $ can be redefined
such that the second term of the first line of (\ref{rewrittenL}) is pure
auxiliary and can also be deleted. Then one is left with a \Ka\ manifold
with the \Ka\ potential ${\cal K}$. We are interested in the \Ka ian
action for $n$ independent complex scalars $z_i$ on the submanifold of the
$(n+1)$-complex-dimensional manifold defined by a constant value of
$\Nphi$. This is a real condition, but the $U(1)$ invariance implies that
another real variable disappears.\vspace{2mm}

\noindent \textbf{Application.}

In practice, the fields may have a further gauge-connection. In that case
the $\partial _\mu $ which we write here, can be replaced by a suitably
covariant derivative.

In $N=1$, the matrix $\Nphi_I{}^J$ emerges from the superconformal tensor
calculus as the second derivative of a scalar function $\Nphi$, and
(\ref{relNphi})  is the last equation of (\ref{N}). In $N=2$ this
treatment is appropriate in all cases where a prepotential $F(X)$
exists\footnote{When the prepotential does not exist \cite{CDFVP}, the
model is dual to one where the prepotential does exist \cite{whatspk},
although that does not guarantee equivalence for the classical action.}.
The matrix $\Nphi_I{}^J$ is the imaginary part of the second derivative
of the holomorphic prepotential, whose third derivative satisfies $X^K
F_{IJK}=0$ \cite{dWVPspec}. Note that the notation which we adapt here is
not the usual one for that case, but identifying further $\Nphi _I{}^J$
with $-2\rmi(F_{IJ}-F^*_{IJ})$, our formulation is also applicable to
$N=2$ as to $N=1$.

We will adopt the dilatational gauge-fixing condition
\begin{equation}
  \Nphi= -3M_P^2\,.
\label{gaugeDNPhi}
\end{equation}
Its normalization is chosen appropriately for $N=1$. For $N=2$ the
normalization changes due to contributions of a second compensating
multiplet to the $eR$ term in the action \cite{dWLVP}.

\vspace{2mm}

\noindent \textbf{Construction.}

We first combine the terms which include the auxiliary field $A_\mu $.
Writing the action (\ref{LscInPhi}) as $ {\cal L}(A_\mu )$, we have
\begin{eqnarray}
{\cal L}(A_\mu )&=& -\ft19 \Nphi (A_\mu -\tilde A_\mu )^2 + {\cal
L}(\tilde A_\mu )\nonumber\\ \tilde A_\mu &\equiv
&\frac{3\rmi}{2\Nphi}\left[ X^I\Nphi_I{}^J(\partial _\mu X_J)-
  (\partial _\mu X^I)\Nphi_I{}^J X_J\right]\,.
\label{Amubos}
\end{eqnarray}

With the coordinates as in (\ref{Xrhoapp}), the last part of
(\ref{relNphi}) implies that $\Nphi_I{}^J$ depends only on $z^i$ and
$z_i$, and not on $\Yrho $. The first condition (\ref{relNphi}) can be
used to obtain
\begin{eqnarray}
  \partial^i {\cal K} &=&-3\frac{x^I\Nphi_I{}^J
  \partial ^ix_J}{x^K\Nphi_K{}^Lx_L}\nonumber\\
  \tilde A_\mu &=&-\frac{\rmi}{2}\left( \partial ^i{\cal K}\,\partial _\mu z_i -
  \partial _i{\cal K}\,\partial _\mu z^i\right) +\frac{3\rmi}{2}\partial _\mu
  \ln \frac{\Yrho }{\Yrho ^*}\,.
\label{partialK}
\end{eqnarray}
The definitions (\ref{defNphiK}) imply
\begin{equation}
  \ln (-\Nphi)=-\ft{1}{3}{\cal K}+\ln (3\Yrho \Yrho ^*)
\label{lnN}
\end{equation}
and using $\tilde{\cal D}_\mu $ for (\ref{cDapp}) with $A$ replaced
by $\tilde A$, we have
\begin{equation}
  \tilde {\cal D}_\mu X_I= \Yrho\, \partial _\mu z_i\,  {\cal
  D}^ix_I+\ft12X_I \partial _\mu \ln \Nphi\,, \qquad
  {\cal D}^i x_I\equiv \left[\partial ^i
  +\ft{1}{3}(\partial ^i{\cal K})\right] x_I\,.
\label{DXI}
\end{equation}
Plugging this in the Lagrangian one obtains
\begin{equation}
    {\cal L}= -\ft14 \Nphi^{-1}\left( \partial _\mu \Nphi\right) ^2
 -\ft19\Nphi \left( A_\mu-\tilde A_\mu \right)^2
    -\Yrho \Yrho ^*\Nphi_I{}^J{\cal D}^ix_I\,{\cal D}_jx^J\,
   \partial  _\mu z_i\, \partial ^\mu z^j\,.
\label{LisKa}
\end{equation}
The last term already has the \Ka\ form, and by using (\ref{lnN}) and the
fact that the conditions (\ref{relNphi}) also imply, for example,
\begin{equation}
  \partial ^j\partial _i\left( x^I \Nphi_I{}^Jx_J\right) =
\left(\partial _i x^I\right)  \Nphi_I{}^J \left(  \partial
^jx_J\right)\,, \label{didjxNx}
\end{equation}
one can check that the \Ka\ potential is indeed ${\cal K}$. This finally
leads to (\ref{rewrittenL}).

When one uses the gauge-fixing (\ref{gaugeDNPhi}) and eliminates the
auxiliary field $A_\mu $, we obtain the \Ka\ action
\begin{equation}
   {\cal L}=-M_P^2\left(\partial ^j\partial _i {\cal K}\right)
   \partial  _\mu z^i\, \partial ^\mu    z_j\,.
\label{KahlerL}
\end{equation}
\vspace{2mm}

\noindent \textbf{Positivity}.

Note that there are positivity conditions restricting the domain of
the scalars, and the form of the matrix $\Nphi$. First of all in
order that the gauge condition (\ref{gaugeDNPhi}) can be satisfied,
\begin{equation}
  x^I \Nphi_I{}^J x_J <0 \,,
\label{signN}
\end{equation}
and thus $\Nphi_I{}^J(z,z^*)$ should have at least one negative
eigenvalue for all values of the scalars in the domain.

On the other hand, to have positive kinetic energy of the physical
scalars, one has to impose positivity of
\begin{equation}
\partial ^j\partial _i {\cal K} \propto -\Nphi\,(\partial _i x^I \Nphi_I{}^J\partial
^jx_J) +(\partial _i x^I \Nphi_I{}^K x_K) \, (x^L \Nphi_L{}^J\partial
^jx_J)\,. \label{poskinen}
\end{equation}
For this one needs the non-triviality condition
 that the matrix $\partial ^i X_I$ has to be of rank $n$.

\vspace{2mm}

\noindent \textbf{\Ka\ transformations and connections}.

In the gauge (\ref{choise}), the \Ka\ connection for a quantity depending
on $z$ (and/or $z^*$) depends on the chiral weight of the
corresponding quantity in the conformal approach. Consider an arbitrary
function $ {\cal V}(X,X^*)$. As it is a function of $X$, it does not
transform under the original \Ka\ transformation. Suppose now that ${\cal V}$
has (Weyl,chiral) weight $(w,c)$. We can write it as
$ {\cal V}(X,X^*)=\Yrho ^{w_+}(\Yrho ^*)^{w_-} V(z,z^*)$ where $w_\pm
=\ft12 w\mp \ft32 c$. Then the resulting \Ka\ transformation of
$V(z,z^*)$, taking into account (\ref{remainingU1}), is
\begin{equation}
  V'(z,z^*)=V(z,z^*)\exp\left[-\ft13(w_+\Lambda _\Yrho
  +w_-\Lambda ^*_\Yrho )\right]\,.
\label{V'}
\end{equation}
The covariant derivative of this quantity is
\begin{equation}
  {\cal D}^i V=\partial ^i V+\ft13w_+\,(\partial ^i {\cal K})\,V\,,
\label{defKcovD}
\end{equation}
and has the same weight under the \Ka\ transformations as $V$ in
(\ref{V'}). It satisfies
\begin{equation}
  (\partial ^I{\cal V})\,{\cal D}^ix_I=\Yrho^{w_+-1}(\Yrho^*)^{w_-}{\cal
  D}^i V\,.
\label{defcDV}
\end{equation}
The \Ka\ covariant derivatives also have Christoffel connections and the
covariant quantities are related to those in the superconformal
formulation by the following formulae\footnote{Note that although we
argued here for the definition of the covariant derivatives by using the
$U(1)$ gauge (\ref{choise}), this gauge has not been used, and these
formulae are independent of the $U(1)$ gauge. In the text they have been
used before the choice of gauge.}:
\begin{eqnarray}
\lefteqn{ {\cal D}^i{\cal D}^jx_I\equiv \partial ^i{\cal D}^jx_I+
\ft13  \left( \partial ^i {\cal K}\right) {\cal D}^jx_I -\Gamma ^{ij}_k{\cal
D}^kx_I  } \nonumber\\ && =- \Yrho
\Nphi^{-1}{}_I{}^J\Nphi_J{}^{KL}({\cal D}^ix_K)
  ({\cal D}^jx_L) \nonumber\\
\lefteqn{  {\cal D}^i{\cal D}^jW\equiv \partial ^i{\cal D}^jW+
  \left( \partial ^i {\cal K}\right) {\cal D}^jW-\Gamma ^{ij}_k{\cal D}^kW}
\nonumber\\ &&  =\Yrho ^{-1}M_P^3{\cal W}^{IJ}({\cal D}^ix_I)   ({\cal
D}^jx_J)+\Yrho ^{-2} M_P^3  {\cal W}^I{\cal D}^i{\cal
D}^jx_I\nonumber\\ \lefteqn{M_P^2\left( R_{k\ell }^{ij}-\ft23 g_k
^{(i}g_\ell^{j)}\right)=-\Yrho \Yrho ^*({\cal D}_\ell x^L)\Nphi_L{}^I
\partial _k{\cal D}^i{\cal D}^jx_I}\nonumber\\
&&=
\left(\Nphi^{IJ}_{KL}-\Nphi_{KL}^M\Nphi^{-1}{}_M{}^N\Nphi_N^{IJ}\right)
({\cal D}_\ell x^L)({\cal D}_k x^K)({\cal D}^ix_I)({\cal
D}^jx_J)(\Yrho \Yrho ^*)^2\,. \label{useful_for_transl}
\end{eqnarray}

\section{Calculation of $\mu $}
\label{app:calcmu} We consider here the case with real scalars, and
minimal \Ka\ potential $ {\cal K}= M_P^{-2}\phi \phi ^*$ for 1 chiral multiplet, such that $|A|^2=1$.
The definition of $\mu $ amounts to
\begin{equation}
  \mu =\ft12\rmi A\dot A^*=-\frac{\dot A_1}{2A_2}
   =\frac{1}{2\alpha ^2}(\dot \alpha _2\alpha _1-\dot \alpha _1\alpha _2)\,.
\label{muinalpha}
\end{equation}
With the real scalar $\phi_1 $ we have
\begin{eqnarray}
  m&=& \rme ^{{\cal K}/2}|W|\,, \qquad \dot m=  ({\cal D}_1 m) \dot \phi \equiv
  m_1\dot \phi  \nonumber\\
  \dot m_1&=&   \left( {\cal D}_1{\cal D}_1 m+M_P^{-2}m\right)
  \dot \phi = \left(m_{11}+m_{3/2}\right)\dot \phi
   \,.
\label{notmm1}
\end{eqnarray}
Note the appearance of the $m$ term in the last equation due to the
derivative on $\phi ^*$ in $ {\cal D}W$, or in other words, due to the
\Ka\ curvature, which is~1, implying non-commutativity of holomorphic and
antiholomorphic derivatives.

The expressions in section~\ref{ss:confInspBG} lead to
\begin{equation}
\alpha   =  \dot \phi ^2+m_1^2\,,\qquad \alpha_1 =  \dot
\phi ^2-m_1^2\,,\qquad \alpha _2 =2m_1\dot \phi \,, \label{simplealpha}
\end{equation}
and from this, the equation $\alpha ^2-\alpha _1^2=\alpha _2^2$ is
obvious. Then
\begin{equation}
  \mu  =\frac{-\ddot \phi  m_1+ \left(m_{11}+m_{3/2}\right)\dot \phi ^2}{ \dot \phi ^2+m_1^2} \,.
\label{finmu}
\end{equation}
The $\phi $ field equation~(\ref{scFE}) reduces to
\begin{equation}
\ddot \phi  =  -3\frac{\dot a}{a}\dot \phi  -m_1\left(-2
m_{3/2}+m_{11}\right)\,, \label{fesimpl}
\end{equation}
and we obtain
\begin{equation}
  \mu =\left(m_{11}+m_{3/2}\right)
  +3(H\dot \phi - m_{3/2}m_1)\frac{m_1}{\dot \phi ^2 +m_1^2}\,,
\label{muism2}
\end{equation}
the expression that we gave in \cite{GravProd}. For large $M_P$, and thus
small $m_{3/2}=mM_P^{-2}$, and small Hubble constant, this reduces to
$\mu =m_{11}=\partial _\phi ^2 W$.

\end{document}